\documentclass[11pt]{article}

\usepackage{subfigure}
\usepackage{amssymb}
\usepackage{amsfonts}
\usepackage{bm}
\usepackage{color}
\usepackage{calc}
\usepackage{amsfonts}
\usepackage{amsmath}
\usepackage[english]{babel}
\usepackage[T1]{fontenc}
\usepackage{verbatim}
\usepackage{enumerate}
\usepackage{graphicx}
\usepackage{natbib}
\usepackage{url}
\usepackage{hyperref}
\usepackage{algorithmic,algorithm}
\usepackage{placeins}
\usepackage{hyperref}

\newcommand{\revt}[1]{{\color{black}#1}}

\def\H{{\mathcal{H}}}
\def\O{{\mathcal{O}}}

\def\P{{\mathcal{P}}}
\def\Q{{\mathcal{Q}}}
\def\R{{\mathcal{R}}}
\def\E{{\mathcal{E}}}

\def\K{{\mathcal{K}}}
\def\L{{\mathcal{L}}}

\def\RR{{\mathbb{R}}}

\def\ZZ{{\mathbb{Z}}}

\def\e{{\mathbf e}}

\def\k{{\mathbf k}}
\def\x{{\bm x}}

\def\u{{\bm u}}

\def\x{{\mathbf x}}

\def\z{{\mathbf z}}

\def\u{{\mathbf u}}

\def\0{{\mathbf 0}}

\def\bomega{\boldsymbol{\omega}}
\def\bnabla{\boldsymbol{\nabla}}
\def\bDelta{\boldsymbol{\Delta}}

\def\Bmp#1{ \begin{minipage}{#1} }
\def\Emp{ \end{minipage} }
\def\Bmpc#1{ \begin{minipage}[c]{#1} }
\def\Bmpt#1{ \begin{minipage}[t]{#1} }
\def\Bmpb#1{ \begin{minipage}[b]{#1} }

\def\tTE{\widetilde{T}_{\E_0}}

\newcommand{\uvec}{\mathbf{u}}

\newcommand{\laplacian}{\Delta}
\newcommand{\rot}{\bnabla\times}

\newcommand{\tomega}{\widetilde{\omega}}
\newcommand{\tuE}{\widetilde{\mathbf{u}}_{\E_0}}

\newcommand{\tuET}{\widetilde{\uvec}_{0;\E_0,T}}
\newcommand{\tuEtT}{\widetilde{\uvec}_{0;\E_0,\tTE}}
\newcommand{\uET}{\uvec_{0;\E_0,T}}
\newcommand{\ET}{\E_T(\u_0)}
\newcommand{\M}[1]{\mathcal{S}_{#1}}

\newcommand{\argmax}{\operatorname{argmax}}

\newcommand{\Id}{\operatorname{Id}}

\newtheorem{problem}{Problem}[section]

\begin{document}
\title{Maximum Amplification of Enstrophy in 3D Navier-Stokes Flows}

\author{Di Kang, Dongfang Yun and Bartosz Protas\thanks{Email address for correspondence: bprotas@mcmaster.ca} 
\\ \\ 
Department of Mathematics and Statistics, McMaster University \\
Hamilton, Ontario, L8S 4K1, Canada
}

\date{\today}

\maketitle

\begin{abstract}
  This investigation concerns a systematic search for potentially
  singular behavior in 3D Navier-Stokes flows. Enstrophy serves as a
  convenient indicator of the regularity of solutions to the Navier
  Stokes system --- as long as this quantity remains finite, the
  solutions are guaranteed to be smooth and satisfy the equations in
  the classical (pointwise) sense. However, there are no estimates
  available with finite a priori bounds on the growth of enstrophy and
  hence the regularity problem for the 3D Navier-Stokes system remains
  open. In order to quantify the maximum possible growth of enstrophy,
  we consider a family of PDE optimization problems in which initial
  conditions with prescribed enstrophy $\E_0$ are sought such that the
  enstrophy in the resulting Navier-Stokes flow is maximized at some
  time $T$. Such problems are solved computationally using a
  large-scale adjoint-based gradient approach derived in the
  continuous setting. By solving these problems for a broad range of
  values of $\E_0$ and $T$, we demonstrate that the maximum growth of
  enstrophy is in fact finite and scales in proportion to $\E_0^{3/2}$
  as $\E_0$ becomes large. Thus, in such worst-case scenario the
  enstrophy still remains bounded for all times and there is no
  evidence for formation of singularity in finite time. We also
  analyze properties of the Navier-Stokes flows leading to the extreme
  enstrophy values and show that this behavior is realized by a series
  of vortex reconnection events.
\end{abstract}

\begin{flushleft}
  Keywords: Navier-Stokes equations, Singularity formation; Enstrophy growth; Variational optimization methods; vortex recommenction
\end{flushleft}



\section{Introduction}
\label{sec:intro}

The goal of this study is to assess the largest growth of enstrophy
possible in finite time in viscous incompressible flows in three
dimensions (3D). This problem is motivated by the question whether
solutions to the 3D incompressible Navier-Stokes system on unbounded
or periodic domains corresponding to smooth initial data may develop a
singularity in finite time \citep{d09}. By formation of a
``singularity'' we mean the situation when some norms of the solution
starting from smooth initial data become unbounded after a finite
time. This so-called ``blow-up problem'' is one of the key open
questions in mathematical fluid mechanics and, in fact, its importance
for mathematics in general has been recognized by the Clay Mathematics
Institute as one of its ``millennium problems'' \citep{f00}.  {Should
  such singular behavior indeed be possible in the solutions of the 3D
  Navier-Stokes problem, it would invalidate this system as a model of
  realistic fluid flows.}  Questions concerning global-in-time
existence of smooth solutions remain open also for a number of other
flow models including the 3D Euler equations \citep{gbk08} and some of
the ``active scalar'' equations \citep{k10}.

At the same time, it is known that suitably defined weak solutions,
which need not satisfy the Navier-Stokes system pointwise in space and
time, but rather in a certain integral sense only, exist globally in
time \citep{l34}. An important tool in the study of the global-in-time
regularity of classical (smooth) solutions are the so-called
``conditional regularity results'' stating additional conditions which
must be satisfied by a weak solution in order for it to also be a
smooth solution, i.e., to satisfy the Navier-Stokes system in the
classical sense as well. One of the best known results of this type,
due to \citet{ft89}, is based on the enstrophy $\E$ (see below,
cf.~\eqref{eq:EnsDef_3D}, for a precise definition {of this
  quantity}) of the time-dependent velocity field $\u(t)$ and asserts
that if the uniform bound
\begin{equation}
\label{eq:RegCrit_FoiasTemam}
\mathop{\sup}_{0 \leq t \leq T} \E(\u(t))  < \infty
\end{equation}
holds, then the regularity {and uniqueness of the solution $\u(t)$
  are} guaranteed up to time $T$ (to be precise, the solution remains
in a certain Gevrey class). Other {well-known} conditional regularity
results are given by the Ladyzhenskaya-Prodi-Serrin conditions in
which global-in-time regularity {follows} from certain integrability
criteria imposed in space and in time on the velocity field $\u(t)$
\citep{KisLad57,Prodi1959,Serrin1962}. These results were recently
extended and generalized by \citet{Gibbon2018} who derived analogous
conditions applicable to derivatives of various degrees of the
velocity field. {We add that \citet{Tao2016} recently showed that
  solutions to a certain suitably-averaged version of the
  Navier-Stokes equation may exhibit blow-up in finite time. One of
  the insights from this work is that understanding singular behavior
  in the Navier-Stokes flows will likely require more refined tools
  than the currently available techniques of harmonic analysis.}

From the {practical} point of view, the advantage of using the
conditional regularity result \eqref{eq:RegCrit_FoiasTemam} is that
the quantity it involves, the enstrophy $\E(\u(t))$, is very
convenient to work with, especially in {the context of numerical
  optimization problems. More specifically, since the enstrophy is a
  seminorm on a Hilbert space, formulation of such optimization
  problems for the Navier-Stokes system, which are the main tool to be
  used in this study, is relatively straightforward}. In addition, in
being directly related to vorticity, the enstrophy is also physically
meaningful. Condition \eqref{eq:RegCrit_FoiasTemam} implies that,
should singularity indeed form in finite time, then all Sobolev norms
of order higher than or equal one of the solution must blow up
simultaneously. In the context of the inviscid Euler system a
conditional regularity result similar to \eqref{eq:RegCrit_FoiasTemam}
is given by the Beale-Kato-Majda (BKM) criterion \citep{bkm84}. The
goal of the present investigation is to probe condition
\eqref{eq:RegCrit_FoiasTemam} computationally by constructing flow
evolutions designed to produce the largest possible increase of
enstrophy in some prescribed time $T$. Such worst-case behavior will
be {determined} systematically by solving a family of suitably defined
variational optimization problems.

While the blow-up problem is fundamentally a question in mathematical
analysis, a lot of computational studies have been carried out since
the mid-'80s in order to shed light on the hydrodynamic mechanisms
which might lead to singularity formation in finite time. Given that
such flows evolving near the edge of regularity involve formation of
very small flow structures, these computations typically require the
use of state-of-the-art computational resources available at a given
time. The computational studies focused on the possibility of
finite-time blow-up in the 3D Navier-Stokes and/or Euler system
include \cite{bmonmu83,ps90,b91,k93,p01,bk08,oc08,o08,ghdg08,
  gbk08,h09,opc12,bb12,opmc14,CampolinaMailybaev2018}, all of which
considered problems defined on domains periodic in all three
dimensions. The investigations by \cite{dggkpv13,k13,gdgkpv14,k13b}
focused on the time evolution of vorticity moments and compared it
against bounds on these quantities obtained using rigorous analysis.
{Recent computations by \citet{Kerr2018} considered a ``trefoil''
  configuration meant to be defined on an unbounded domain (although
  the computational domain was always truncated to a finite periodic
  box). A simplified semi-analytic model of vortex reconnection was
  recently developed and analyzed based on the Biot-Savart law and
  asymptotic techniques by
  \citet{MoffattKimura2019a,MoffattKimura2019b}.}  We also mention the
studies by \cite{mbf08} and \cite{sc09}, along with references found
therein, in which various complexified forms of the Euler equation
were investigated. The idea of this approach is that, since the
solutions to complexified equations have singularities in the complex
plane, singularity formation in the real-valued problem is manifested
by the collapse of the complex-plane singularities onto the real axis.
Overall, the outcome of these investigations is rather inconclusive:
while for the Navier-Stokes {system most of the} recent computations
do not offer support for finite-time blow-up, the evidence appears
split in the case of the Euler system.  In particular, the studies by
\cite{bb12} and \cite{opc12} hinted at the possibility of singularity
formation in finite time. In this connection we also highlight the
{computational} investigations by \cite{lh14a,lh14b} in which
blow-up was {documented} in axisymmetric Euler flows on a bounded
(tubular) domain. {Recently, \citet{ElgindiJeong2018} proved
  finite-time singularity formation in 3D axisymmetric Euler flows on
  domains exterior to a boundary with conical shape.}

A common feature of all of the aforementioned investigations was that
the initial data for the Navier-Stokes or Euler system was chosen in
an ad-hoc manner, based on some heuristic arguments. On the other
hand, in the present study we pursue a fundamentally different
approach, proposed originally by \cite{ld08} and employed also by
\cite{ap11a,ap13a,ap13b,ap16,Yun2018} for a range of related problems,
in which the initial data leading to the most singular behaviour is
sought systematically via solution of a suitable variational
optimization problem. In the present investigation we look for the
initial data which, subject to some constraints, will lead to flow
evolution maximizing the enstrophy growth {over} some prescribed time
{interval $[0,T]$, where} $0 < T < \infty$, with the intention of
verifying whether this growth could possibly become unbounded.  Since
the flow evolution is governed by the Navier-Stokes system {of partial
  differential equations (PDEs)}, this leads to a PDE-constrained
optimization problem for the initial data $\u_0$ {which is} amenable
to solution using a gradient {approach with} gradient information
obtained from the solutions of an adjoint system. The motivation for
this investigation comes from our earlier study \citep{ap16}, see also
\citet{ld08}, where families of vortex states maximizing the {\em
  instantaneous} rate of growth of enstrophy were found. Although
these vector fields did saturate the rigorous upper bounds on the
instantaneous rate of growth of enstrophy, this maximal growth was in
fact very rapidly depleted during the subsequent flow evolution,
resulting in a very small only increase of enstrophy {over finite
  times}. The main conclusion from this {result} is that if an
unbounded growth of enstrophy should be possible under the 3D
Navier-Stokes dynamics, it must be associated with ``instantaneously
suboptimal'' initial data which does not maximize the instantaneous
rate of enstrophy production. In the present investigation we embark
on a systematic search for such initial data.

It ought to be emphasized that solution of optimization problems
involving the 3D time-dependent Navier-Stokes system leads to very
challenging computational problems even at moderate Reynolds numbers.
We remark that, in order to establish a direct link with the results
of the mathematical analysis discussed below, in our investigation we
therefore follow a rather different strategy than in most of the
computational studies of extreme Navier-Stokes and Euler flows
referenced above.  While these earlier studies relied on data from a
relatively small number of simulations performed at a high (at the
given time) resolution, in the present investigation we explore a
broad range of cases, each of which is however computed at a more
moderate resolution (or, equivalently, Reynolds number).  With such an
approach to the use of available computational resources, we are able
to reveal trends resulting from the variation of {key} parameters
which otherwise would be hard to detect.  Systematic computations
conducted in this way thus allow us {to establish sharpness of
  various a priori estimates relevant for a given problem; if these
  estimates turn out not to be sharp, or are not available, then such
  results can help formulate ``targets'' for what can potentially be
  proved.}

By addressing the question about the maximum growth of enstrophy
possible in finite time in the 3D Navier-Stokes system, the present
investigation represents an important milestone in our long-term
research program in which analogous questions have also been
considered in the context of more tractable problems involving the
one-dimensional (1D) Burgers equation and the two-dimensional (2D)
Navier-Stokes {system}.  Although global-in-time existence of
classical (smooth) solutions is well known for both these problems
\citep{kl04}, questions concerning the sharpness of the corresponding
estimates for the instantaneous and finite-time growth of various
enstrophy-like quantities are relevant, because these estimates are
obtained using essentially the same methods as employed to derive
their 3D counterparts. Since in 2D flows on unbounded or periodic
domains the enstrophy may not increase ($d\E/dt \leq 0$), the relevant
quantity in this case is the palinstrophy $\P(\u(t)) :=
\frac{1}{2}\int_\Omega | \bnabla\bomega(t,\x) |^2 \,d\x$, where
$\bomega :=\rot\u$ is the vorticity (which reduces to a pseudo-scalar
in 2D). Questions concerning sharpness of the different estimates
{obtained with energy-type methods and} considered in this research
program are summarized together with the results obtained to date in
Table \ref{tab:estimates}. We remark that {for the 1D Burgers problem
  the maximum growth of enstrophy in finite time found as a function
  of the initial enstrophy $\E_0$ by solving a suitable constrained
  PDE optimization problem does not saturate the upper bound in the
  corresponding estimate {which states that $\max_{t>0} \E(u(t)) <
    \O(\E_0^3)$,} indicating that this estimate may be improved}
\citep{ap11a}. {We note that sharper bounds were independently
  obtained by \citet{Biryuk2001} and \cite{p12} using different
  techniques not relying on energy methods. They predict the maximum
  finite-time growth of enstrophy to scale as $\O(\E_0^{3/2})$, which
  is the behavior actually observed in computations by \citet{ap11a},
  but impose more stringent assumptions on the regularity of the
  initial data.}  On the other hand, in 2D the bounds on both the
\emph{instantaneous} and \emph{finite-time} growth of palinstrophy
were found to be sharp and, somewhat surprisingly, both estimates were
realized by the same family of incompressible vector fields
parameterized by energy $\K$ and palinstrophy $\P$, obtained as the
solution of an {\em instantaneous} optimization problem
\citep{ap13a,ayala_doering_simon_2018}.  Thus, somewhat paradoxically,
the results currently available for the 2D Navier-Stokes system are in
fact more satisfactory than the results available for the 1D Burgers
system.  We add that what distinguishes the 2D problem in regard to
both the instantaneous and finite-time bounds is that the right-hand
sides (RHS) of these bounds are expressed in terms of two quantities,
namely, energy $\K$ and {palinstrophy $\P$}, in contrast to the
enstrophy alone appearing in the 1D and 3D estimates. As a result, the
2D instantaneous optimization problem had to be solved subject to {\em
  two} constraints. Insights concerning the maximum growth of
enstrophy in the 1D Burgers equation in the presence of stochastic
excitations were provided by \citet{PocasProtas2018}. Bounds on the
instantaneous rate of growth of enstrophy in the 1D fractional Burgers
equation, which is known to exhibit a finite-time singularity
formation in the supercritical regime \citep{kns08}, were derived by
\citet{Yun2018} who also analyzed the sharpness of these bounds.

\begin{table}
  \begin{center}
    \hspace*{-1.1cm}
    \begin{tabular}{l|c|c}      
      &  \Bmp{3.0cm} \small \begin{center} {\sc Estimate} \\ \smallskip \end{center} \Emp   
      & \Bmp{3.5cm} \small \begin{center} {\sc Realizability }  \end{center} \Emp \\  
      \hline
      \Bmp{2.5cm}  \small {\begin{center} \smallskip 1D Burgers  \\ instantaneous \smallskip \end{center}} \Emp &  
      \small {$\frac{d\E}{dt} \leq \frac{3}{2}\left(\frac{1}{\pi^2\nu}\right)^{1/3}\E^{5/3}$}  & 
      \Bmp{3.5cm} \footnotesize {\begin{center} \smallskip {\sc Yes} \\ \citep{ld08}  \smallskip  \end{center}} \Emp \\ 
      \hline 
      \Bmp{3.0cm} \small {\begin{center} \smallskip 1D Burgers  \\ finite-time \smallskip \end{center}} \Emp &  
      \small {$\max_{t \in [0,T]} \E(u(t)) \leq \left[\E_0^{1/3} + \frac{1}{16}\left(\frac{1}{\pi^2 \nu}\right)^{4/3}\E_0\right]^{3}$} &  
      \Bmp{3.5cm} \small {\begin{center} \smallskip {\sc No} \\ \citep{ap11a} \smallskip  \end{center}} \Emp \\ 
      \hline 
      \Bmp{3.0cm} \small {\begin{center} \smallskip 2D Navier-Stokes  \\ instantaneous \smallskip\end{center}} \Emp &  
      \Bmp{7.0cm} \smallskip \centering \small $\frac{d\P}{dt}  \le -\nu\frac{\P^2}{\E} + \frac{C_1}{\nu} \E\,\P$ \\ 
      \smallskip $\frac{d\P}{dt} \le C_2\sqrt{\log\left(\K^{1/2}/\nu\right)}\,  \P^{3/2}$ \smallskip \Emp& 
      \Bmp{3.5cm} \small {\begin{center} \smallskip {\sc Yes} \\ \citep{ap13a,ayala_doering_simon_2018}  \smallskip  \end{center}} \Emp  \\ 
      \hline 
      \Bmp{3.0cm} \small \begin{center} \smallskip 2D Navier-Stokes \\ finite-time \smallskip\end{center} \Emp &  
      \Bmp{7.0cm} \smallskip \centering \small $\max_{t>0}  \P(\u(t)) \le \P_0 + \frac{C_1}{2\nu^2}\E_0^2$ \\ 
      \smallskip $\max_{t>0} \P(\u(t)) \le \left(\P_0^{1/2} + \frac{C_2}{4\nu^2}\K_0^{1/2}\E_0\right)^2$ \smallskip \Emp  & 
      \Bmp{3.5cm} \small {\begin{center} \smallskip {\sc Yes} \\ \citep{ap13a,ayala_doering_simon_2018}  \smallskip  \end{center}} \Emp \\ 
      \hline 
      \Bmp{3.0cm} \small {\begin{center} \smallskip 3D Navier-Stokes  \\ instantaneous  \smallskip \end{center}} \Emp &  
      \small {$\frac{d\E}{dt} \leq \frac{27}{8\,\pi^4\,\nu^3} \E^3$} & \Bmp{3.5cm} \small {\begin{center} \smallskip {\sc Yes} \\ \citep{ld08} \smallskip  \end{center}} \Emp  \\ 
      \hline 
      \Bmp{3.0cm} \small \begin{center} \smallskip\smallskip 3D Navier-Stokes  \\ finite-time \smallskip \end{center} \Emp &  
      \Bmp{7.0cm} \smallskip \centering \small  $\E(\u(t)) \le \frac{\E_0}{\sqrt{1 - 4 \frac{C \E_0^2}{\nu^3} t}}$  \Emp & \Bmp{3.5cm} \centering {???} \smallskip\smallskip \Emp \\
\hline
    \end{tabular}
  \end{center}
  \caption{Summary of selected estimates for the instantaneous 
    rate of growth and the growth over finite time of enstrophy \revt{$\E$} 
    and palinstrophy \revt{$\P$} in 1D Burgers, 2D and 3D Navier-Stokes systems. 
{All of these estimates are obtained using similar energy-type methods.} The quantities $\K$ 
\revt{(kinetic energy)} and $\E$ are defined in \eqref{eq:EnerDef_3D} and \eqref{eq:EnsDef_3D}, {respectively}.}
  \label{tab:estimates}
\end{table}

We remark that in the research program outlined above we seek to
systematically identify ``extreme'' solutions which may saturate the
different bounds given in Table \ref{tab:estimates}. However, a
complementary approach to quantify extreme behavior of a broad class
of dynamical systems was recently developed as a generalization of the
``background method'' of \citet{DoeringConstantin1992}. It relies on
computation of an optimal Lyapunov functional which under the
sum-of-squares approximation reduces to solution of a convex
semidefinite optimization problem \citep{Chernyshenkoetal2014}. To
date, this approach has been used to obtain new results concerning the
average and extreme behavior of some simple models, both in finite and
infinite dimension
\citep{Tobascoetal2018,Goluskin2018,Goluskin2019,FantuzziGoluskin2019}.

In this study we construct {two families} of optimal initial data
parameterized by the initial enstrophy $\E_0$ and the length of the
time window $T$ for the Navier-Stokes system on a 3D periodic domain
which produce the largest possible {growth of} enstrophy $\E(\u(T))$
at the prescribed time $0 < T < \infty$. {The two families are
  associated with symmetric and asymmetric states and dominate in
  terms of the enstrophy growth, respectively, for small and large
  initial enstrophies $\E_0$.}  Our computations based on solution of
the corresponding PDE-constrained optimization problems demonstrate
that for a given value of $\E_0$, there exists an optimal time $\tTE$
such that the maximum growth $\max_{0\le t \le \tTE} \E(\u(t))$ is
largest and, when the initial enstrophy $\E_0$ is sufficiently large,
$\tTE$ decreases with $\E_0$.  Moreover, the maximum (``worst-case'')
growth of enstrophy {realized by asymmetric initial conditions} scales
as $\max_{T>0} \max_{0\le t \le T} \E(\u(t)) \sim C \, \E_0^{3/2}$ for
a broad range of initial enstrophy values $\E_0 \in [100, 1000]$ and
some constant $C$, suggesting global boundedness of this quantity,
cf.~condition \eqref{eq:RegCrit_FoiasTemam}, and, consequently, global
existence of smooth solutions \citep{ft89}. In the limit of large
initial enstrophy $\E_0$, the initial conditions responsible for the
worst-case growth of enstrophy have the form of three perpendicular
pairs of anti-parallel vortex tubes, {whereas the corresponding flow
  evolutions feature a sequence of reconnection events.}

The structure of the paper is as follows: in the next section we
review key estimates characterizing the growth of enstrophy, both
instantaneously and in finite time, {in 3D Navier-Stokes flows}
emphasizing the relation of these bounds to the question of global
{existence of smooth solutions},
cf.~\eqref{eq:RegCrit_FoiasTemam}; then, in \S \ref{sec:optim} we
formulate a variational optimization problem designed to probe the
{worst-case} growth of enstrophy in finite time and a numerical
approach to solve this problem is introduced in \S \ref{sec:numer};
our computational results are presented in \S \ref{sec:results},
whereas final comments and conclusions are deferred to \S
\ref{sec:final}.

\section{Bounds on the Growth of Enstrophy}
\label{sec:bounds}

We consider the incompressible Navier-Stokes system defined on the 3D
unit cube $\Omega = [0,1]^3$ with periodic boundary conditions
\begin{subequations}\label{eq:NSE3D}
\begin{alignat}{2}
\partial_t\u + \u\cdot\bnabla\u + \bnabla p - \nu\laplacian\u & = 0 & &\qquad\mbox{in} \,\,\Omega\times(0,T], \\
\bnabla\cdot\u & = 0 & & \qquad\mbox{in} \,\,\Omega\times[0,T], \\
\u(0) & = \u_0, &   &
\end{alignat}
\end{subequations}
where the vector $\u = {[u_1, u_2, u_3]^T}$ is the velocity
field, $p$ is the pressure and $\nu>0$ is the coefficient of kinematic
viscosity (hereafter we will set $\nu=0.01$ which is the same value as
used in earlier studies of closely-related problems
\citep{ld08,ap16}). The velocity gradient $\bnabla\u$ is the tensor
with components $[\bnabla\u]_{ij} = \partial_j u_i$, $i,j=1,2,3$.
The fluid density $\rho$ is assumed constant and equal to unity
($\rho=1$). The relevant properties of solutions to system
\eqref{eq:NSE3D} can be studied using energy methods, with the energy
$\K(\u(t))$ and its rate of growth given by
\begin{eqnarray}
  \K(\u(t)) & := & \frac{1}{2}\int_\Omega |\u(t,\x)|^2 \,d\x, \label{eq:EnerDef_3D}\\
  \frac{d\K(\u(t))}{dt} & = & -\nu\int_\Omega |\nabla\u(t,\x)|^2 \, d\x, \label{eq:dK/dt_3D}
\end{eqnarray}
where ``$:=$'' means ``equal to by definition''. The enstrophy
$\E(\u(t))$ and its rate of growth are given by\footnote{{We note
    that unlike energy, cf.~\eqref{eq:EnerDef_3D}, enstrophy is often
    defined without the factor of 1/2. However, for consistency with
    earlier studies belonging to this research program
    \citep{ap11a,ap13a,ap13b,ap16,Yun2018}, we choose to retain this
    factor here.}}
\begin{eqnarray}
\E(\u(t)) & := & \frac{1}{2}\int_\Omega | \rot\u(t,\x) |^2 \,d\x, \label{eq:EnsDef_3D}\\
\frac{d\E(\u(t))}{dt} & = & -\nu\int_\Omega |\laplacian\u|^2\,d\x  + 
\int_{\Omega} \u\cdot\nabla\u\cdot\laplacian\u\, d\x =: \R(\u(t)) . \label{eq:dEdt}
\end{eqnarray}
For incompressible flows with periodic boundary conditions we also
have the following identity \citep{dg95}
\begin{equation}
\int_{\Omega} |\rot\u|^2\,d\x = \int_{\Omega} |\nabla\u|^2\,d\x.
\label{eq:duL2}
\end{equation}
Hence, combining \eqref{eq:EnerDef_3D}--\eqref{eq:duL2}, {the
  following system of ordinary differential equations is obtained for
  energy and enstrophy}
\begin{subequations}
\label{eq:dKdtdEdt}
\begin{align}
\frac{d\K(\u(t))}{dt} & =  -2\nu \E(\u(t)), \label{eq:dKdt_system}\\
\frac{d\E(\u(t))}{dt} & =  \R(\u(t)). \label{eq:dEdt_system}
\end{align}
\end{subequations}

A standard approach at this point is to try to bound $d\E / dt$ and
using {classical} techniques of functional analysis it is
possible to obtain the following well-known estimate in terms of $\K$
and $\E$ \citep{d09}
\begin{equation}\label{eq:dEdt_estimate_KE}
\frac{d\E}{dt} \leq -\nu \frac{\E^2}{\K} + \frac{c}{\nu^3} \E^3,
\end{equation} 
{where $c$ is} {a known} constant. A related estimate expressed
entirely in terms of the enstrophy $\E$ is given by
\begin{equation}
\frac{d\E}{dt} \leq \frac{27}{8\,\pi^4\,\nu^3} \E^3. 
\label{eq:dEdt_estimate_E}
\end{equation} 
By simply integrating the differential inequality in
\eqref{eq:dEdt_estimate_E} with respect to time we obtain the
finite-time bound
\begin{equation}
\E(\u(t)) \leq \frac{\E_0}{\sqrt{1 - \frac{27}{4\,\pi^4\,\nu^3}\,\E_0^2\, t}} 
\label{eq:Et_estimate_E0}
\end{equation}
which clearly becomes infinite at time $t_0 = 4\,\pi^4\,\nu^3 / (27\,
\E_0^2)$. Thus, based on estimate \eqref{eq:Et_estimate_E0}, it is not
possible to establish the boundedness of the enstrophy $\E(\u(t))$
{and hence also the regularity of solutions} globally in time.
Therefore, the question about the finite-time singularity formation
can be recast in terms of whether or not {there exists initial
  data $\u_0$ with enstrophy $\E_0 < \infty$ such that in the
  resulting flow evolution the enstrophy $\E(\u(t))$ becomes unbounded
  in finite time, as allowed by estimate \eqref{eq:Et_estimate_E0}.}
A systematic search for such {worst-case} initial data using
variational optimization methods is the {main} theme of this study. We
add that while the analysis presented here was carried out based on
the vorticity and enstrophy, an inequality analogous to
\eqref{eq:dEdt_estimate_E} can also be obtained in terms of strain,
i.e., the symmetric part of the velocity gradient {$\bnabla\u$},
resulting in a smaller value of the constant prefactor
\citep{Miller2017}.

\begin{figure}
\begin{center}
\mbox{
\Bmp{0.45\textwidth}
\subfigure[]{\includegraphics[width=1.0\textwidth]{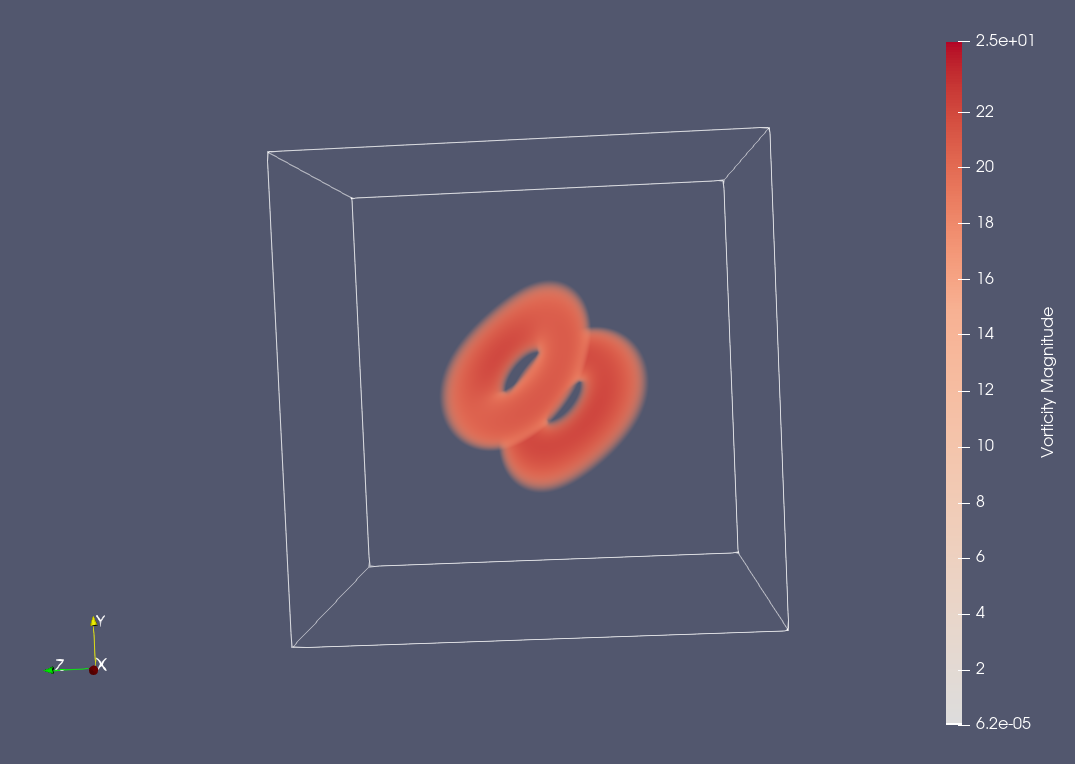}}
\Emp
\Bmp{0.525\textwidth}
\subfigure[]{\includegraphics[width=1.0\textwidth]{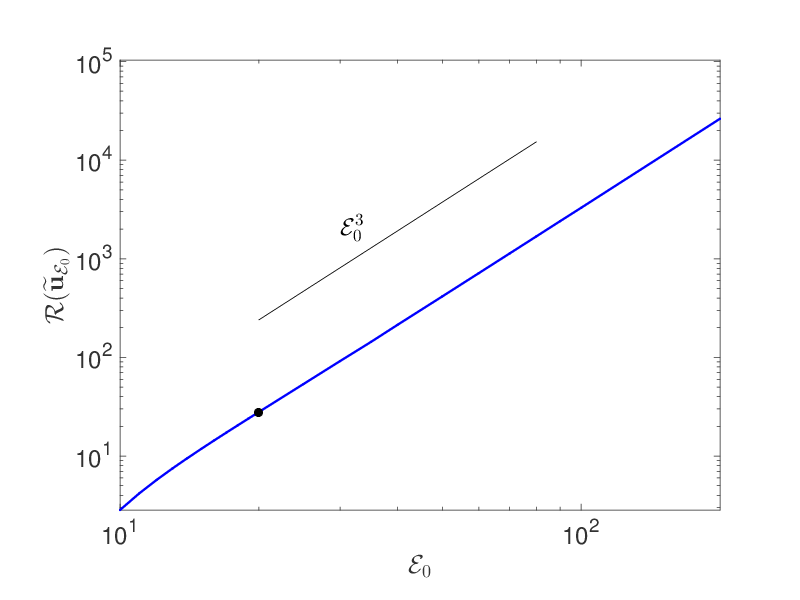}}
\Emp
}
\caption{(a) Extreme vortex state $\tuE$ obtained by \citet{ap16},
  {see also \citet{ld08}, as a solution of the instantaneous
    optimization Problem \ref{pb:maxdEdt_E} for $\E_0 = 20$}; shades
  of red correspond to the magnitude of the vorticity $|\left(\bnabla
    \times \tuE\right)(\x)|$ (see the color bar). (b) {The}
  maximum rate of growth of enstrophy $\R(\tuE)$ {as a function
    of the enstrophy $\E_0$ (the black solid symbol corresponds to the
    value of $\E_0$ characterizing the extreme vortex state shown in
    panel (a))}.}
\label{fig:tuE}
\end{center}
\end{figure}

The question of sharpness of the instantaneous bound
\eqref{eq:dEdt_estimate_E} was addressed in the seminal study by
\cite{ld08}, see also \cite{l06}, and further elaborated by
\citet{ap16}, who constructed a family of divergence-free velocity
fields $\tuE$ parameterized by the enstrophy $\E_0$ which saturate
this bound. These fields were obtained by {numerically solving the
  following variational optimization problem}
\begin{problem}\label{pb:maxdEdt_E}
  Given $\E_0\in\mathbb{R}_+$ and the objective functional $\R(\u)$
  from equation \eqref{eq:dEdt}, find
\begin{align*}
  \tuE & =  \mathop{\arg\max}_{\u\in\M{\E_0}} \,  \R(\u), \quad \text{where} \\
  \M{\E_0} & = \left\{\u\in H^2(\Omega)\,\colon\,\bnabla\cdot\u = 0,
    \; \E(\u) = \E_0 \right\}
\end{align*} 
\end{problem}
\noindent
for the enstrophy $\E_0$ spanning a broad range of values. {Since
  Problem \ref{pb:maxdEdt_E} is in general non-convex,
  ``$\mathop{\arg\max}$'' represents a {\em local} maximizer, which
  might also be global. The symbol} $H^2(\Omega)$ denotes the Sobolev
space of functions with square-integrable second derivatives endowed
with the inner product \citep{af05}
\begin{equation}
\forall\,\mathbf{z}_1, \mathbf{z}_2 \in H^2(\Omega) \qquad 
\Big\langle \mathbf{z}_1, \mathbf{z}_2 \Big\rangle_{H^2(\Omega)}
= \int_{\Omega} \mathbf{z}_1 \cdot \mathbf{z}_2 
+ \bnabla \mathbf{z}_1 \colon \bnabla \mathbf{z}_2
+ \Delta \mathbf{z}_1 \cdot \Delta \mathbf{z}_2  \, d\x
\label{eq:ipH2} 
\end{equation}
{which, for simplicity, was chosen in a somewhat nonstandard form
  as it does not involve mixed derivatives in the last term.}  For
sufficiently large values of $\E_0$ the thus obtained instantaneously
optimal fields $\tuE$ have the form of a pair of colliding vortex
rings, cf.~figure \ref{fig:tuE}(a). The corresponding rate of growth
{of enstrophy} $d\E/dt$ was found to be proportional to $\E^3$,
cf.~figure \ref{fig:tuE}(b), demonstrating that estimate
\eqref{eq:dEdt_estimate_E} is sharp up to a numerical prefactor.
However, the sharpness of the instantaneous estimate alone does not
allow us to conclude about the possibility of singularity formation,
because for this situation to occur, a sufficiently large enstrophy
growth rate would need to be sustained over a {\em finite} time window
$[0,t_0)$. At the same time, it was shown by \citet{ap16} that when
the extreme vortex states $\tuE$ are used as the initial data $\u_0$
in the Navier-Stokes system \eqref{eq:NSE3D}, the initially maximal
rate of growth of enstrophy is immediately depleted producing only a
very modest increase $\max_{t \ge 0}\E(\u(t)) - \E_0$ during
subsequent evolution. We note that in the limit $\E_0 \rightarrow
\infty$ the extreme vortex states $\tuE$ are given by pairs of
axisymmetric vortex rings without swirl and such flows are known to be
globally well-posed in the classical sense \citep{Feng2015}. The fact
that singularity {cannot} form in such axisymmetric configurations
with no vorticity on the axis can also be deduced from the celebrated
Caffarelli-Kohn-Nirenberg theorem \citep{CaffarelliKohnNirenberg1982}.

{In addition to sharpness under worst-case conditions, another
  question pertaining to inequality \eqref{eq:dEdt_estimate_E} is
  whether the upper bound on its RHS can also be realized under
  generic conditions in turbulent flows. This problem was studied by
  \citet{Schumacher2010} who demonstrated that in turbulent flows the
  rate of change of ensemble-averaged squared vorticity grows at most
  as $\frac{d}{dt}\langle \bomega^2 \rangle \sim \langle \bomega^2
  \rangle^{3/2}$, where $\langle \cdot \rangle$ denotes ensemble
  averaging. This observation is consistent with the statistical
  theory of turbulence, more specifically, the K{\'a}rm{\'a}n-Howarth
  equation \citep{davidson:turbulence}.}

The key conclusion from the results recalled above is that if a
significant, let alone unbounded, growth of enstrophy is to be
achieved in finite time, it must be associated with initial data
$\u_0$ other than the extreme vortex states $\tuE$ saturating the
upper bound in estimate \eqref{eq:dEdt_estimate_E} on the
instantaneous rate of growth of enstrophy, cf.~figure \ref{fig:tuE}.
More {specifically}, assuming the instantaneous rate of growth of
enstrophy in the form $d\E / dt = C \, \E^{\alpha}$ {for some
  prefactor} $C>0$, any exponent $\alpha > 2$ will {lead to} blow-up
of $\E(\u(t))$ {at some} finite time {$t_0 = t_0(\alpha)$}
if this rate of growth is sustained over the interval $[0,t_0)$.  The
fact that there is no blow-up {when} $1 < \alpha \le 2$ follows from
the {observation} that one factor of $\E$ in
\eqref{eq:dEdt_estimate_E} can be bounded in terms of the initial
energy $\K_0 := \K(\u_0)$ using \eqref{eq:dKdt_system} as follows
\begin{equation}
\int_0^t \E(\u(s))\, ds = \frac{1}{2\nu} \left[ \K_0 - \K(\u(t))\right] \leq \frac{1}{2\nu} \K_0,
\label{eq:Kt}
\end{equation}
{which upon employing Gr\"onwall's lemma yields the bound
\begin{equation}
\max_{0 \le t \le T} \E(\u(t))  \le  \E_0\, \exp\left[\int_0^T  \E(\u(s))\, ds\right] \le \E_0 \, \exp\left[\frac{1}{2\nu} \K_0\right]
\label{eq:Gronwall}
\end{equation}
valid for $0 \le \alpha \le 2$.}  Evidently, as the rate of growth of
enstrophy slows down {and} $\alpha \rightarrow 2^+$, {for blow-up to
  occur a certain minimum growth rate} must be sustained over windows
of time with increasing length, i.e., $t_0 \rightarrow \infty$. {To
  assess the feasibility of such a scenario,} in the next section we
introduce an optimization approach which will allow us to
systematically search for worst-case flow evolutions {producing the
  maximum possible growth of enstrophy in finite time}.

\section{Maximization Problem}
\label{sec:optim}

In order to probe the upper bound in estimate
\eqref{eq:Et_estimate_E0} for realizability, our objective in this
section is to construct initial data $\tuET$ for the Navier-Stokes
system \eqref{eq:NSE3D} with {the} prescribed enstrophy $\E_0 > 0$,
such that at {the given} time $T>0$ the corresponding flow evolution
{will produce the maximum possible} value of enstrophy $\E(\u(T))$
{under the assumption that the Navier-Stokes system admits smooth
  (classical) solutions on the time interval $(0,T]$}.  Defining the
objective function $\E_T \; : \; H^1(\Omega) \rightarrow \RR_+$, where
\begin{equation}
\ET := \E(\u(T)) = \frac{1}{2}\int_\Omega | \rot\u(T,\x) |^2 \,d\x,
\label{eq:ET}
\end{equation}
we thus arrive at the following
\begin{problem}\label{pb:maxET}
  Given $\E_0, T \in\mathbb{R}_+$ and the objective functional $\ET$ from
  equation \eqref{eq:ET}, find
\begin{align*}
\tuET & =  \mathop{\arg\max}_{\u_0 \in \Q_{\E_0}} \, \ET, \quad \text{where} \\
\Q_{\E_0} & =  \left\{\u\in H^1(\Omega)\,\colon\,\bnabla\cdot\u = 0, \; \E(\u) = \E_0 \right\},
\end{align*} 
\end{problem}
\noindent
where $H^1(\Omega)$ denotes the Sobolev space of functions with
square-integrable derivatives endowed with the inner product
\citep{af05}
\begin{equation}
\forall\,\mathbf{z}_1, \mathbf{z}_2 \in H^1(\Omega) \qquad 
\Big\langle \mathbf{z}_1, \mathbf{z}_2 \Big\rangle_{H^1(\Omega)}
= \int_{\Omega} \mathbf{z}_1 \cdot \mathbf{z}_2 
+ \ell_1^2 \,\bnabla \mathbf{z}_1 \colon \bnabla \mathbf{z}_2 \, d\x,  
\label{eq:ipH1} 
\end{equation}
where $\ell_1\in \RR_+$ is a parameter with the meaning of a
{characteristic} length scale (the reasons for introducing this
parameter in the definition of the inner product will become clear
below). The inner product in the space $L^2(\Omega)$ is obtained from
\eqref{eq:ipH1} by setting $\ell_1 = 0$. The maximizers in Problem
\ref{pb:maxET} are constrained to belong to the manifold $\Q_{\E_0}$
which represents the intersection of the subspace of divergence-free
vector fields $(\bnabla\cdot\u = 0)$ and a nonlinear manifold defined
by the enstrophy constraint $(\E(\u) = \E_0)$ in the Sobolev space
$H^1(\Omega)$, where the smoothness requirement is necessary to ensure
that the initial enstrophy $\E(\u_0)$ is well defined. Without loss of
generality, we assume that $\int_{\Omega} \u_0 \, d\x = \0$, a
property which is also invariant during the flow evolution. {The
  fact that Problem \ref{pb:maxET} admits solutions is a consequence
  of the assumption that with the given parameters $\E_0$ and $T$
  solutions of the Navier-Stokes system \eqref{eq:NSE3D} are smooth on
  the time interval $[0,T]$.}

The key insight we seek to deduce is how the maximum growth of
enstrophy {$\max_{T>0}\, \E_T(\tuET)$} obtained for time intervals
$(0,T]$ with different lengths $0 < T < \infty$ scales with the
initial enstrophy $\E_0$. In order to evaluate {this quantity} for a
given value of the initial enstrophy $\E_0$, we thus need to solve
Problem \ref{pb:maxET} for different values of {$T \in (0,
  T_{\text{max}}]$, where $T_{\text{max}} < \infty$ is the maximum
  considered length of the time interval,} and with fixed $\E_0$, so
that the maximum with respect to $T$ can be evaluated. {We note
  that this approach is justified by the existence of bounds
  (expressed in terms of norms of the initial data $\u_0$ and
  viscosity $\nu$) on the largest time $T_{\text{max}}$ when a
  singularity might occur \citep{Ohkitani2016}.} Assuming that the
optimal initial data $\tuET$ {obtained as solution of Problem
  \ref{pb:maxET}} has (at least piecewise) continuous dependence on
the parameter $T$, we will refer to the mapping $T \longmapsto \tuET$
with $\E_0$ fixed as a ``maximizing branch''.

In order to solve Problem \ref{pb:maxET} for given values of $\E_0$
and $T$ we adopt an ``optimize-then-discretize'' approach \citep{g03}
in which a gradient method is first formulated in {the}
infinite-dimensional (continuous) setting and only then the resulting
{equations and} expressions are discretized for the purpose of
numerical solution. An essentially identical approach was used by
\citet{ap11a} to solve an optimization problem analogous to Problem
\ref{pb:maxET}, but formulated for the 1D Burgers equation. The
maximizer $\tuET$ can be found as $\tuET = \lim_{n\rightarrow \infty}
\uET^{(n)}$ using the following iterative procedure representing a
discretization of a gradient flow projected on
{$\mathcal{Q}_{\E_0}$}
\begin{equation}
\begin{aligned}
\uET^{(n+1)} & =  \mathbb{P}_{\mathcal{Q}_{\E_0}}\left(\;\uET^{(n)} + \tau_n \nabla\E_T\left(\uET^{(n)}\right)\;\right), \\ 
\uET^{(1)} & =  \u^0,
\end{aligned}
\label{eq:desc}
\end{equation}
where $\uET^{(n)}$ is an approximation of the maximizer obtained at
the $n$-th iteration, $\mathbb{P}_{\mathcal{Q}_{\E_0}} \; : \;
H^1(\Omega) \rightarrow \Q_{\E_0}$ is the projection operator, $\u^0$
is the initial guess and $\tau_n$ is the length of the step.  A key
element of the iterative procedure \eqref{eq:desc} is the evaluation
of the gradient $\nabla\ET$ of the objective functional $\ET$, cf.
\eqref{eq:ET}, representing its (infinite-dimensional) sensitivity to
perturbations of the initial data $\u_0$ {in the governing system
  \eqref{eq:NSE3D}. We emphasize that it is essential for the gradient
  to possess} the required regularity, namely, $\nabla\E_T(\u_0) \in
H^1(\Omega)$.

The first step to determine the gradient $\nabla\ET$ is to consider
the G\^{a}teaux (directional) differential $\E'_T(\u_0;\cdot) \; : \;
H^1(\Omega) \rightarrow \RR$ of the objective functional $\ET$ defined
as $\E'_T(\u_0;\u_0') = \lim_{\epsilon \rightarrow 0}
\epsilon^{-1}\left[\E_T(\u_0+\epsilon \u_0') - \ET\right]$ for some
arbitrary perturbation $\u_0' \in H^1(\Omega)$. The gradient
$\nabla\ET$ can {then} be extracted from the G\^{a}teaux
differential $\E'_T(\u_0;\u_0')$ recognizing that, when viewed as a
function of its second argument, this differential is a bounded linear
functional on the space $H^1(\Omega)$ and we can therefore invoke the
Riesz representation theorem \citep{l69}
\begin{equation}
\E'_T(\u_0;\u_0')
= \Big\langle \nabla^{L_2}\ET, \u_0' \Big\rangle_{L^2(\Omega)} = \Big\langle \nabla\ET, \u_0' \Big\rangle_{H^1(\Omega)},
\label{eq:riesz}
\end{equation}
where the gradient $\nabla\ET$ is the Riesz representer in the
function space $H^1(\Omega)$. In \eqref{eq:riesz} we also formally
defined the gradient $\nabla^{L_2}\ET$ computed with respect to the
$L^2$ topology as it will be useful in subsequent computations. Given
the definition of the objective functional in \eqref{eq:ET}, its
G\^{a}teaux differential can be expressed as
\begin{equation}
\E'_T(\u_0;\u_0') = \int_\Omega (\rot\u(T,\x))\cdot(\rot\u'(T,\x))  \,d\x 
= \int_\Omega \bDelta\u(T,\x))\cdot\u'(T,\x)  \,d\x,
\label{eq:dET}
\end{equation}
where the last equality follows from integration by parts and the
vector identity $\bnabla\times(\bnabla\times\z) =
\bnabla(\bnabla\cdot\z) - \bDelta\z$, whereas the perturbation field
$\u' = \u'(t,\x)$ is a solution of the Navier-Stokes system linearized
around the trajectory corresponding to the initial data $\u_0$
\citep{g03}, i.e.,
\begin{subequations}
\label{eq:lNSE3D}
\begin{align}
 \L\begin{bmatrix} \u' \\ p' \end{bmatrix} := 
& \begin{bmatrix}
\partial_{t}\u'+\u'\cdot\bnabla\u+\u\cdot\bnabla\u'+\bnabla p'-\nu\bDelta\u' \\
\bnabla\cdot\u'
\end{bmatrix} = \begin{bmatrix} \mathbf{0} \\ 0\end{bmatrix}, \label{eq:lNSE3Da} \\
 \u'(0)= &\u_0' \label{eq:lNSE3Db}
\end{align}
\end{subequations}
which is subject to the periodic boundary conditions and where $p'$ is
the perturbation pressure.

We note that expression \eqref{eq:dET} for the G\^{a}teaux
differential is not consistent with the Riesz form \eqref{eq:riesz},
because the perturbation $\u_0'$ of the initial data does not appear
in it explicitly as a factor, but is instead hidden as the initial
{condition} in the linearized problem, cf.~\eqref{eq:lNSE3Db}. In
order to transform \eqref{eq:dET} to the Riesz form, we introduce the
{\em adjoint states} $\u^* \; : \; [0,T]\times\Omega \rightarrow
\RR^3$ and $p^* \; : \; [0,T]\times\Omega \rightarrow \RR$, and the
following duality-pairing relation
\begin{equation}
\begin{aligned}
\left( \L\begin{bmatrix} \u' \\ p' \end{bmatrix}, \begin{bmatrix} \u^* \\ p^* \end{bmatrix} \right)
:= & \int_0^T \int_{\Omega} \L\begin{bmatrix} \u' \\ p' \end{bmatrix} \cdot \begin{bmatrix} \u^* \\ p^* \end{bmatrix} \, d\x \, dt 
= \left( \begin{bmatrix} \u' \\ p' \end{bmatrix}, \L^*\begin{bmatrix} \u^* \\ p^* \end{bmatrix}\right) + \\
\phantom{=} & {\underbrace{\int_\Omega \u'(T,\x)\cdot\u^*(T,\x)  \,d\x}_{\E'_T(\u_0;\u_0')}} - 
\int_\Omega \u'(0,\x)\cdot\u^*(0,\x)  \,d\x = 0,
\end{aligned}
\label{eq:dual}
\end{equation}
where ``$\cdot$'' in the first integrand expression denotes the
Euclidean dot product evaluated at $(t,\x)$. {Performing}
integration by parts with respect to {both space and time then}
allows us to define the {\em adjoint system} as
\begin{subequations}
\label{eq:aNSE3D}
\begin{align}
 \L^*\begin{bmatrix} \u^* \\ p^* \end{bmatrix} := 
& \begin{bmatrix}
-\partial_{t}\u^*-\left[\bnabla\u^*+\left(\bnabla\u^{*}\right)^T\right]\u-\bnabla p^*-\nu\bDelta\u^* \\
-\bnabla\cdot\u^*
\end{bmatrix}  = \begin{bmatrix} {\mathbf{0}}  \\ 0\end{bmatrix}, \label{eq:aNSE3Da} \\
 \u^*(T)= & {\bDelta\u}  \label{eq:aNSE3Db}
\end{align}
\end{subequations}
which is also subject to the periodic boundary conditions. We note
that in identity \eqref{eq:dual} all boundary terms resulting from
integration by parts {with respect to the space variables} vanish
due to the periodic boundary conditions. The term $\int_\Omega
\u'(T,\x)\cdot\u^*(T,\x) \,d\x$ {resulting from integration by
  parts with respect to time is equal to the G\^ateaux differential
  \eqref{eq:dET} due to the judicious} choice of the terminal
condition \eqref{eq:aNSE3Db}, {such that} identity
\eqref{eq:dual} implies
\begin{equation}
\E'_T(\u_0;\u_0') = \int_\Omega \u'_0(\x)\cdot\u^*(0,\x)  \,d\x.
\label{eq:dET2}
\end{equation}
Applying the first equality in Riesz relations \eqref{eq:riesz} to
\eqref{eq:dET2} we obtain the $L^2$ gradient as
\begin{equation}
\nabla^{L^2}\ET = \u^*(0).
\label{eq:gradL2}
\end{equation}

In order to obtain the required Sobolev $H^1$ gradient $\nabla\ET$, we
identify the G\^{a}teaux differential in \eqref{eq:dET2} with the $H^1$
inner product, cf.~\eqref{eq:ipH1}. {Then, recognizing that the
perturbations $\u_0'$ are arbitrary, we obtain the following elliptic
boundary-value problem \citep{pbh04}}
\begin{equation}
\left[ \Id \, - \,\ell_1^2 \,\bDelta \right] \nabla\E_T(\u_0)
= \nabla^{L_2} \E_T(\u_0)   \qquad \text{in} \ \Omega 
\label{eq:gradH1}
\end{equation}
subject to the periodic boundary conditions, which must be solved to
determine $\nabla\ET$.  The gradient fields $\nabla^{L_2}\ET$ and
$\nabla\ET$ can be interpreted as infinite-dimensional sensitivities
of the objective functional $\ET$, cf.~\eqref{eq:ET}, with respect to
perturbations of the initial data $\u_0$. While these two gradients
may point towards the same local maximizer, they represent distinct
``directions'', since they are defined with respect to different
{norms} ($L^2$ vs.~$H^1$). As shown by \citet{pbh04}, extraction
of gradients in spaces of smoother functions such as $H^1(\Omega)$ can
be interpreted as low-pass filtering of the $L^2$ gradients with the
parameter $\ell_1$ acting as the cut-off length-scale. {Although
  Sobolev gradients obtained with different $0 < \ell_1 < \infty$ are
  equivalent, in the precise sense of norm equivalence \citep{b77},
  the value of $\ell_1$ tends to have a significant effect on the rate
  of convergence of gradient iterations \eqref{eq:desc} \citep{pbh04}}
and the choice of its numerical value will be discussed in \S
\ref{sec:numer}. We emphasize that, while the $H^1$ gradient is used
exclusively in the actual computations, cf.~\eqref{eq:desc}, the $L^2$
gradient is computed first merely as an intermediate step.

Evaluation of the $L^2$ gradient at a given iteration {via
  \eqref{eq:gradL2}} requires solution of the Navier-Stokes system
\eqref{eq:NSE3D} followed by solution of the adjoint system
\eqref{eq:aNSE3D}. We note that this system is a linear problem with
coefficients and the terminal condition determined by the solution of
the Navier-Stokes system obtained earlier during the iteration. The
adjoint system \eqref{eq:aNSE3D} is a {\em terminal} value problem,
implying that it must be integrated {\em backwards} in time from $t=T$
to $t=0$ (since the term with the time derivative has a negative sign,
this problem is well posed).  Once the $L^2$ gradient is determined
using \eqref{eq:gradL2}, the corresponding Sobolev $H^1$ gradient can
be obtained by solving problem \eqref{eq:gradH1}. We add that the thus
computed gradient satisfies the divergence-free condition by
construction, i.e., $\bnabla\cdot(\nabla\ET) = 0$.

As regards the fixed-enstrophy constraint $\E(\u_0) = \E_0$, this
property is enforced in iterations \eqref{eq:desc} using the
projection operator $\mathbb{P}_{\mathcal{Q}_{\E_0}}$ defined as the
normalization
\begin{equation}
\label{eq:PQ}
\mathbb{P}_{\Q_{\E_0}}(\u_0) := \sqrt{\frac{\E_0}{\E(\u_0)}}\,\u_0
\end{equation}
which {clearly} preserves the divergence-free property of the
argument.

The step size $\tau_n$ in algorithm \eqref{eq:desc} is computed
as
\begin{equation}\label{eq:tau_n}
\tau_n = \mathop{\argmax}_{\tau>0} \left\{ \E_T\left[\mathbb{P}_{\Q_{\E_0}}
\left( \;\uET^{(n)} + \tau\,\nabla\E_T(\uET^{(n)}) \;\right)\right] \right\}
\end{equation}
which is done using a suitable derivative-free line-search {approach,
  such as a variant of Brent's algorithm \citep{nw00,numRecipes}}.
Equation \eqref{eq:tau_n} can be interpreted as a modification of {the
  standard line-search problem} where optimization is performed
following an arc (a geodesic in the limit of infinitesimal step sizes)
lying on the constraint manifold $\mathcal{S}_{\E_0}$, rather than
{along} a straight line. This approach was already successfully
employed to solve similar problems in \citet{ap11a,ap13a,ap16}.

Maximizing branches are computed using a continuation approach by
fixing one parameter, e.g., $\E_0$, {and then} solving Problem
\ref{pb:maxET} with procedure \eqref{eq:desc} {repeatedly} for
increasing values of $T$. In this process the maximizer $\tuET$
{obtained for some $\E_0$ and $T$} is employed as the initial guess
$\u^0$ in \eqref{eq:desc} to compute the maximizer
$\widetilde{\u}_{0;\E_0,T+\Delta T}$ on a larger time interval
$[0,T+\Delta T]$, {or $\widetilde{\u}_{0;\E_0+\Delta\E_0,T}$ for a
  larger initial enstrophy $\E_0+\Delta\E_0$,} for some sufficiently
small $\Delta T$ {or $\Delta\E_0$}. Since in the limit $T \rightarrow
0$ solutions of the finite-time optimization Problem \ref{pb:maxET}
coincide with the solutions of the instantaneous optimization Problem
\ref{pb:maxdEdt_E}, {for small initial enstrophy values $\E_0$} the
instantaneous maximizers $\tuE$, {cf.~figure \ref{fig:tuE},} are used
as ``seeds'' to initiate the computation of the maximizing branch for
{the} given value of $\E_0$, i.e., as the initial guess for
$\widetilde{\u}_{0;\E_0,\Delta T}$. The procedure outlined above is
summarized as Algorithm \ref{alg:optimAlg}. {For larger initial
  enstrophy values it is also convenient to perform continuation with
  respect to $\E_0$ with $T$ fixed and in such case} an essentially
the same procedure applies, except that the order of the two outermost
``repeat'' loops in Algorithm \ref{alg:optimAlg} is reversed. While
there exist alternatives to the continuation approach, provided
$\Delta \E_0$ and $\Delta T$ are sufficiently small, this technique in
fact results in the fastest convergence of iterations \eqref{eq:desc}
and also ensures that computed optimal initial data lie on a
maximizing branch.

\begin{algorithm}[h!]
\begin{algorithmic}
\STATE
\STATE set $\E_0 = 0$, $T = 0$
\REPEAT 
\STATE \COMMENT{------------------------ loop over increasing enstrophy values $\E_0$ ------------------------}
\STATE $\E_0 = \E_0 + \Delta \E$
\STATE compute $\tuE$ {by solving Problem \ref{pb:maxdEdt_E}}, as described in \citet{ap16}
\STATE $\uET^{(0)} = \tuE$
\REPEAT 
\STATE \COMMENT{------------------------ loop over {expanding} time intervals $T$ --------------------------}
\STATE $T = T + \Delta T$
\STATE $n = 0$
\STATE compute $e_0 = \E_T\left(\uET^{(0)}\right)$
\REPEAT 

\STATE \COMMENT{--------------------------- optimization iterations \eqref{eq:desc}  --------------------------------}

\STATE solve the Navier-Stokes system with initial condition $\uET^{(n)}$, see equation \eqref{eq:NSE3D}

\STATE solve the adjoint system to obtain {$\u^*$ and $p^*$}, see equation \eqref{eq:aNSE3D}

\STATE compute the $L^2$ gradient $\nabla^{L_2}\E_T\left(\uET^{(n)}\right)$, see equation \eqref{eq:gradL2}

\STATE compute the Sobolev gradient $\nabla\E_T\left(\uET^{(n)}\right)$, see equation \eqref{eq:gradH1}

\STATE compute the {optimal} step size $\tau_n$, see equation \eqref{eq:tau_n}

\STATE set $\uET^{(n+1)} = \mathbb{P}_{\Q_{\E_0}}\left(\;\uET^{(n)} + \tau_n \nabla\E_T\left(\uET^{(n)}\right)\;\right)$, see equations \eqref{eq:PQ}

\STATE set $e_1 = \E_T\left(\uET^{(n+1)}\right)$

\STATE compute \texttt{relative\_change} $ = (e_1 - e_0)/e_0$

\STATE set $e_0 = e_1$

\STATE set $n=n+1$

\UNTIL{ \ \texttt{relative\_change} $<$ $\epsilon$}
\STATE $\tuET = \uET^{(n+1)}$
\STATE $\uET^{(0)} = \tuET$
\UNTIL {\ $T > T_{\text{max}}$}
\UNTIL {\ $\E_0 > \E_{\text{max}}$}

\end{algorithmic}
\caption{
 Computation of maximizing branches {parameterized by $T$ for different $\E_0$} using continuation approach.  \newline
     \textbf{Input:} \newline
 \hspace*{0.22cm} $\E_{\text{max}}$ --- maximum enstrophy \newline
 \hspace*{0.22cm} $T_{\text{max}}$ --- maximum time interval \newline
 \hspace*{0.22cm}    $\Delta \E$ --- (adjustable) increment of  enstrophy  \newline
 \hspace*{0.22cm}    $\Delta T$ --- (adjustable) increment of the length of the time interval \newline
 \hspace*{0.22cm}    $\epsilon$ --- tolerance in the solution of optimization problem  \ref{pb:maxET} via iterations \eqref{eq:desc} \newline
 \hspace*{0.22cm}    $\ell_1$ --- adjustable length scale defining inner product \eqref{eq:ipH1}, see also \eqref{eq:gradH1} \newline
 \textbf{Output:} \newline
 \hspace*{0.22cm}    branches of optimal initial data $\tuET$, \ $0 \le \E_0 \le \E_{\text{max}}$, $0 \le T \le T_{\text{max}}$, 
}
\label{alg:optimAlg}
\end{algorithm}

Finally, we remark that the ``optimize-then-discretize'' approach
{adopted} here has the key advantage that in such a continuous
formulation the expressions representing the sensitivity of the
objective functional $\ET$, i.e., the gradients $\nabla^{L_2}\ET$ and
$\nabla\ET$, are independent of the specific discretization approach
chosen to evaluate them.  This should be contrasted with the discrete
(``discretize-then-optimize'') formulation, where a change of the
discretization method would require rederivation of the gradient
expressions.  In addition, the continuous formulation allows us to
strictly enforce the regularity of maximizers required in Problem
\ref{pb:maxET}. Last, but not least, the continuous formulation is
also arguably better adapted to study problems motivated by questions
in mathematical analysis {such as the problems considered here}.
Our strategy for the numerical implementation of the different
elements of Algorithm \ref{alg:optimAlg} is presented in the next
section.

\section{Numerical Approach}
\label{sec:numer}

In this section we briefly describe the key elements of the numerical
approach used to implement Algorithm \ref{alg:optimAlg} and also
comment on the validation strategies we employed. Evaluation of the
objective functional \eqref{eq:ET} requires solution of the
Navier-Stokes system \eqref{eq:NSE3D} on the time interval $[0,T]$
with the given initial data $\u_0$. This system is solved numerically
with an approach combining a pseudo-spectral approximation of spatial
derivatives with a {fourth-order} semi-implicit Runge-Kutta method
\citep{NumRenaissance} used to discretize the problem in time.  {In
  the evaluation of the nonlinear term dealiasing is performed using
  the Gaussian filtering approach proposed by \cite{hl07}.}  Massively
parallel implementation based on MPI and using the {\tt fftw} routines
\citep{fftw} to perform Fourier transforms allowed us to {employ}
resolutions varying from $128^3$ to $512^3$ in the low-enstrophy and
high-enstrophy cases, respectively. Solution of optimization problem
\ref{pb:maxET} for a large initial enstrophy $\E_0 \lessapprox 1000$
and an intermediate length $T$ of the time interval typically
required a computational time of $\O(10^2)$ hours on $\O(10^2)$ CPU
cores.  A number of different diagnostics were checked to ensure that
all flow solutions discussed in \S \ref{sec:results} are well
resolved. We refer the reader to the dissertation by \cite{a14} for
additional details and a validation of this approach.

In addition to the Navier-Stokes system \eqref{eq:NSE3D}, evaluation
of the $L^2$ gradient \eqref{eq:gradL2} also requires solution of the
adjoint system \eqref{eq:aNSE3D}. This problem is solved using
essentially the same numerical approach as used to solve the
Navier-Stokes system \eqref{eq:NSE3D}. The velocity field $\u =
\u(t,\x)$ needed to evaluate the coefficients and the terminal
condition in the adjoint system is saved at discrete time levels
during solution of the Navier-Stokes system (since the considered time
intervals $[0,T]$ are not very long, {data for} entire flow evolutions
could be stored with our available temporary storage resources). The
accuracy {of the evaluation} of the gradient $\nabla^{L^2}\ET$
determined as described in \S \ref{sec:optim} is verified by examining
the quantity $\kappa(\epsilon) :=
\epsilon^{-1}\left[\E_T(\u_0+\epsilon \u_0') - \ET\right] / \langle
\nabla^{L_2}\ET, \u_0' \rangle_{L^2(\Omega)}$ which represents the
ratio of a forward finite-difference approximation of the G\^{a}teaux
differential $\E'_T(\u_0;\u_0')$ and its expression given in terms of
the Riesz formula \eqref{eq:riesz} with the gradient given in
\eqref{eq:gradL2}.  The evaluation of gradients is validated when
$\kappa(\epsilon) \approx 1$ for intermediate values of $\epsilon$ and
different choices of $\u_0$ and $\u_0'$. {For such intermediate values
  of $\epsilon$ we observe that, as expected, $|\kappa(\epsilon) - 1|$
  is reduced as the numerical discretization parameters are refined
  (when $\epsilon \rightarrow 0$, $|\kappa(\epsilon)|$ becomes
  unbounded as a result of round-off errors due to subtractive
  cancellation, whereas when $\epsilon$ is large $\kappa(\epsilon)$
  deviates from unity because of truncation errors in the
  finite-difference approximation of the G\^{a}teaux differential,
  both of which are well-known effects \citep{a14})}.

As regards the computation of the Sobolev $H^1$ gradients,
cf.~\eqref{eq:gradH1}, the parameter $\ell_1$ {is} adjusted during the
optimization iterations and {is} chosen so that $\ell_1 \in
[\ell_{\min}, \ell_{\max}]$, where $\ell_{\min}$ is the length scale
associated with the spatial resolution {$N = 128, 256, 512$} used for
computations and $\ell_{\max}$ is the characteristic length scale of
the domain $\Omega$, that is, $\ell_{\min} \sim \O( 1/N) $ and
$\ell_{\max} \sim \O(1)$. We remark that, given the equivalence of the
{Sobolev} inner products \eqref{eq:ipH1} corresponding to different
values of $\ell_1$ (as long as $\ell_1 \neq 0$), these choices do not
affect the maximizers found, but only how rapidly they are approached
by iterations \eqref{eq:desc}. {The computational results
  presented in the next section have been thoroughly validated to
  ensure they are converged with respect to refinement of the
  different numerical parameters discussed above.}

\section{Computational Results}
\label{sec:results}

In this section we proceed to present our computational results
obtained by solving the finite-time optimization problem
\ref{pb:maxET} for a broad range of initial enstrophies $\E_0$ and
lengths $T$ of the time interval. The ultimate goal is to understand
what is the largest growth of enstrophy possible under the 3D
Navier-Stokes dynamics, in particular, whether this growth could
saturate estimate \eqref{eq:Et_estimate_E0} and become unbounded in
finite time, then how this maximum growth depends on the initial
enstrophy $\E_0$ and, finally, what is the flow mechanism realizing
this extreme behavior.  To address these questions we compute branches
of maximizing solutions corresponding to different fixed values of the
initial enstrophy $\E_0$ and increasing lengths $T$ of the time
window. For smaller values of $\E_0$ this can be done using Algorithm
\ref{alg:optimAlg} in which solutions of the instantaneous
maximization problem \ref{pb:maxdEdt_E} are used to initialize the
computation of the maximizing branches for short optimization times
$T$.  Since for large initial enstrophy values ($\E_0 > 100$), the
instantaneous problem \ref{pb:maxdEdt_E} is harder to solve because of
increased resolution requirements (more on this below), in such cases
it is more convenient to initiate computation of the new branches
corresponding to increased values of $\E_0$ by performing continuation
with respect to $\E_0$ for some intermediate values of $T$. Only then
we can again perform continuation with respect to $T$ at a new fixed
value of $\E_0$.

\begin{figure}
\begin{center}
\mbox{
\Bmp{0.5\textwidth}
\subfigure[$\E_0 = 50$, $T = 0.2,0.3,0.4$]{\includegraphics[width=1.0\textwidth]{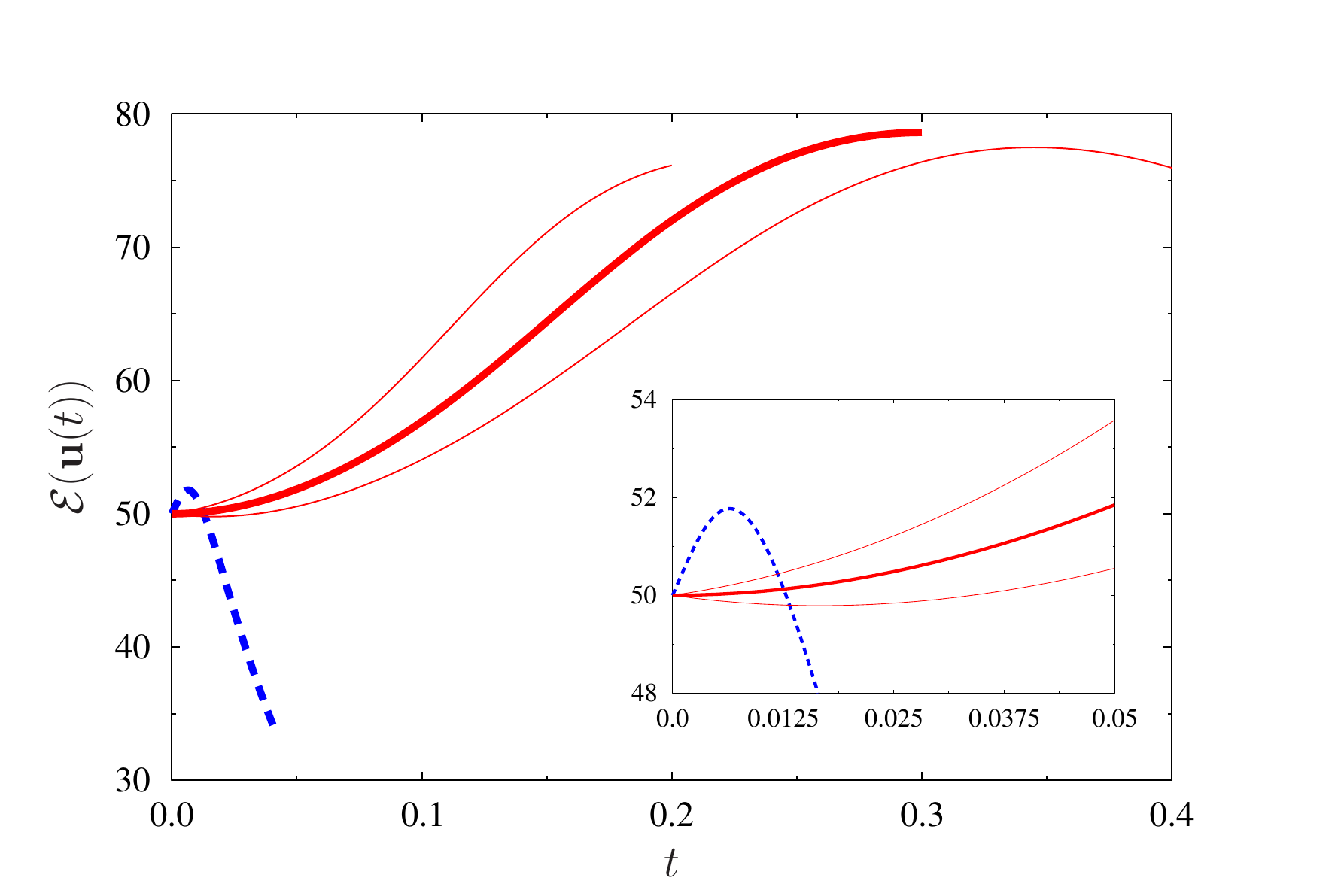}}
\Emp
\Bmp{0.5\textwidth}
\subfigure[$\E_0 = 200$, $T = 0.15,0.23,0.3$]{\includegraphics[width=1.0\textwidth]{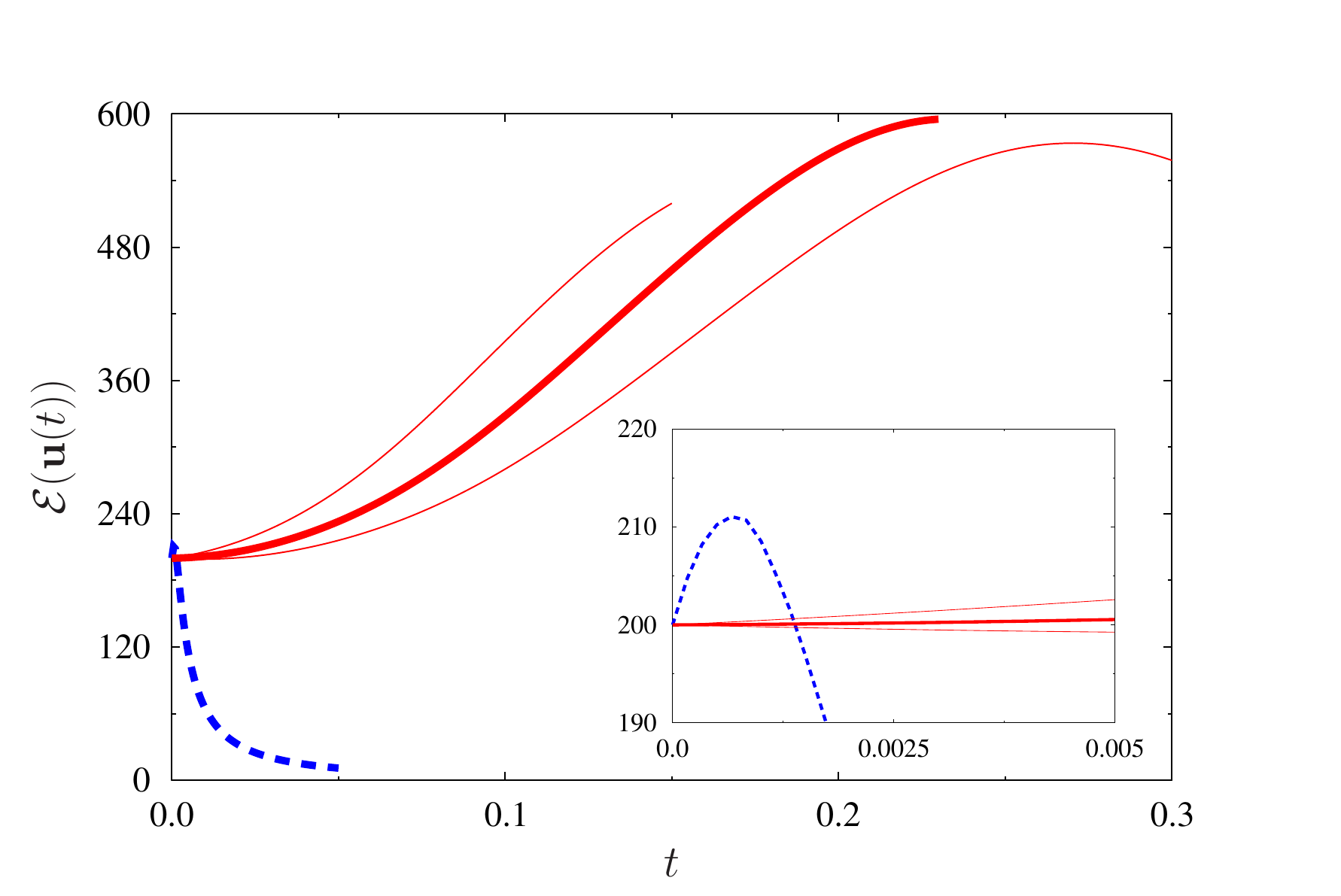}}
\Emp
}
\caption{Enstrophy $\E(\u(t))$ as a function of time $t$ obtained from
  the solution of the Navier-Stokes system \eqref{eq:NSE3D} with the
  initial condition $\u_0$ given by (blue dashed line) the maximizer
  $\tuE$ of the instantaneous optimization problem \ref{pb:maxdEdt_E}
  and (red solid lines) the maximizers $\tuET$ of the finite-time
  optimization problem \ref{pb:maxET} for the indicated values of
  $\E_0$ and $T$ (the curves corresponding to the optimal lengths of
  the time window, {$\tTE = 0.3$ for $\E_0 = 50$ and $\tTE = 0.23$ for
    $\E_0 = 200$}, are marked with thick lines whereas the insets
  represent magnifications of the initial stages of evolution).}
\label{fig:Et}
\end{center}
\end{figure}

A typical time evolution of enstrophy $\E(\u(t))$ is presented in
figure \ref{fig:Et} where we show the results produced by solving the
Navier-Stokes system \eqref{eq:NSE3D} with the initial data $\tuE$ and
$\tuET$ obtained as the solutions of the instantaneous and finite-time
optimization problems \ref{pb:maxdEdt_E} and \ref{pb:maxET} for $\E_0
= 50$ and $\E_0 = 200$, and different $T$.  We note that when the
instantaneously optimal initial data $\tuE$ is used, then the
enstrophy grows very rapidly for short times which is followed by an
immediate depletion of its growth, as already analyzed by
\citet{ap16}. On the other hand, when the optimal initial data $\tuET$
obtained as solution of the finite-time optimization problem
\ref{pb:maxET} with ``long'' time windows $T$ is used, then the
enstrophy grows very slowly at first (or even decreases when $T$ is
sufficiently large), but eventually a much larger growth is achieved
at the end of the time window, i.e., at $t = T$. As is also evident
from figure \ref{fig:Et}, when $T$ is sufficiently large, the
enstrophy achieves a well-defined maximum as a function of time and
there exists a value of $T$ for which the maximum enstrophy is
achieved precisely at the end of the optimization window, i.e., for $t
= T$. Clearly, this value, which we shall denote $\tTE :=
\mathop{\arg\max}_{T > 0} \E_T(\tuET)$, defines the optimal length of
the optimization interval in Problem \ref{pb:maxET} for which the
largest maximum enstrophy is achieved for a given $\E_0$.  Then,
$\E_T(\tuEtT) = \max_{T > 0} \E_T(\tuET)$ is the largest enstrophy
achievable using an initial condition with enstrophy $\E_0$.
Approximating the optimal length of the time interval $\tTE$ for a
given value of the initial enstrophy $\E_0$ requires solution of the
finite-time optimization problem \ref{pb:maxET} for several different
values of $T$.

\begin{figure}
\begin{center}
\mbox{
\Bmp{0.5\textwidth}
\subfigure[$\E_0 = 100$, $\tTE = 0.27$ (symmetric)]{\includegraphics[width=1.0\textwidth]{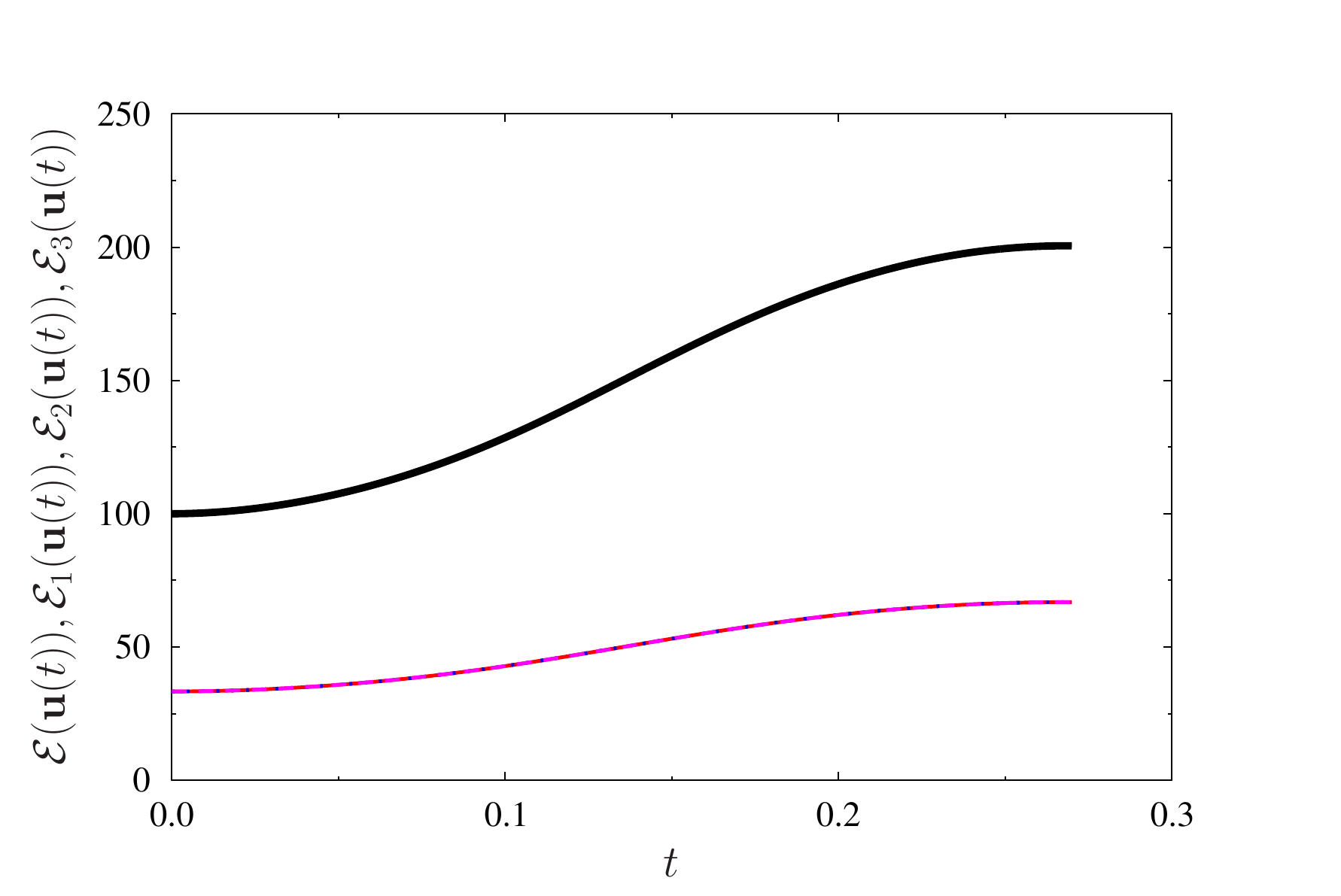}}
\Emp
\Bmp{0.5\textwidth}
\subfigure[$\E_0 = 100$, $\tTE = 0.27$ (asymmetric)]{\includegraphics[width=1.0\textwidth]{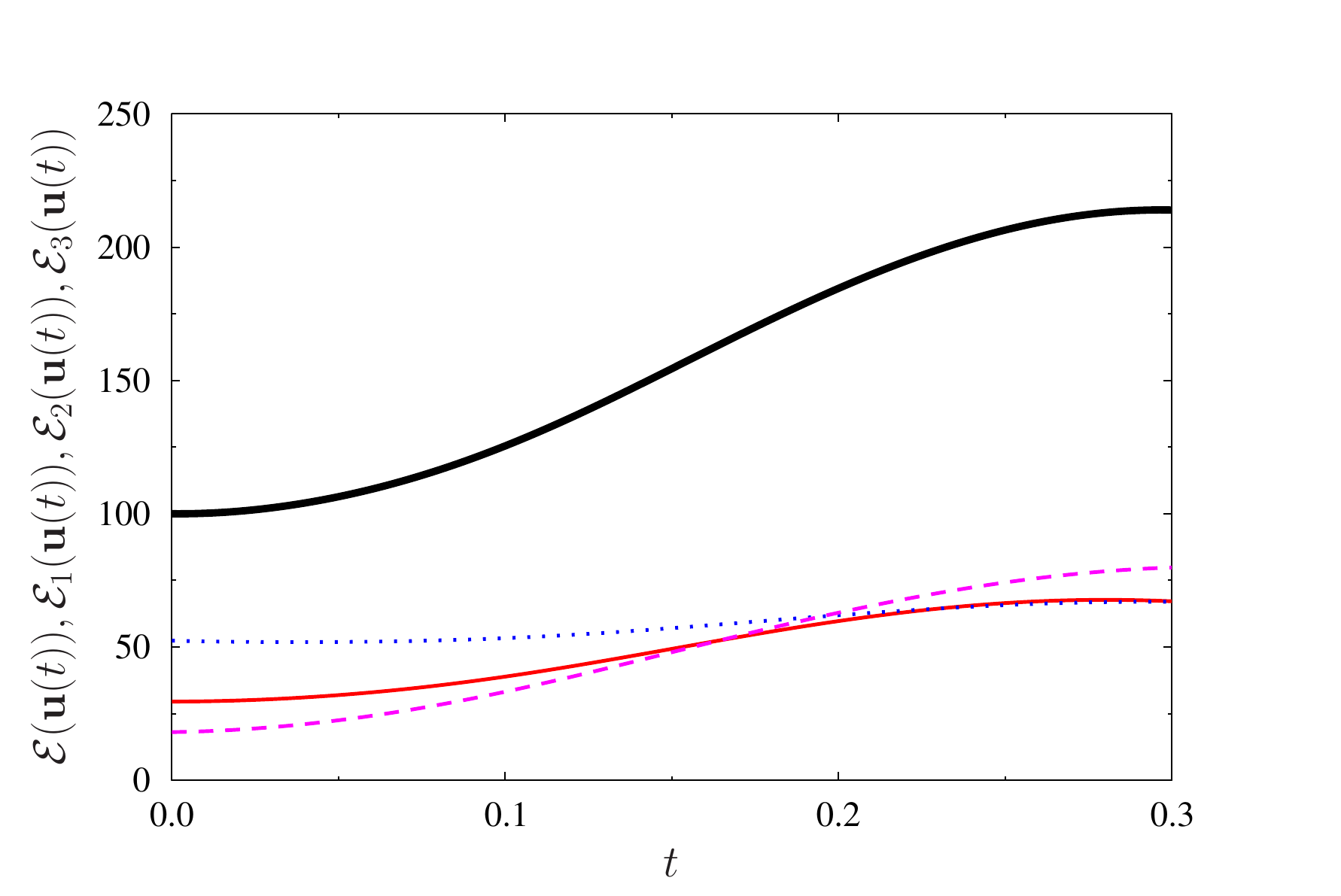}}
\Emp
}
\mbox{
\Bmp{0.5\textwidth}
\subfigure[$\E_0 = 500$, $\tTE = 0.17$  (asymmetric)]{\includegraphics[width=1.0\textwidth]{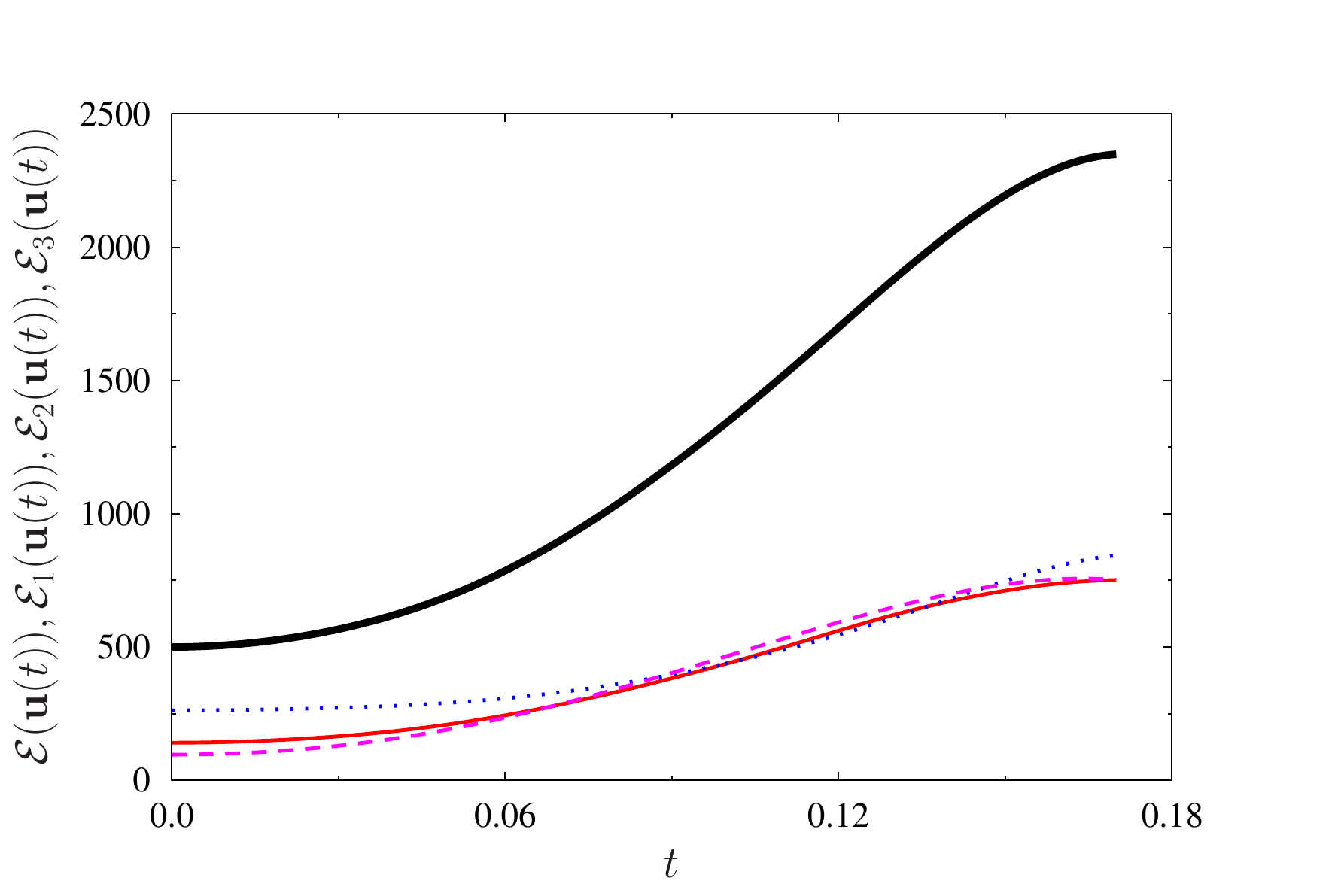}}
\Emp
\Bmp{0.5\textwidth}
\subfigure[$\E_0 = 1000$, $\tTE = 0.12$  (asymmetric)]{\includegraphics[width=1.0\textwidth]{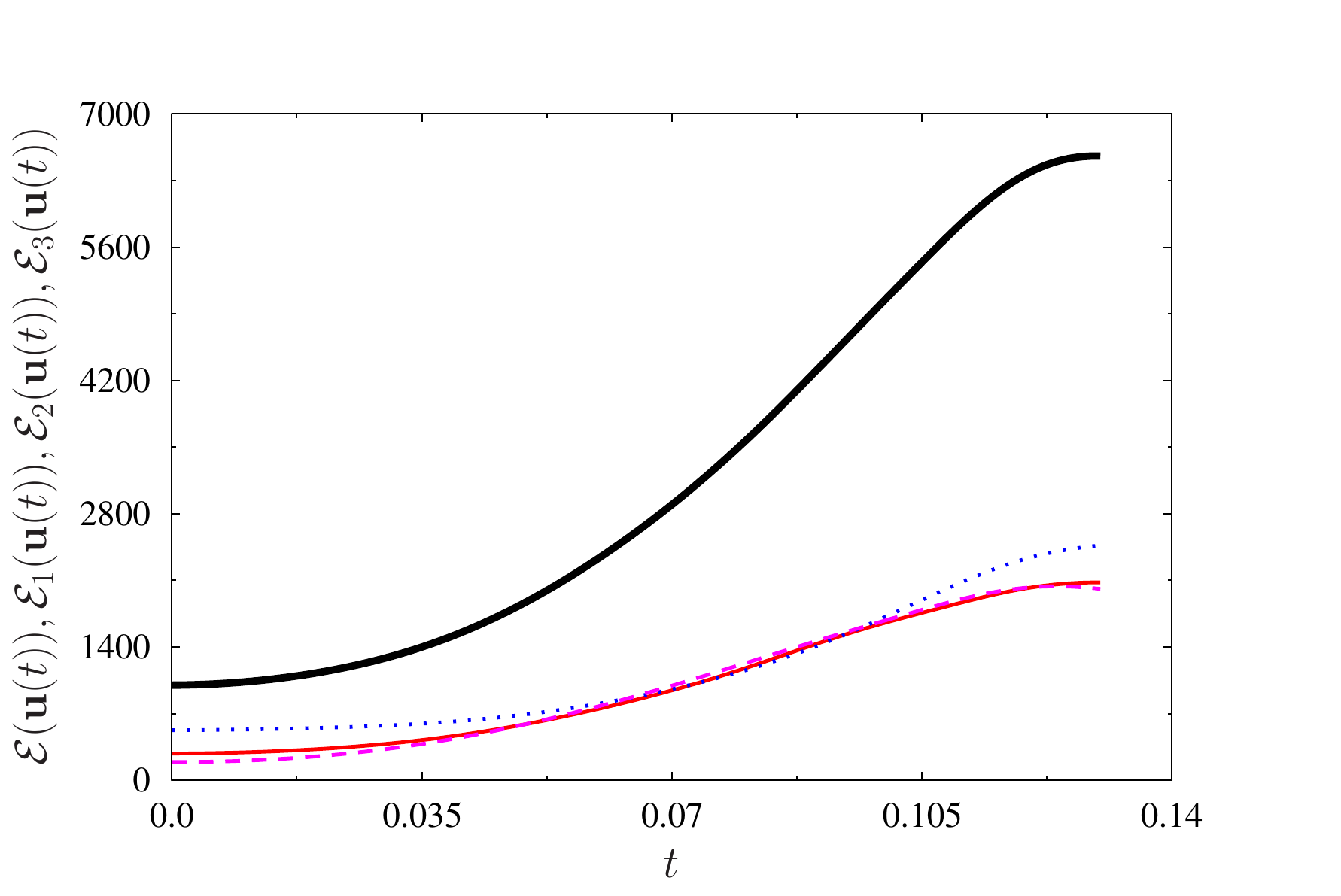}}
\Emp
}
\caption{Evolution of (black thick solid lines) the total enstrophy
  $\E(\u(t))$ and its components (red thin solid lines) $\E_1(\u(t))$,
  (blue dotted lines) $\E_2(\u(t))$ and (pink dashed lines)
  $\E_3(\u(t))$ in the solution of the Navier-Stokes system
  \eqref{eq:NSE3D} with the optimal initial conditions $\tuET$
  obtained by solving the finite-time optimization problem
  \ref{pb:maxET} for the indicated values of $\E_0$ and $T$. Panels
  (a) and (b--d) correspond to the optimal initial conditions $\tuET$
  from the symmetric and asymmetric branch, respectively.}
\label{fig:E123}
\end{center}
\end{figure}

Closer inspection of the solutions to the Navier-Stokes system
\eqref{eq:NSE3D} corresponding to optimal initial condition $\tuET$
obtained by solving problem \ref{pb:maxET} for $\E_0 \lessapprox 100$
and $\E_0 \gtrapprox 100$ indicates that these solutions have in fact
distinct properties in terms of symmetry. To analyze this issue, in
figure \ref{fig:E123} we show the time histories of the enstrophy
components $\E_i(\u(t))$, $i=1,2,3$, associated with the three
coordinate directions defined as
\begin{equation}
\E_i(\u(t)) := \int_{\Omega} \left| \left(\bnabla \times \u(t) \right) \cdot \e_i \right|^2\, d\x, \quad i=1,2,3,
\label{eq:Ei}
\end{equation}
where $\e_1$, $\e_2$, $\e_3$ are the unit vectors of the Cartesian
coordinate system and we have the obvious identity $\forall t \ \
\E(\u(t)) = \sum_{i=1}^3 \E_i(\u(t))$.  In solutions corresponding to
optimal initial conditions $\tuET$ obtained for $\E_0 \lessapprox 100$
we have $\forall t \in [0, \tTE] \quad \E_1(\u(t)) = \E_2(\u(t)) =
\E_3(\u(t)) = (1/3) \E(\u(t))$, indicating that the enstrophy is
equipartitioned among the three coordinate directions, cf.~figure
\ref{fig:E123}(a) for the case $\E_0 = 100$ and $\tTE = 0.27$. On the
other hand, this special property is absent in the solutions
corresponding to optimal initial conditions $\tuET$ obtained for $\E_0
\gtrapprox 100$, cf.~figure \ref{fig:E123}(b--d), in which the three
enstrophy components all contain different fractions of the total
enstrophy.  Moreover, in these cases the ordering of $\E_1(\u(t))$,
$\E_2(\u(t))$ and $\E_3(\u(t))$ changes during the flow evolution. For
example, for $\E_0 = 1000$ and $\tTE = 0.12$, cf.~figure
\ref{fig:E123}(d), while we have $\E_3(\u(0)) < \E_1(\u(0)) <
\E_2(\u(0))$ and $\E_3(\u(\tTE)) < \E_1(\u(\tTE)) < \E_2(\u(\tTE))$ at
the initial and final instant of time, this ordering changes six times
and is completely reversed during the flow evolution. Such transfer of
enstrophy between different vorticity components is known to signal
the phenomenon of reconnection of vortex lines
\citep{HussainDuraisamy2011,VelascoFuentes2017}. {We add that in
  all the flow evolutions discussed above the helicity $\H(\u(t)) :=
  \int_{\Omega} \u(t) \cdot \left(\bnabla \times \u(t) \right) \, d\x$
  remains identically equal to zero, i.e., $\forall t \in [0,T]$
  $\H(\u(t)) = 0$.}

\begin{figure}
\begin{center}
\mbox{
\Bmp{0.5\textwidth}
\subfigure[]{\includegraphics[width=1.0\textwidth]{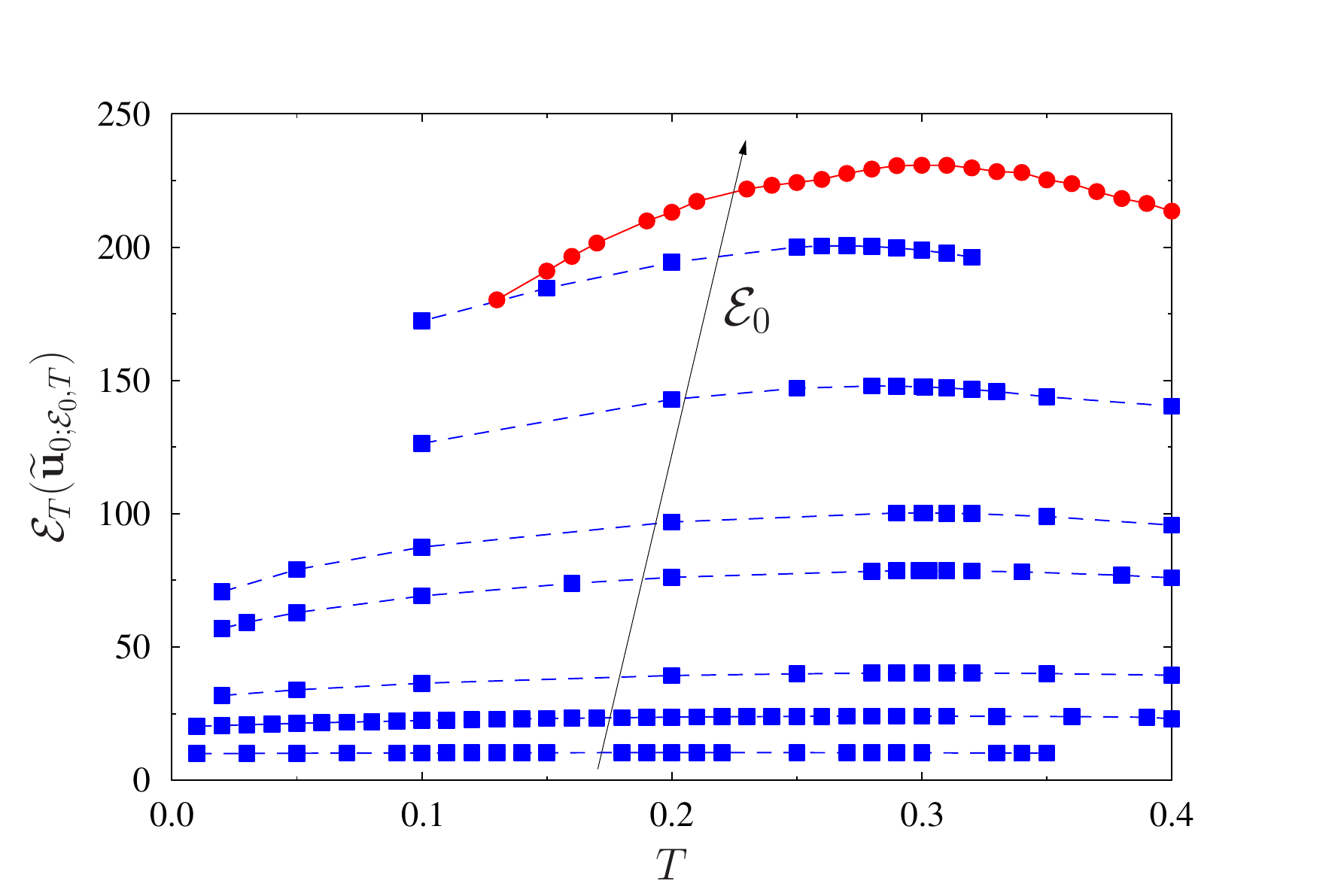}}
\Emp
\Bmp{0.5\textwidth}
\subfigure[]{\includegraphics[width=1.0\textwidth]{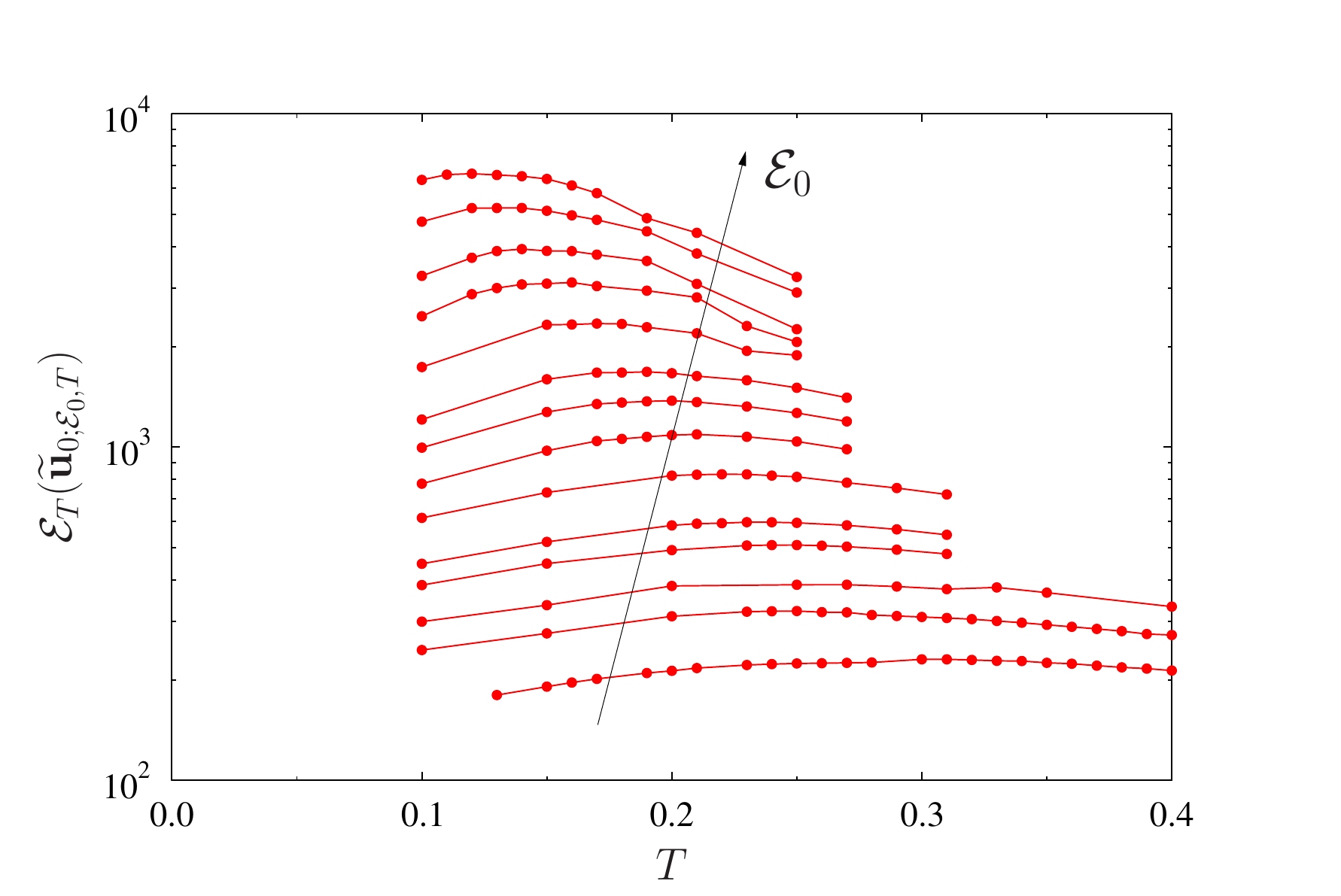}}
\Emp
}
\caption{Maximum attained enstrophy $\E_T(\tuET)$ as a function of the
  length $T$ of the time interval over which maximization is performed
  in Problem \ref{pb:maxET} {for initial enstrophies (a) $0 < \E_0 \le
    100$ and (b) $100 < \E_0 \le 1000$. Blue dashed and red solid
    curves correspond to flow evolutions starting from, respectively,
    symmetric and asymmetric optimal initial conditions $\tuET$ with
    the same value of the initial enstrophy} and different $T$ (the
  trends with the increase of $\E_0$ are indicated with arrows,
  whereas solid symbols represent the values of $\E_0$ and $T$ for
  which Problem \ref{pb:maxET} was solved).}
\label{fig:maxEt_vs_T}
\end{center}
\end{figure}

{We thus have evidence for the existence of maximizers $\tuET$ of two
  distinct types: those for which the flow evolution is characterized
  by the properties $\forall t \in [0, \tTE] \quad \E_1(\u(t)) =
  \E_2(\u(t)) = \E_3(\u(t))= (1/3) \E(\u(t))$ and $\H(\u(t)) = 0$, and
  those for which these properties do not hold. We will refer to these
  families of maximizers as ``symmetric'' and ``asymmetric'',
  respectively. The symmetric maximizers are normally found when the
  finite-time optimization problem \ref{pb:maxET} is solved for $\E_0
  < 100$, whereas the asymmetric maximizers are found when this
  problem is solved for $\E_0 > 100$. However, maximizers of both
  types were found for $\E_0 \approx 100$. These results are
  illustrated in figures \ref{fig:maxEt_vs_T}(a) and
  \ref{fig:maxEt_vs_T}(b) where we show the dependence of the maximum
  attained enstrophy $\E_T(\tuET)$ on the length $T$ of the time
  interval over which optimization was performed in Problem
  \ref{pb:maxET} for initial enstrophies $\E_0 \le 100$ and $\E >
  100$, respectively. In figure \ref{fig:maxEt_vs_T}(a) we see that
  indeed two distinct branches of symmetric and asymmetric maximizers
  are simultaneously present when $\E_0 = 100$, thus demonstrating
  that these maximizers are nonunique. There are indications that
  symmetric maximizers also exist for $\E_0 > 100$ and asymmetric ones
  exist for $\E_0 < 100$, however, since they correspond to suboptimal
  local maxima, it is very difficult to compute them with the gradient
  approach \eqref{eq:desc}. It is possible to capture both the
  symmetric and the asymmetric branch for $\E_0 = 100$, because for
  this value of the initial enstrophy the maximum attained enstrophy
  $\E_T(\tuET)$ is comparable for both branches. The asymmetry of
  optimal initial data $\tuET$ on the asymmetric branch vanishes as $T
  \rightarrow 0.13$ which indicates that for $\E_0 = 100$ the
  asymmetric branch bifurcates off the symmetric branch at $T \approx
  0.13$.  In figure \ref{fig:maxEt_vs_T}(b)} a significant, albeit
finite, increase of the enstrophy growth is observed as a function of
$\E_0$ {when asymmetric maximizers $\tuET$ are used as the initial
  data for system \eqref{eq:NSE3D} and this} figure also confirms that
for each value of the initial enstrophy $\E_0$ the maximum enstrophy
$\E_T(\tuET)$ achieves a well-defined maximum for a certain optimal
time $\tTE$. We note that the accuracy with which $\tTE$ can be
determined for given $\E_0$ is limited by the number of time intervals
$[0,T]$ for which the finite-time optimization problem \ref{pb:maxET}
can be solved, which is a matter of the computational cost.  However,
we use spline interpolation to improve the accuracy with which the
quantities $\max_{T > 0} \E_T(\tuET)$ and $\tTE$ are determined for a
given initial enstrophy $\E_0$.

\begin{figure}
\begin{center}
\mbox{
\Bmp{0.5\textwidth}
\subfigure[]{\includegraphics[width=1.0\textwidth]{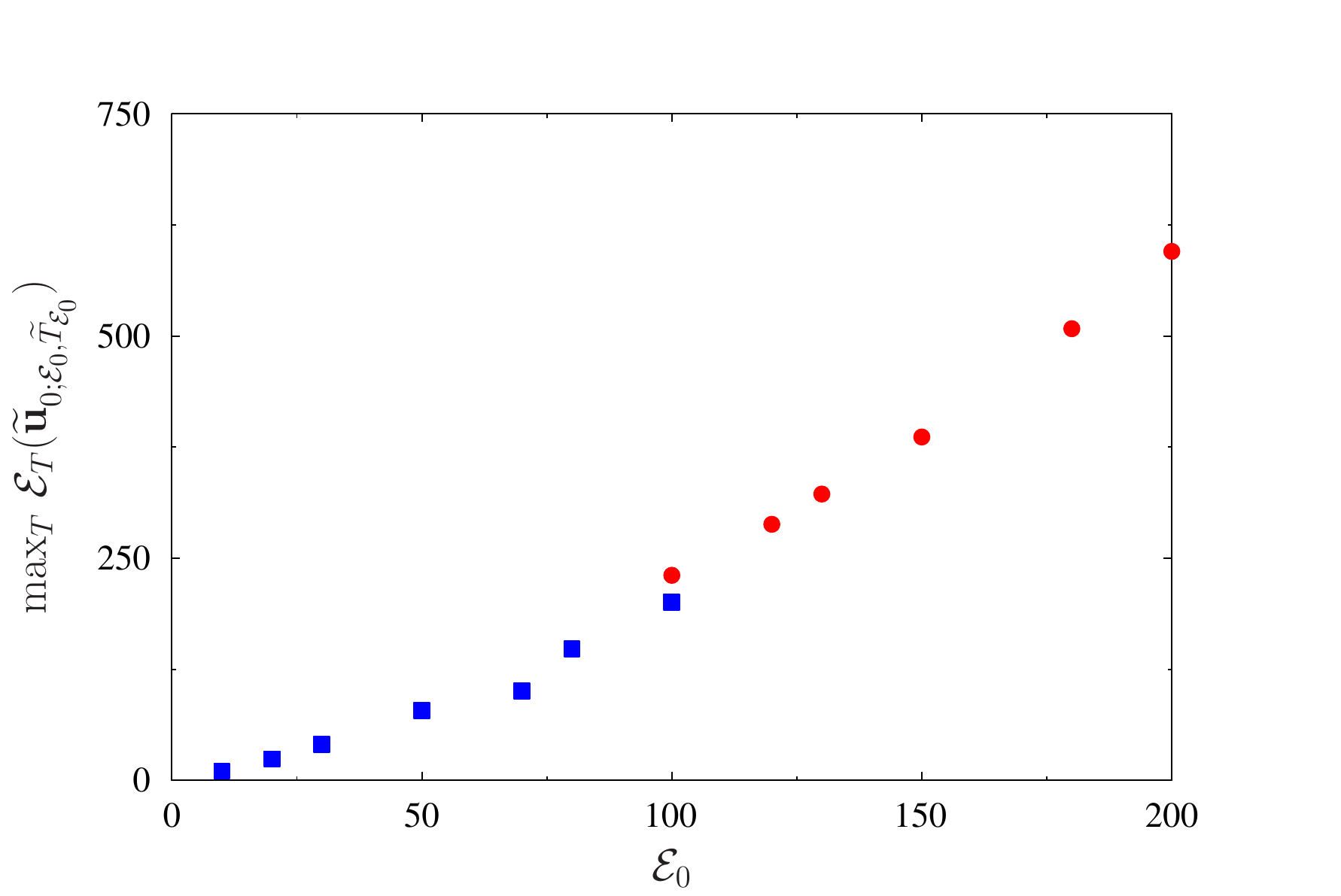}}
\Emp
\Bmp{0.5\textwidth}
\subfigure[]{\includegraphics[width=1.0\textwidth]{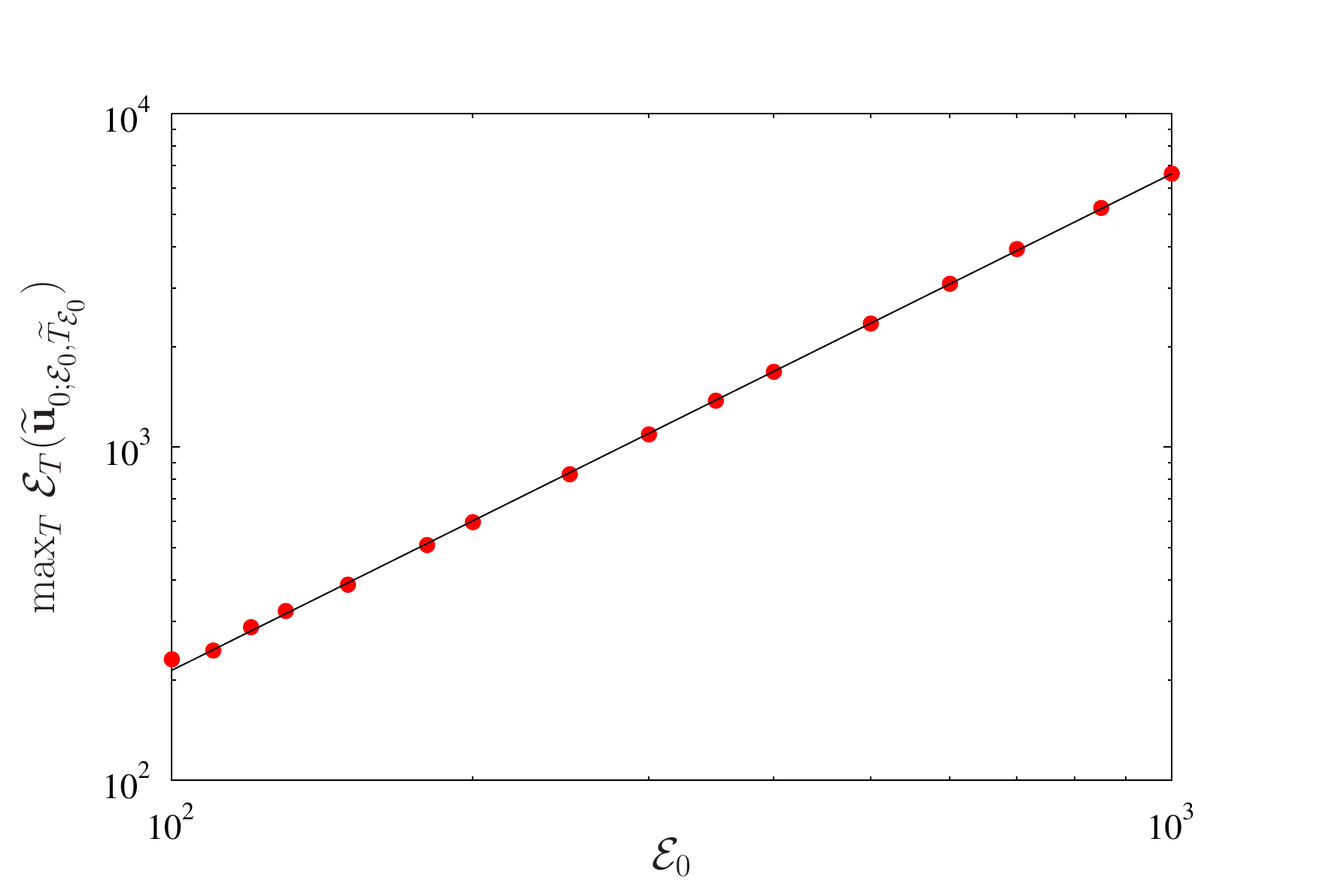}}
\Emp
}
\caption{Dependence of the maximum enstrophy growth $\max_{T > 0} \,
  \E_T(\tuET)$ on the initial enstrophy using (a) linear scaling {for
    small $\E_0$} and (b) logarithmic scaling {for large $\E_0$}. {The
    blue squares and red circles correspond to, respectively, the
    symmetric and asymmetric branches of} solutions of the finite-time
  optimization problem \ref{pb:maxET} for different $\E_0$. {The black
    solid line in panel (b)} represents the least-squares fit
  \eqref{eq:maxEt_vs_E0}.}
\label{fig:maxEt_vs_E0}
\end{center}
\end{figure}

\begin{figure}
\begin{center}
\includegraphics[width=0.6\textwidth]{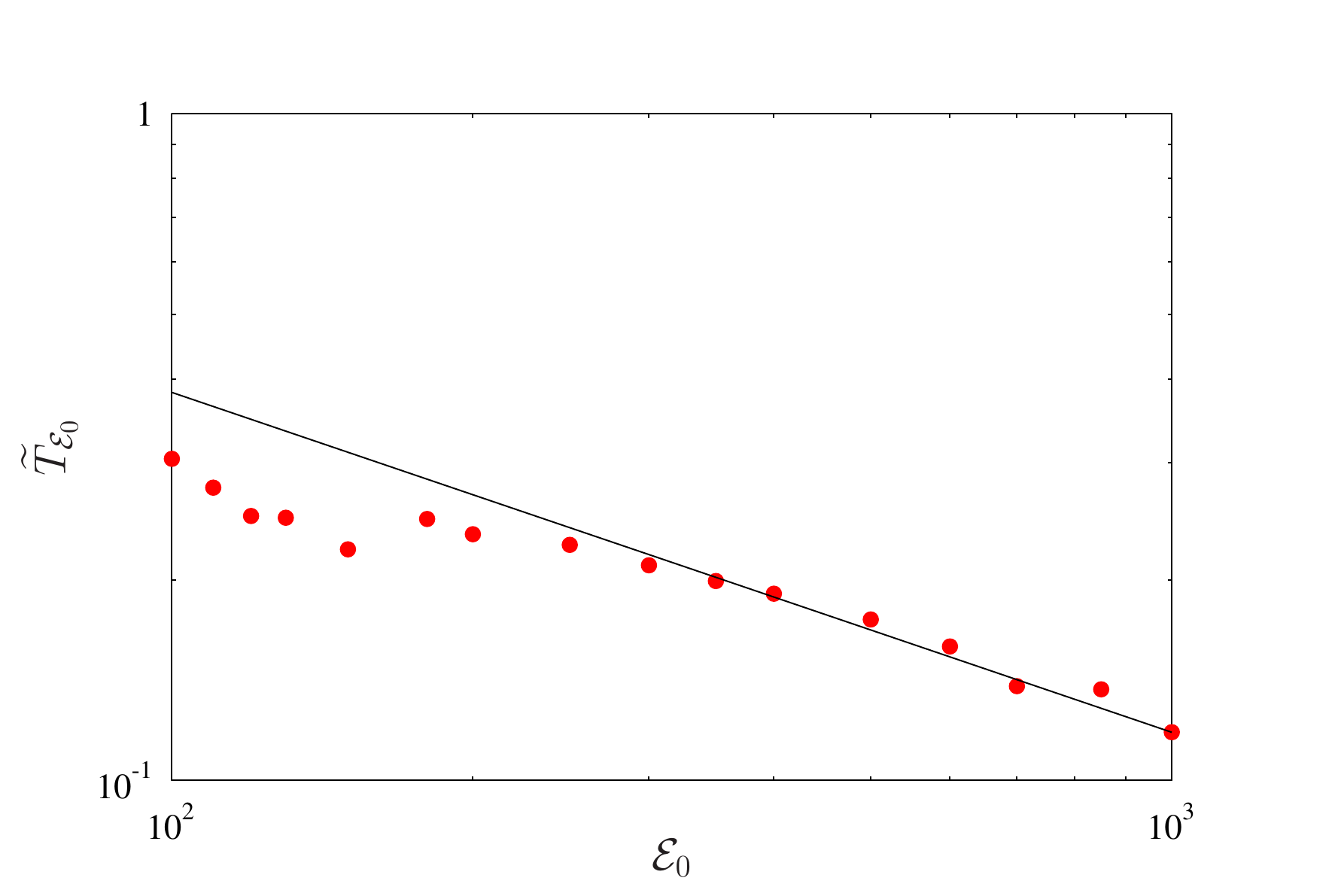}
\caption{Dependence of the time $\tTE$ when the maximum enstrophy
  $\max_{T > 0} \, \E_T(\tuET)$ is achieved {(with asymmetric
    maximizers)} on the initial enstrophy $\E_0$. The red circles
  represent the interpolated data corresponding to solutions of the
  finite-time optimization problem \ref{pb:maxET} for different
  $\E_0$, whereas the black solid line represents the least-squares
  fit \eqref{eq:Tmax_vs_E0}.}
\label{fig:Tmax_vs_E0}
\end{center}
\end{figure}

As the central result of this study, the maximum growth of enstrophy
$\max_{T > 0} \E_T(\tuET)$ achieved using initial conditions with
enstrophy $\E_0$ is shown in figure \ref{fig:maxEt_vs_E0} as a
function of $\E_0$. {In figure \ref{fig:maxEt_vs_E0}(a)
  corresponding to small values of $\E_0$ we see that the dependence
  of $\max_{T > 0} \E_T(\tuET)$ on $\E_0$ appears slightly different
  for the symmetric and asymmetric branches, i.e., for $\E_0 < 100$
  and $\E_0 > 100$.}  As is evident from figure
\ref{fig:maxEt_vs_E0}(b), {for asymmetric maximizers} this growth
follows a well-defined power-law relation
\begin{equation}
\max_{T > 0} \E_T(\tuET)  \ \sim \ \left( 0.224 \ \pm 0.006 \right) \, \E_0^{1.490 \, \pm 0.004} 
\label{eq:maxEt_vs_E0}
\end{equation}
obtained by performing a least-squares fit of the relation between
$\log_{10} \left[\max_{T > 0} \E_T(\tuET) \right]$ and $\log_{10}
\E_0$ for $100 \le \E_0 \le 1000$.  The fit is performed in the
logarithmic coordinates in order for the least-squares error not to be
dominated by contributions from the data corresponding to large values
of $\E_0$.  As regards the times when the maxima are achieved for
different values of $\E_0$, when {asymmetric maximizers are
  considered and} the initial enstrophy is sufficiently large $(\E_0
\gtrapprox 400$), figure \ref{fig:Tmax_vs_E0} shows that $\tTE$ also
exhibits a power-law dependence on $\E_0$, described by the relation
\begin{equation}
\tTE  \ \sim \ \left( 4.0 \, \pm 1.1 \right) \, \E_0^{-0.51 \, \pm 0.04} 
\label{eq:Tmax_vs_E0}
\end{equation}
obtained from a least-squares fit of the relation between $\E_0$ and
$\tTE$. We observe that for the maximum enstrophy $\max_{T > 0}
\E_T(\tuET)$ the power-law dependence on $\E_0$ becomes evident
starting from smaller values of $\E_0$ and the fit is characterized by
smaller error bars than for $\tTE$, cf.~figures
\ref{fig:maxEt_vs_E0}(b) and \ref{fig:Tmax_vs_E0}. {We also
  remark that the dependence of $\tTE$ on $\E_0$ given in
  \eqref{eq:Tmax_vs_E0} appears different from the form of the upper
  bounds on the maximum blow-up time obtained by \citet{Ohkitani2016}.
  We add that the times $t_0$ when the upper bound in estimate
  \eqref{eq:Et_estimate_E0} becomes unbounded are about 8--10 orders
  of magnitude shorter than the times $\tTE$ shown in figure
  \ref{fig:Tmax_vs_E0}.}

\begin{figure}
\begin{center}
\includegraphics[width=0.6\textwidth]{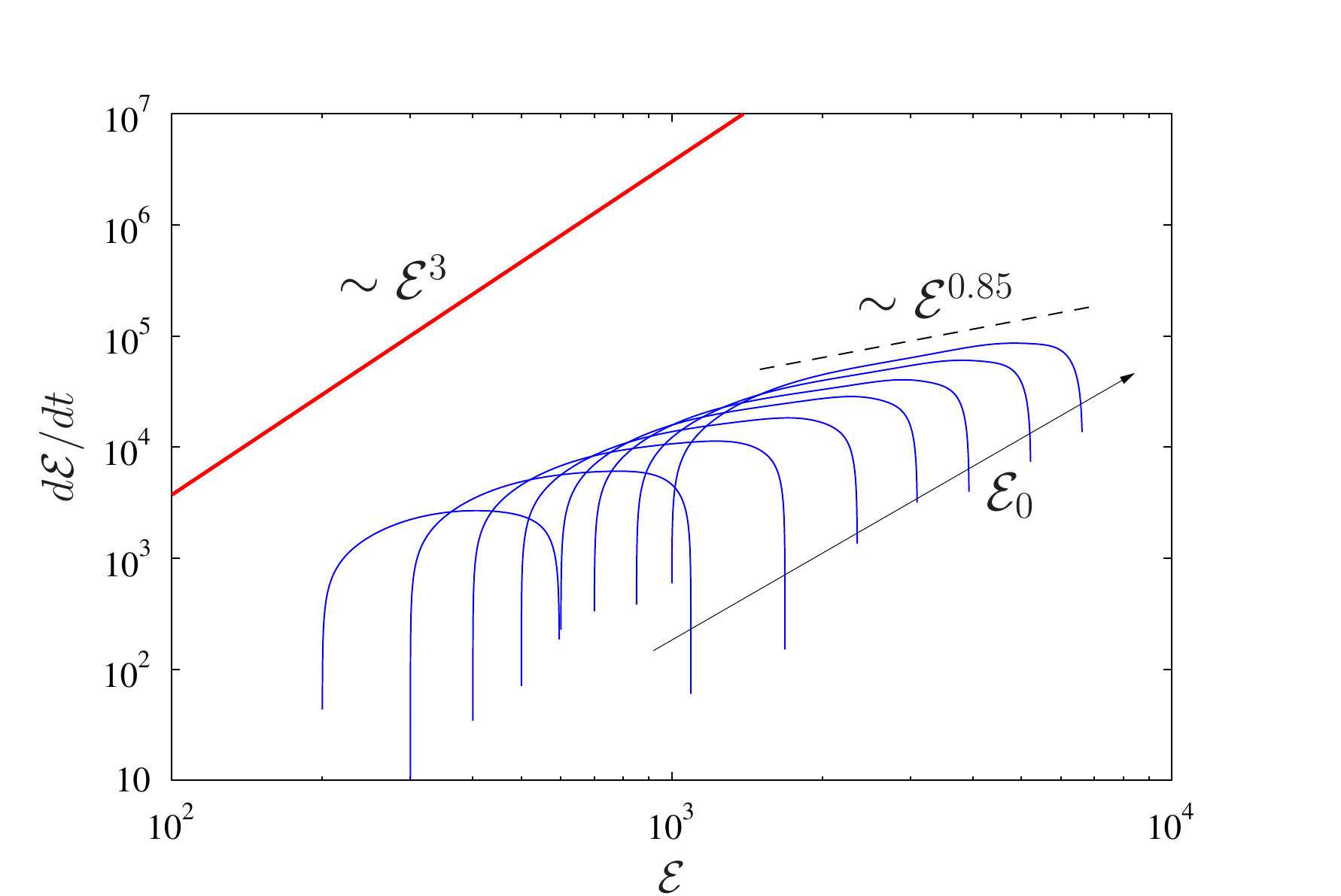}
\caption{{Flow trajectories corresponding to the optimal initial
    data $\tuEtT$ {with different $\E_0 \in [100,1000]$,
      cf.~figure \ref{fig:E123}(b--c),} shown using the coordinates
    $\{ \E, d\E/dt\}$ (blue solid lines with the arrow indicating the
    trend with the increase of $\E_0$). The thick red line represents
    the relation $d\E/dt = 1.72\cdot 10^{-3} \, \E^3$, cf.~figure
    \ref{fig:tuE}, whereas the dashed black line represents the
    relation $d\E/dt = 10^{2} \, \E^{0.85}$.}}
\label{fig:dEvsE}
\end{center}
\end{figure}
{In order to understand how close the flow evolutions
  corresponding to the optimal initial data $\tuEtT$ come to
  saturating a priori bounds on the rate of growth of enstrophy,
  cf.~\eqref{eq:dEdt_estimate_E}, in figure \ref{fig:dEvsE} we plot
  the corresponding trajectories using the coordinates $\{ \E,
  d\E/dt\}$, such that each trajectory is parameterized by time $t$
  (since the logarithmic scale is used, initial parts of the
  trajectories when $d\E/dt \lessapprox 0$ are not shown). The slope
  of the tangent to each of the curves thus represents the exponent
  $\alpha$ characterizing the instantaneous rate of enstrophy
  production $d\E/dt \sim \E^{\alpha}$. In figure \ref{fig:dEvsE} we
  also indicate the relation $d\E/dt = 1.72\cdot 10^{-3} \, \E^3$
  describing the maximum rate of enstrophy growth realized by the
  solutions of the instantaneous optimization problem
  \ref{pb:maxdEdt_E} \citep{ap16}, cf.~figure \ref{fig:tuE} (we note
  that the prefactor in this relation is approximately 7 orders of
  magnitude smaller than the prefactor in estimate
  \eqref{eq:dEdt_estimate_E}). We observe that the rate of growth of
  enstrophy realized by the trajectories corresponding to the optimal
  initial conditions $\tuEtT$ is at all times and for all values of
  $\E_0$ several orders of magnitude smaller than the maximum rate of
  growth achieved by the instantaneous maximizers $\tuE$, cf.~figure
  \ref{fig:tuE}. On the other hand, in figure \ref{fig:dEvsE} we also
  note that the exponent $\alpha$ characterizing the growth of
  enstrophy can be much larger than 3 at intermediate stages along
  these trajectories (we recall from the discussion in \S
  \ref{sec:bounds} that the rate of enstrophy production $d\E/dt \sim
  \E^{\alpha}$ must be sustained with $\alpha > 2$ over a sufficiently
  long time interval for enstrophy to become unbounded in finite
  time). We add that at the final stages of the flow evolutions before
  the enstrophy maximum is reached at $t = \tTE$ the enstrophy is
  amplified at the approximate rate $d\E/dt \sim \E^{0.85}$.}

\begin{figure}
\begin{center}
\mbox{\subfigure[instantaneous]{\includegraphics[width=0.45\textwidth]{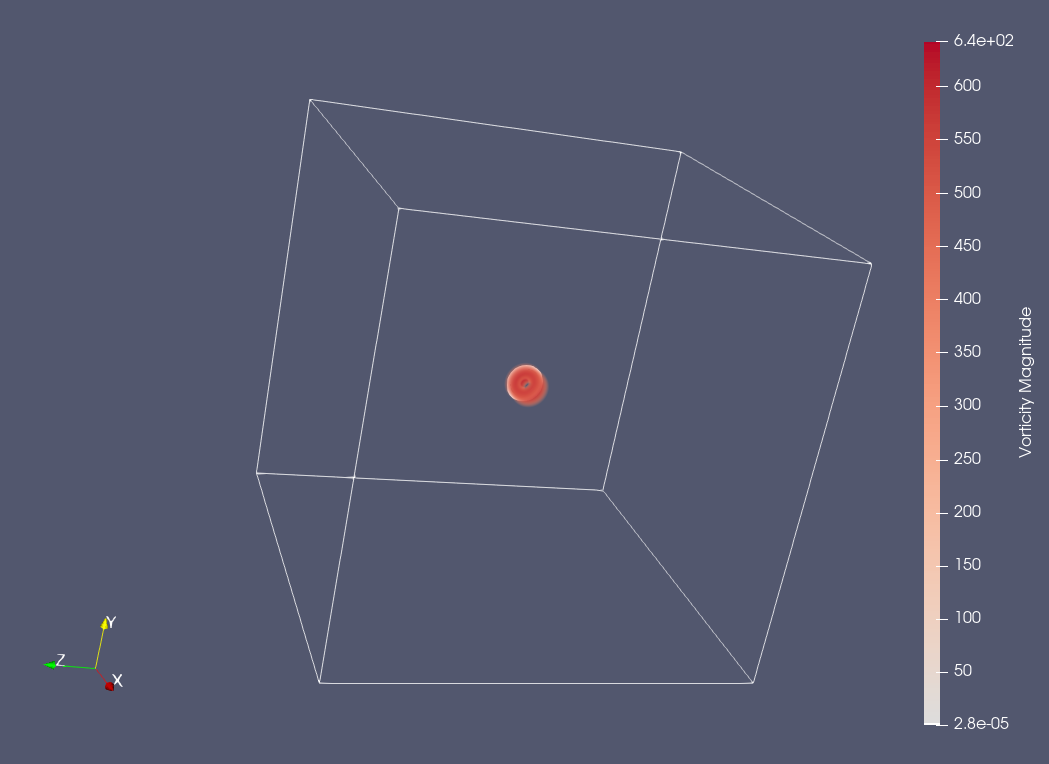}}\qquad\qquad
\subfigure[$T=0.1$ (symmetric)]{\includegraphics[width=0.45\textwidth]{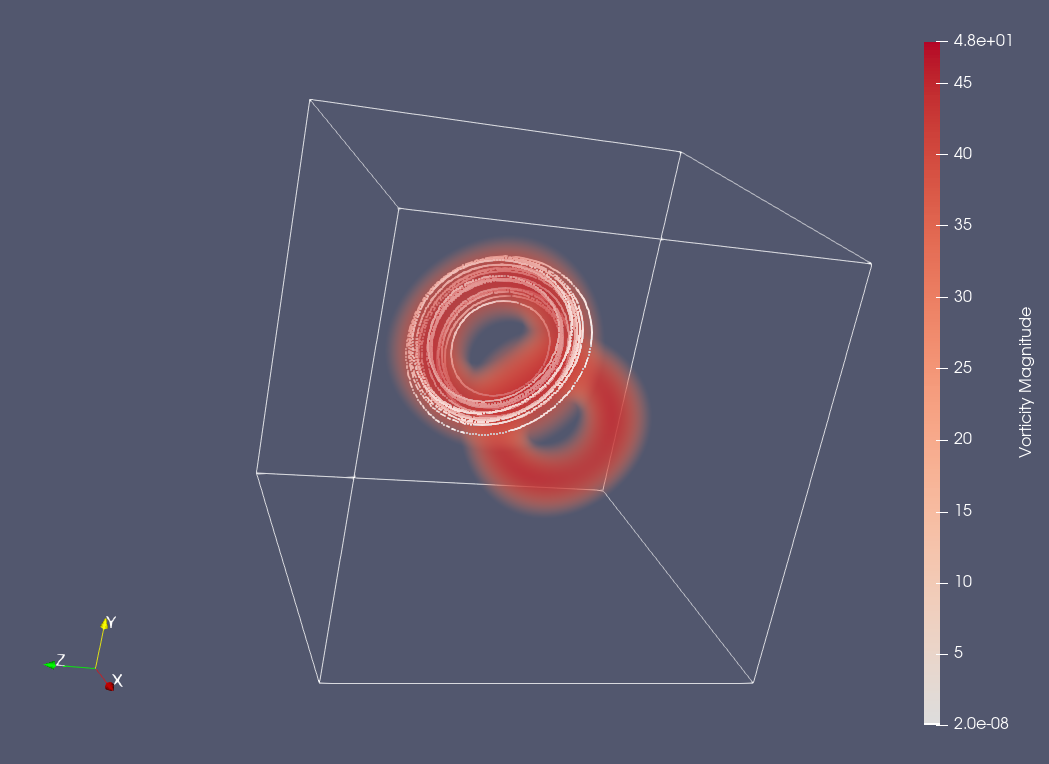}}}
\mbox{\subfigure[$T=0.2$ (symmetric)]{\includegraphics[width=0.45\textwidth]{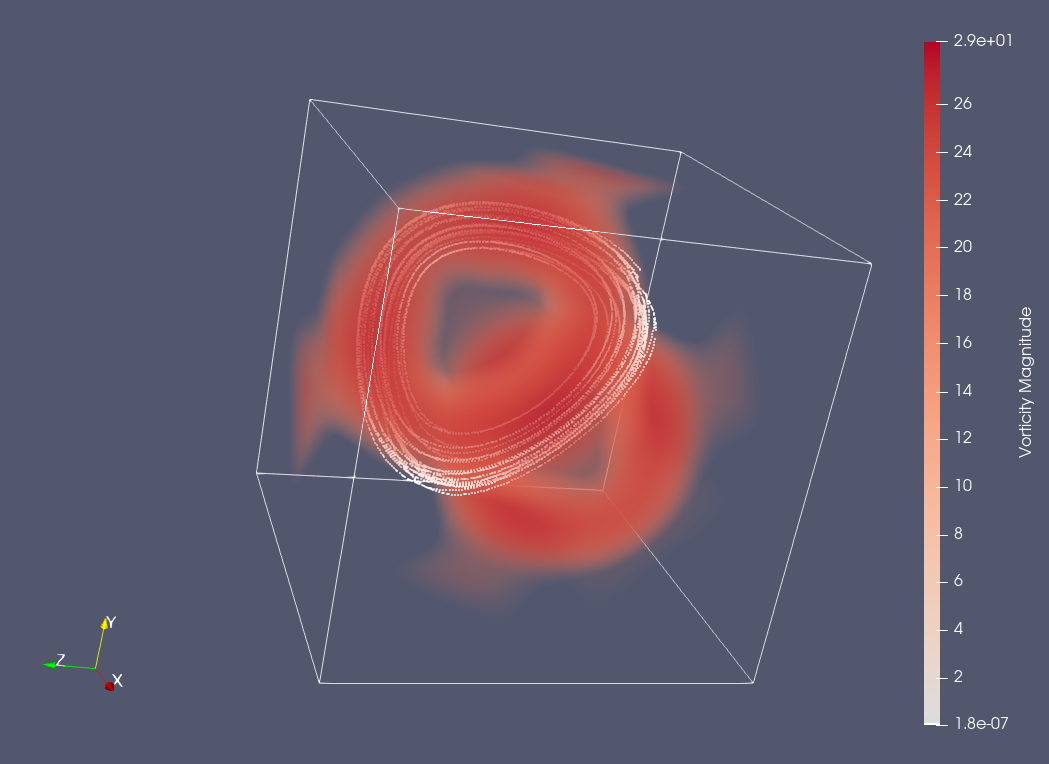}}\qquad\qquad
\subfigure[$T=0.2$ (asymmetric)]{\includegraphics[width=0.45\textwidth]{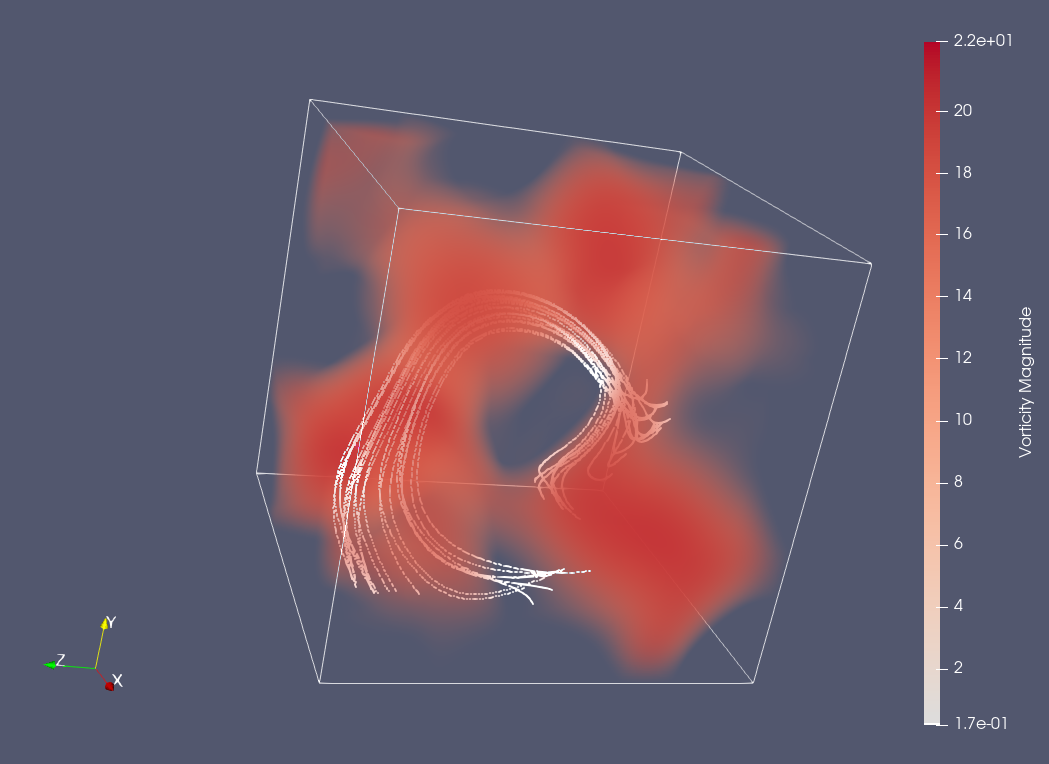}}} 
\mbox{\subfigure[$T=0.3$ (symmetric) ]{\includegraphics[width=0.45\textwidth]{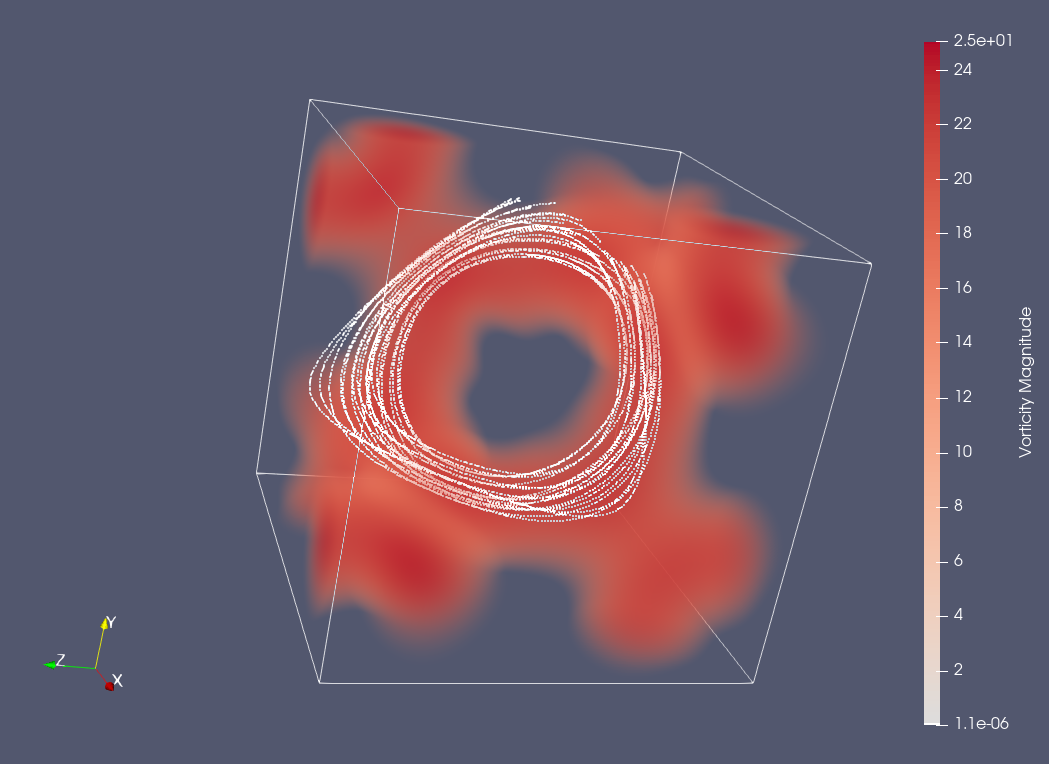}}\qquad\qquad
\subfigure[$T=0.3 = \tTE$ (asymmetric)]{\includegraphics[width=0.45\textwidth]{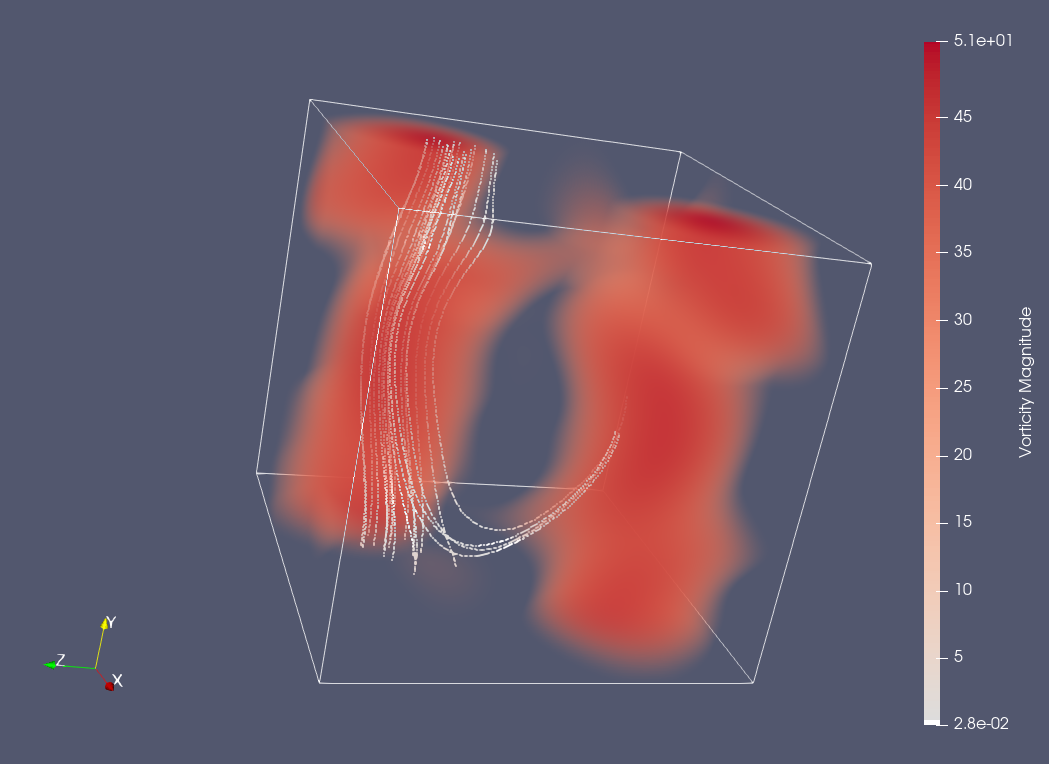}}}
\mbox{\subfigure[$T=0.33$ (symmetric) ]{\includegraphics[width=0.45\textwidth]{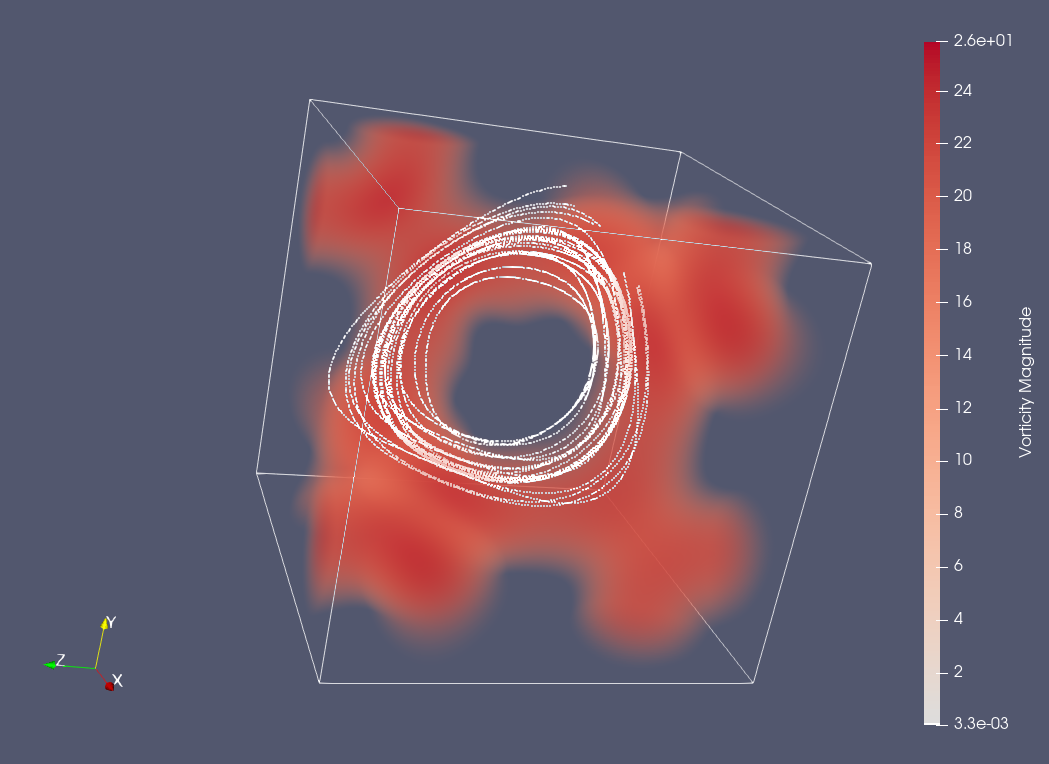}}\qquad\qquad
\subfigure[$T=0.4$ (asymmetric)]{\includegraphics[width=0.45\textwidth]{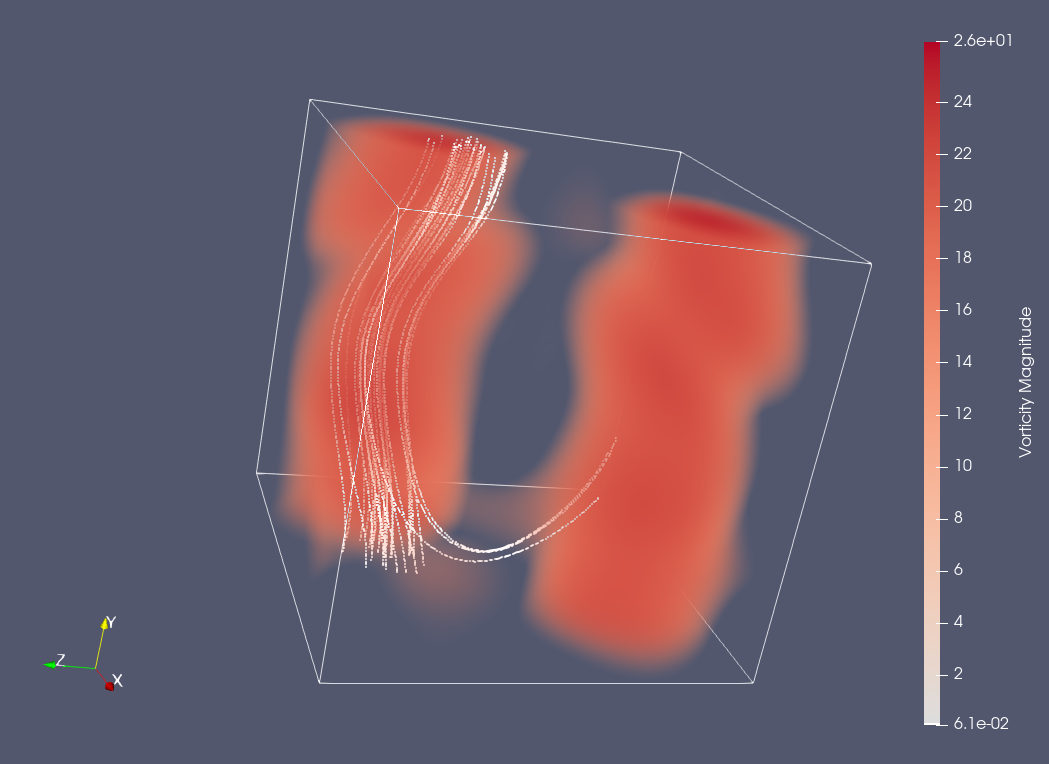}}}
\caption{Optimal initial conditions (a) $\tuE$ obtained by solving the
  instantaneous optimization problem \ref{pb:maxdEdt_E} and (b--h)
  $\tuET$ obtained by solving the finite-time optimization problem
  \ref{pb:maxET} for the initial enstrophy $\E_0 = 100$ and indicated
  lengths $T$ of the time interval.  {Optimal initial conditions from
    both the symmetric and asymmetric branch are shown.} Shades of red
  correspond to the magnitude of the vorticity $|\left(\bnabla \times
    \tuE\right)(\x)|$ or $|\left(\bnabla \times \tuET\right)(\x)|$
  (see the color bars), whereas white curves represent vortex lines
  chosen to pass through regions with strong vorticity.}
\label{fig:tuE0_100}
\end{center}
\end{figure}

\begin{figure}
\begin{center}
\includegraphics[width=0.6\textwidth]{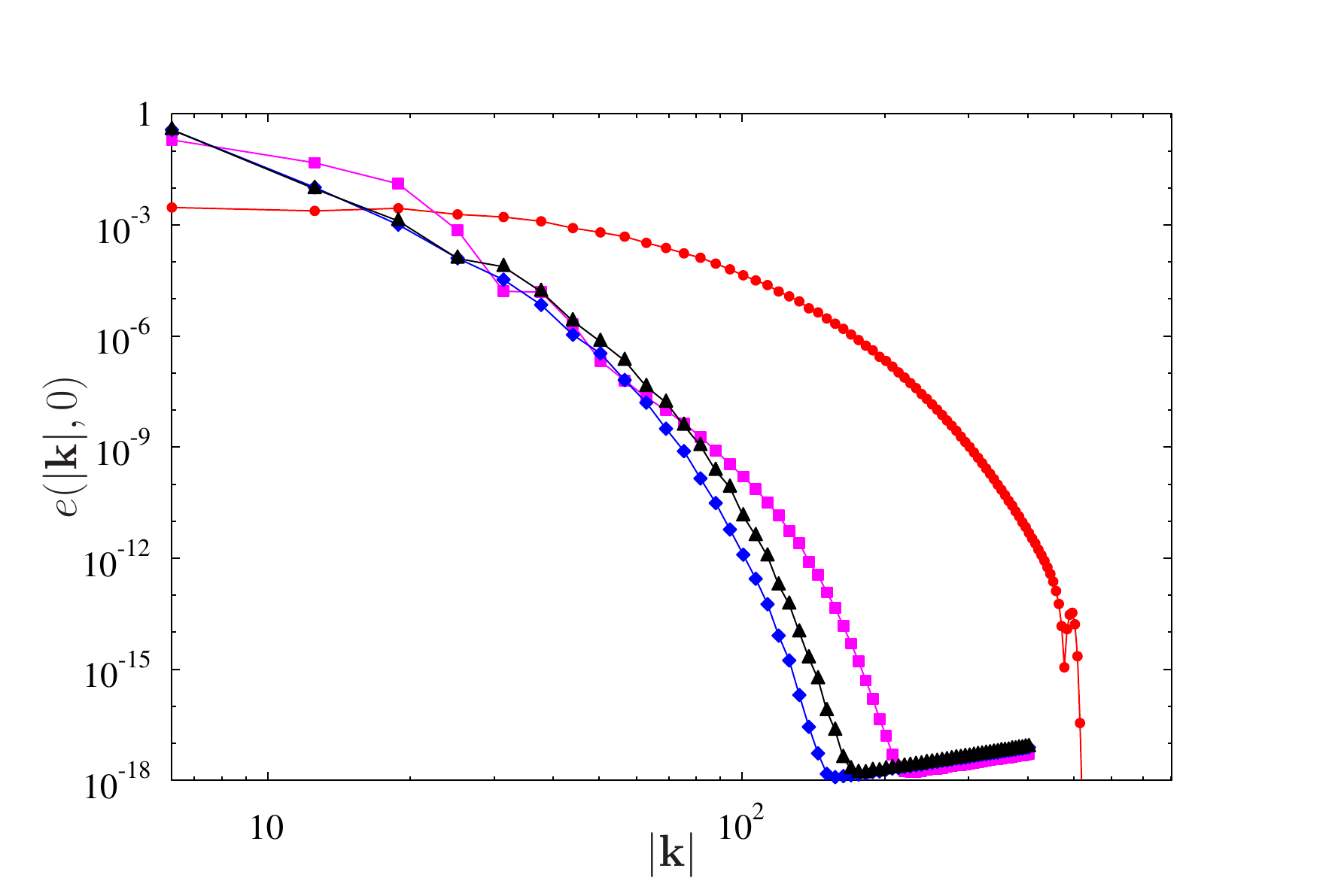}
\caption{Energy spectra of the optimal initial conditions $\tuE$ and
  $\tuET$ obtained by solving the instantaneous optimization problem
  \ref{pb:maxdEdt_E} (red circles) and the finite-time optimization
  problem \ref{pb:maxET} for the initial enstrophy $\E_0 = 100$ and $T
  = 0.1$ (pink squares), $T = 0.2$ (blue diamonds) and $T = 0.3$
  (black triangles). In the latter case the optimal initial conditions
  $\tuET$ come from the asymmetric branch.}
\label{fig:spec_u0_T_E0_100}
\end{center}
\end{figure}

We now move on to characterize the structure of the optimal initial
conditions which give rise to the maximum enstrophy growth reported in
figures \ref{fig:Et}, \ref{fig:maxEt_vs_T} and \ref{fig:maxEt_vs_E0}.
First, in figure \ref{fig:tuE0_100} we analyze the structure of
{both the symmetric and asymmetric} initial conditions $\tuET$
obtained by solving the finite-time optimization problem
\ref{pb:maxET} for a fixed $\E_0 = 100$ and different $T$, and also
show the instantaneously optimal initial condition $\tuE$ obtained as
a solution of Problem \ref{pb:maxdEdt_E} for the same value of $\E_0$
\citep{ap16}. These initial conditions are presented in figure
\ref{fig:tuE0_100} in terms of their vorticity $\bnabla \times \tuET$
and $\bnabla \times \tuE$, whose magnitude is visualized via volume
rendering and, in addition, we also show a number of vortex lines
(i.e., lines everywhere tangent to the vorticity field) chosen to pass
through regions with strong vorticity.  Such an approach allows us to
simultaneously assess both the intensity and the structure of the
vorticity field. In figure \ref{fig:tuE0_100} we see that as $T$
increases the structure of the optimal initial condition gradually
changes from two colliding vortex rings characterizing the
instantaneous maximizers $\tuE$ \citep{ld08,ap16} to a more complex
vorticity distribution filling the entire flow domain. {There are
  also evident differences between the optimal initial conditions
  belonging to the symmetric and asymmetric branches, cf.~figures
  \ref{fig:tuE0_100}(b,c,e,g) versus figures \ref{fig:tuE0_100}(d,f,h)
  --- while the former retain the structure of deformed rings, the
  latter develop a tubular form. Since the asymmetric optimal initial
  conditions $\tuET$ give rise to a much larger growth of enstrophy,
  cf.~figure \ref{fig:maxEt_vs_E0}, hereafter we will exclusively
  focus on this case.}

The fact that the optimal initial conditions $\tuET$ obtained by
solving the finite-time optimization problem \ref{pb:maxET} are less
localized than the maximizers of the instantaneous optimization
problem \ref{pb:maxdEdt_E} is also evident from figure
\ref{fig:spec_u0_T_E0_100} showing their energy spectra $e(|\k|,0)$
defined as
\begin{equation}
e(|\k|,t) := \int_{S_{|\k|}} |\k|^2 \left|\left[\widehat{{\mathbf{u}}}(t)\right]_{\k}\right|^2 \, d\sigma,
\label{eq:e}
\end{equation}
where $\left[\widehat{{\mathbf{u}}}(t)\right]_{\k}$ are the Fourier
coefficients of the velocity field $\u(t)$, $\sigma$ is the solid
angle in the wavenumber space and $S_{|\k|}$ denotes the sphere of
radius $|\k|$ in this space, such that $\K(\u(t)) = \int_0^{\infty}
e(k,t)\, dk$ (with some abuse of notation justified by simplicity,
here we have treated the wavevector $\k$ as a continuous rather than
discrete variable). In figure \ref{fig:spec_u0_T_E0_100} we see that
while all the fields $\tuET$ remain real-analytic (their energy
spectra vanish exponentially fast for high wavenumbers $|\k|$),
interestingly, the ones corresponding to intermediate lengths $T
\approx 0.2$ of the optimization interval are in fact the smoothest in
the sense that their energy spectra $e(|\k|,0)$ decay fastest with
$|\k|$. In particular, the energy spectrum of the instantaneously
optimal initial condition $\tuE$ decays much slower.  These
observations indicate that the finite-time optimization problems
\ref{pb:maxET} corresponding to intermediate and longer time intervals
$T$ are in fact less demanding in terms of space resolution $N$, which
is why for larger values of the initial enstrophy $\E_0$ it is easier
to perform continuation with respect to $\E_0$ at some fixed
intermediate $T$, rather than the other way round starting from the
instantaneous maximizers $\tuE$ which are hard to compute when $\E_0 >
100$.

\begin{figure}
\begin{center}
\mbox{\subfigure[$\E_0=150$, $\tTE=0.27$]{\includegraphics[width=0.3\textwidth]{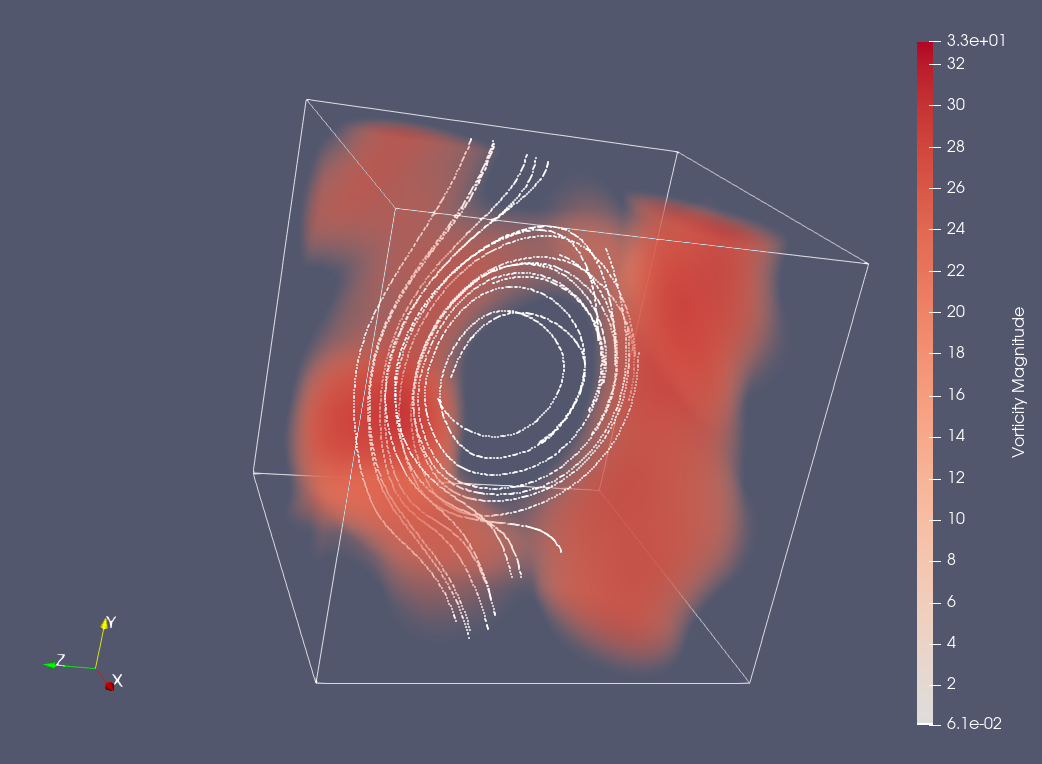}}\qquad
\subfigure[$\E_0=200$, $\tTE=0.23$]{\includegraphics[width=0.3\textwidth]{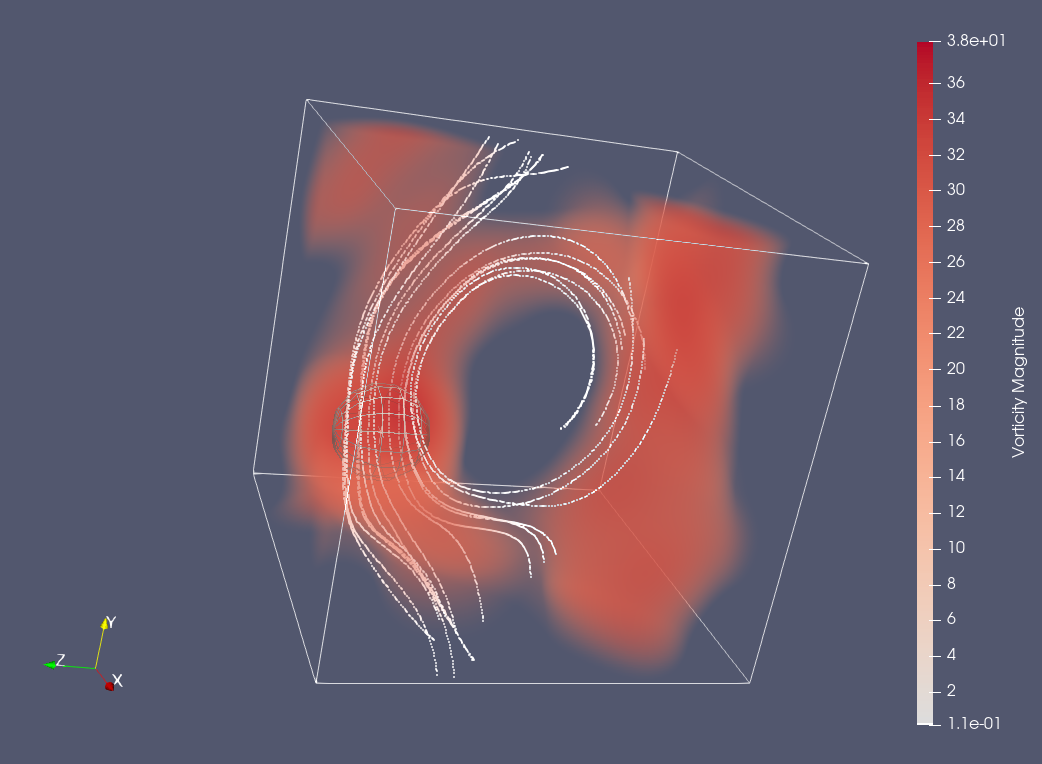}}\qquad
\subfigure[$\E_0=300$, $\tTE=0.21$]{\includegraphics[width=0.3\textwidth]{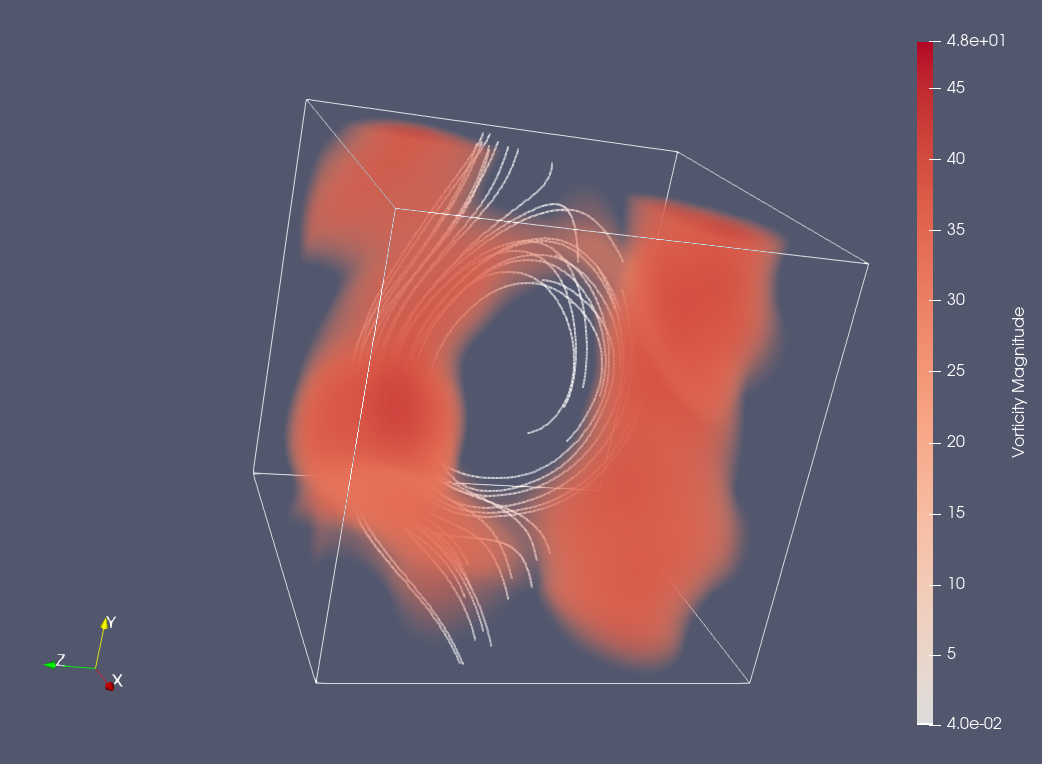}}}
\mbox{\subfigure[$\E_0=400$, $\tTE=0.19$]{\includegraphics[width=0.3\textwidth]{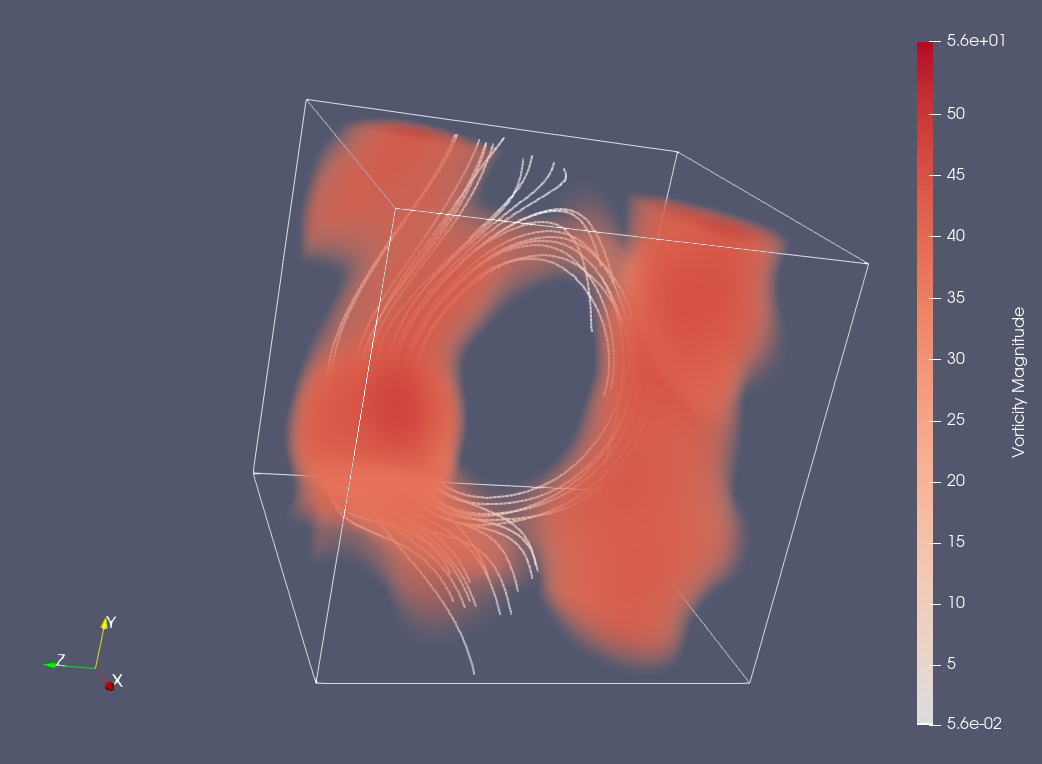}}\qquad
\subfigure[$\E_0=500$, $\tTE=0.17$]{\includegraphics[width=0.3\textwidth]{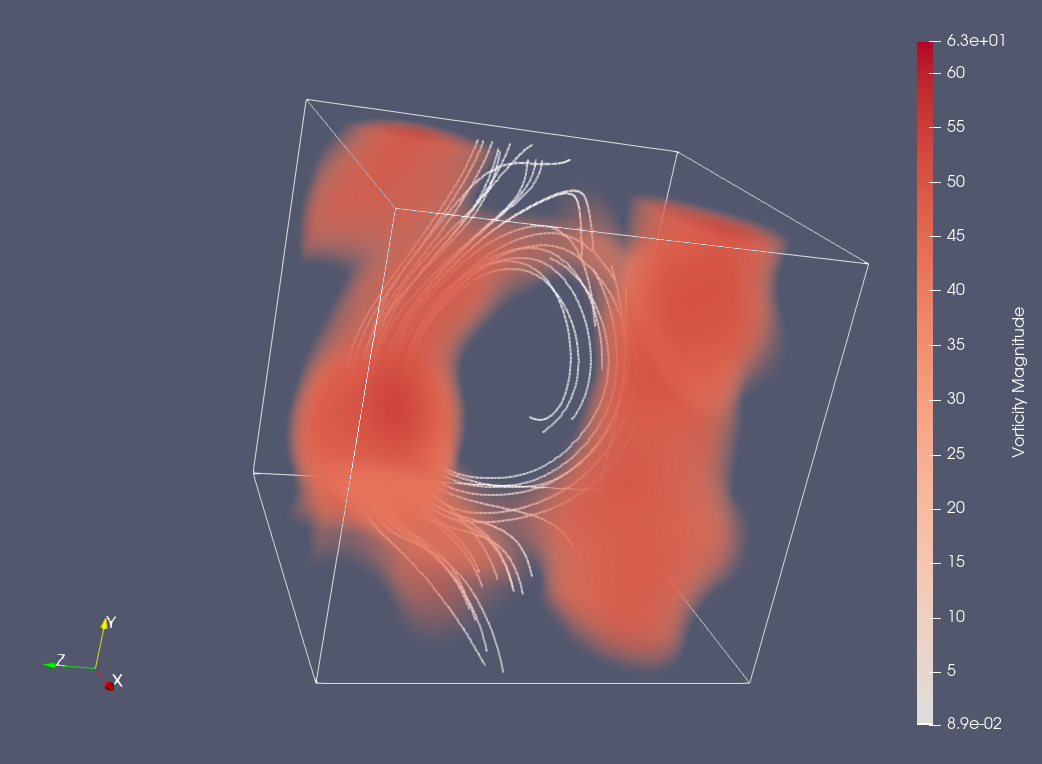}}\qquad
\subfigure[$\E_0=600$, $\tTE=0.15$]{\includegraphics[width=0.3\textwidth]{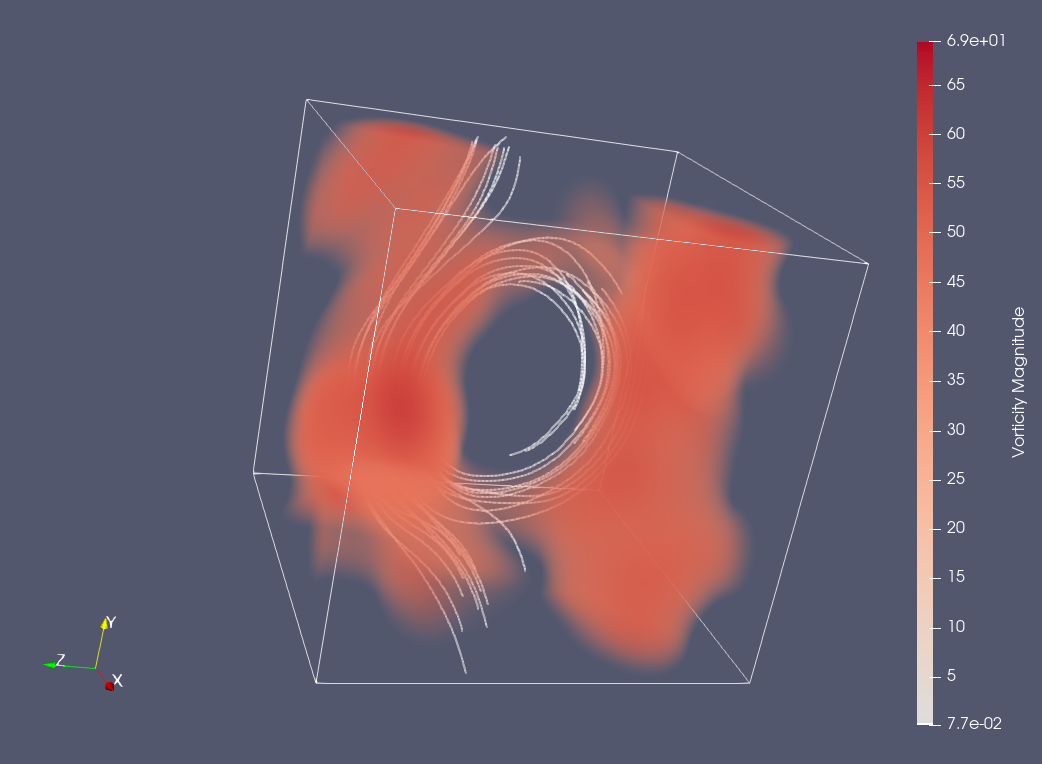}}} 
\mbox{\subfigure[$\E_0=700$, $\tTE=0.14$]{\includegraphics[width=0.3\textwidth]{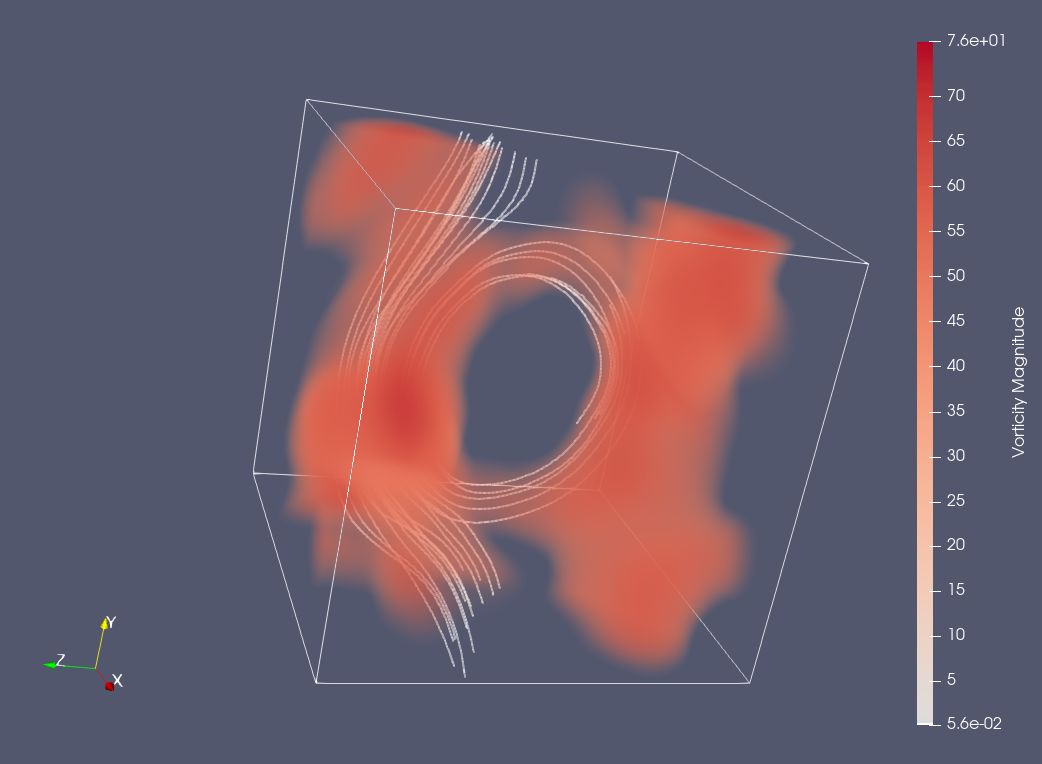}}\qquad
\subfigure[$\E_0=850$, $\tTE=0.13$]{\includegraphics[width=0.3\textwidth]{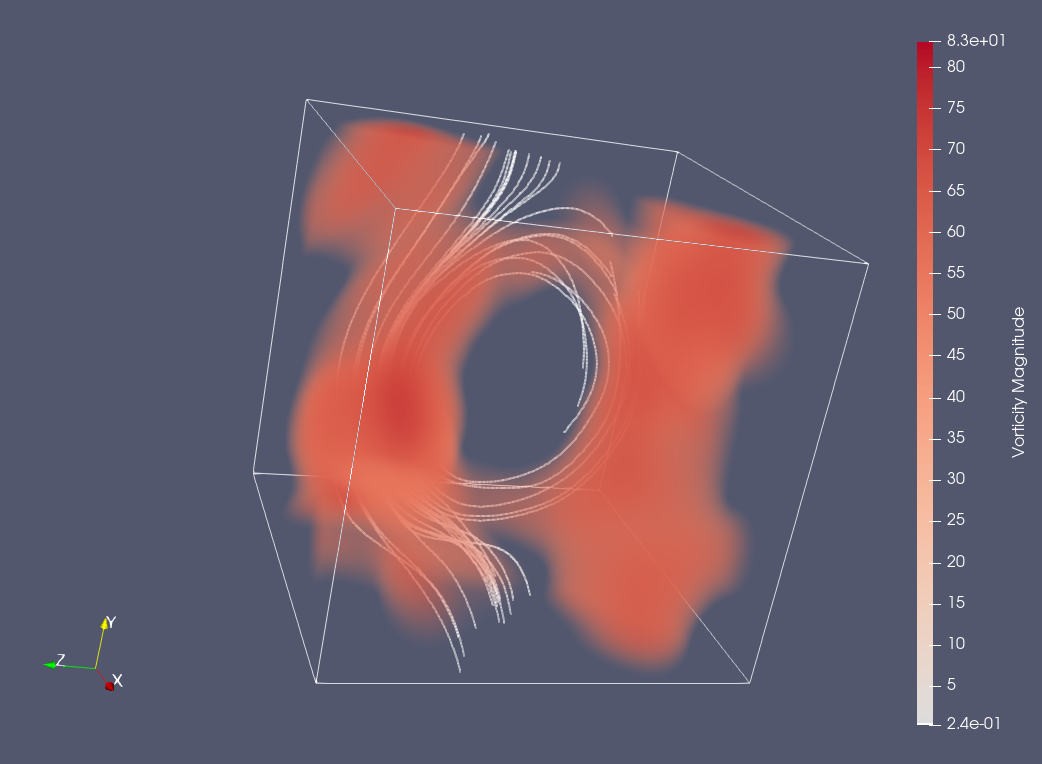}}\qquad
\subfigure[$\E_0=1000$, $\tTE=0.12$]{\includegraphics[width=0.3\textwidth]{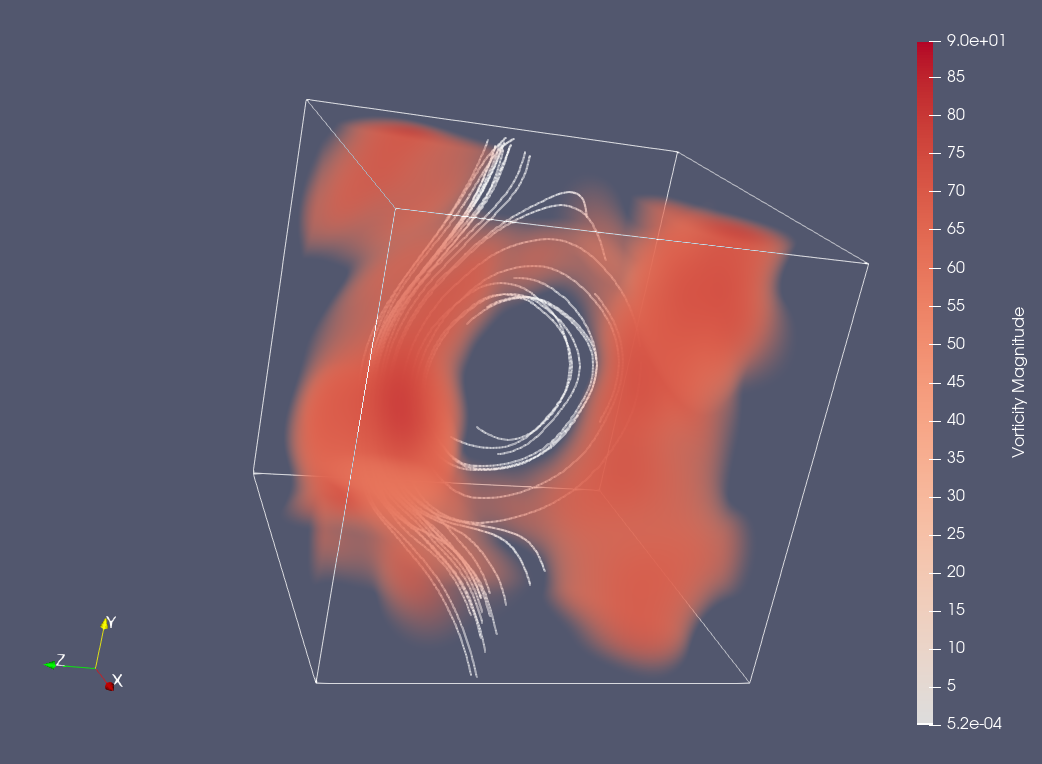}}}
\caption{{Asymmetric} optimal initial conditions $\tuEtT$ obtained by
  solving the finite-time optimization problem \ref{pb:maxET} for
  different indicated values of the initial enstrophy $\E_0$ and the
  corresponding optimal lengths $\tTE$ of the time interval
  (cf.~figure \ref{fig:Tmax_vs_E0}). Shades of red correspond to the
  magnitude of the vorticity $\left|\left(\bnabla \times
      \tuEtT\right)(\x)\right|$ (see the color bars), whereas white
  curves represent vortex lines chosen to pass through regions with
  strong vorticity.}
\label{fig:tuE0T}
\end{center}
\end{figure}

\begin{figure}
\begin{center}
\mbox{\subfigure[$\tomega_1$]{\includegraphics[width=0.3\textwidth]{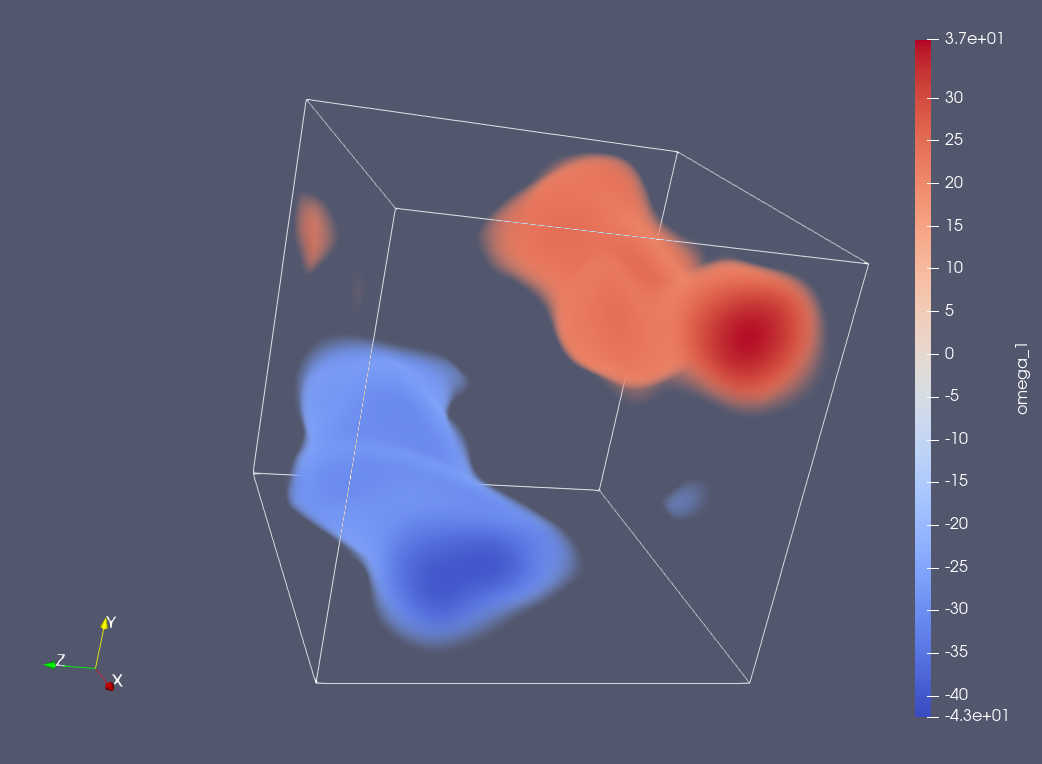}}\qquad
\subfigure[$\tomega_2$]{\includegraphics[width=0.3\textwidth]{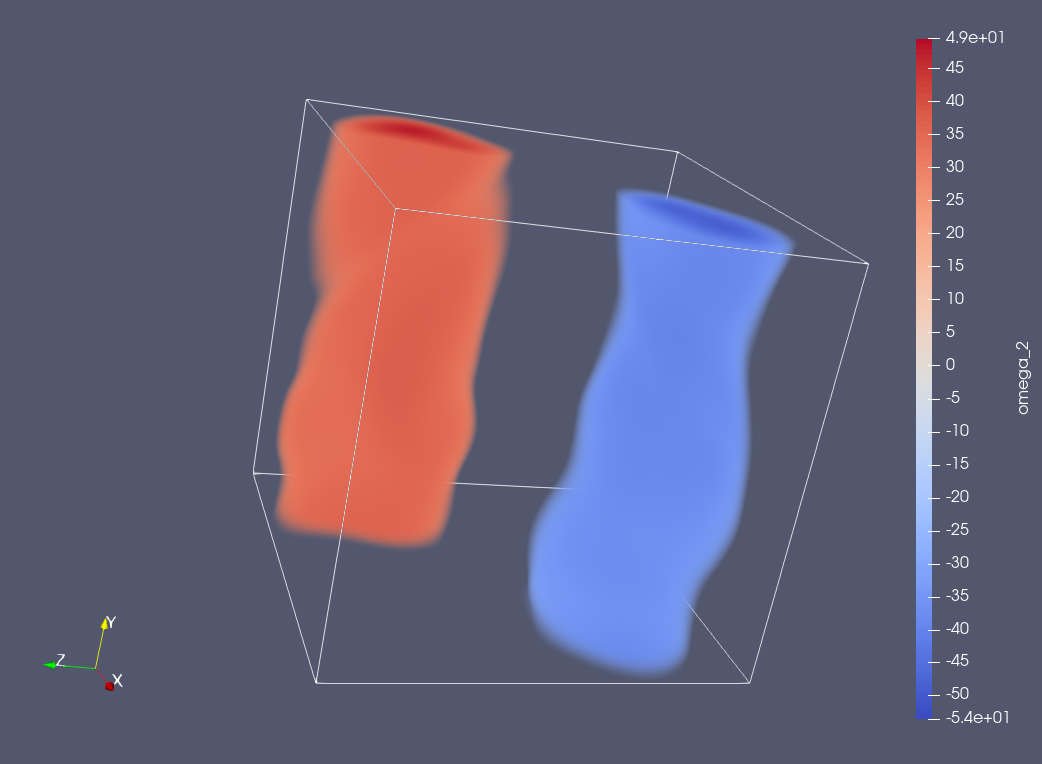}}\qquad
\subfigure[$\tomega_3$]{\includegraphics[width=0.3\textwidth]{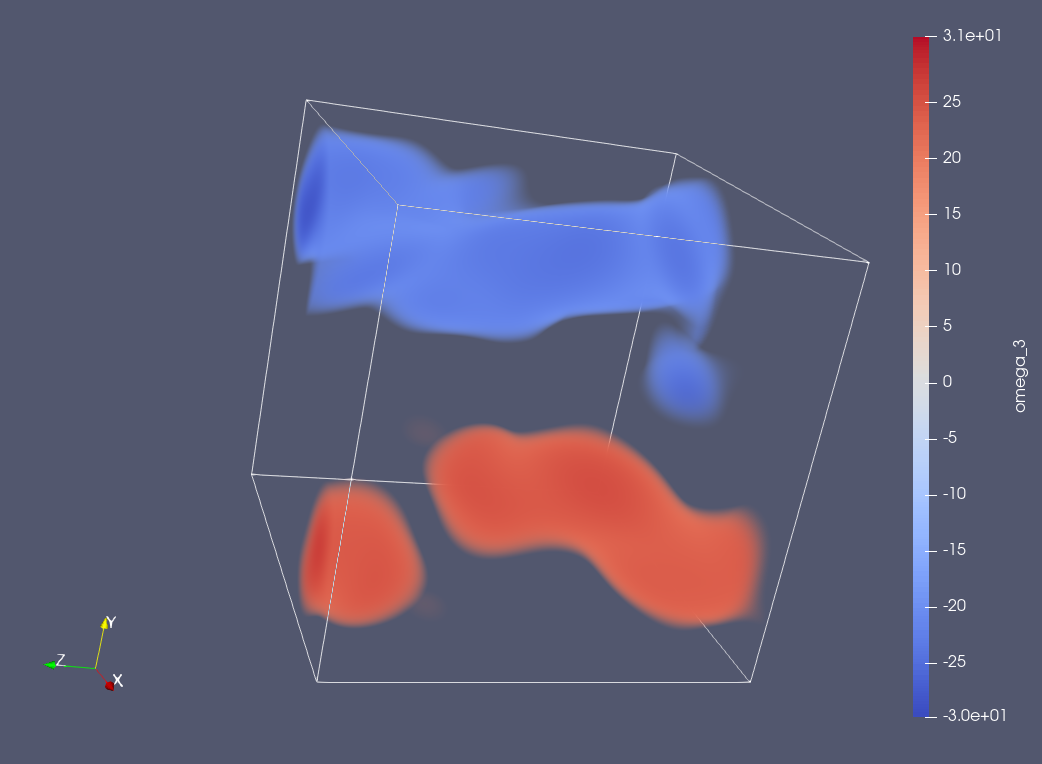}}}
\caption{Vorticity components of the {asymmetric} optimal initial
  condition $\tuEtT$ obtained by solving the finite-time optimization
  problem \ref{pb:maxET} for the initial enstrophy $\E_0 = 500$ and
  the corresponding optimal length $\tTE = 0.17$ of the time interval,
  cf.~figure \ref{fig:tuE0T}(e).}
\label{fig:tubes}
\end{center}
\end{figure}

The optimal {asymmetric} initial conditions $\tuEtT$ obtained for the
optimal time intervals $[0, \tTE]$ are shown in figure \ref{fig:tuE0T}
for increasing values of the initial enstrophy $\E_0$.  We note that
the structure of these fields does not much change with $\E_0$, except
that, as expected, the vorticity magnitude $\left|\left(\bnabla \times
    \tuEtT\right)(\x)\right|$ in the vortex regions increases.  In
order to shed additional light on the structure of these optimal
initial conditions, the field $\tuEtT$ corresponding to the initial
enstrophy $\E_0 = 500$, cf.~figure \ref{fig:tuE0T}(e), is further
analyzed in figure \ref{fig:tubes} in terms of its different vorticity
components $[\tomega_1(t), \tomega_2(t), \tomega_3(t)]^T := \bnabla
\times \u(t)$. We observe that the optimal initial condition has in
fact the form of three perpendicular pairs of anti-parallel vortex
tubes perturbed near the regions where they intersect (these
perturbations allow the vorticity field to satisfy the divergence-free
condition \mbox{$\bnabla\cdot\left(\bnabla \times \tuEtT \right) =
  0$}). The relative magnitudes of the three vorticity components are
given by the approximate relations $0.76 \, : \, 1 \, : \, 0.63$
representing the ratios $\tomega_1 \, : \; \tomega_2 \, : \,
\tomega_3$.

\begin{figure}
\begin{center}
\includegraphics[width=0.6\textwidth]{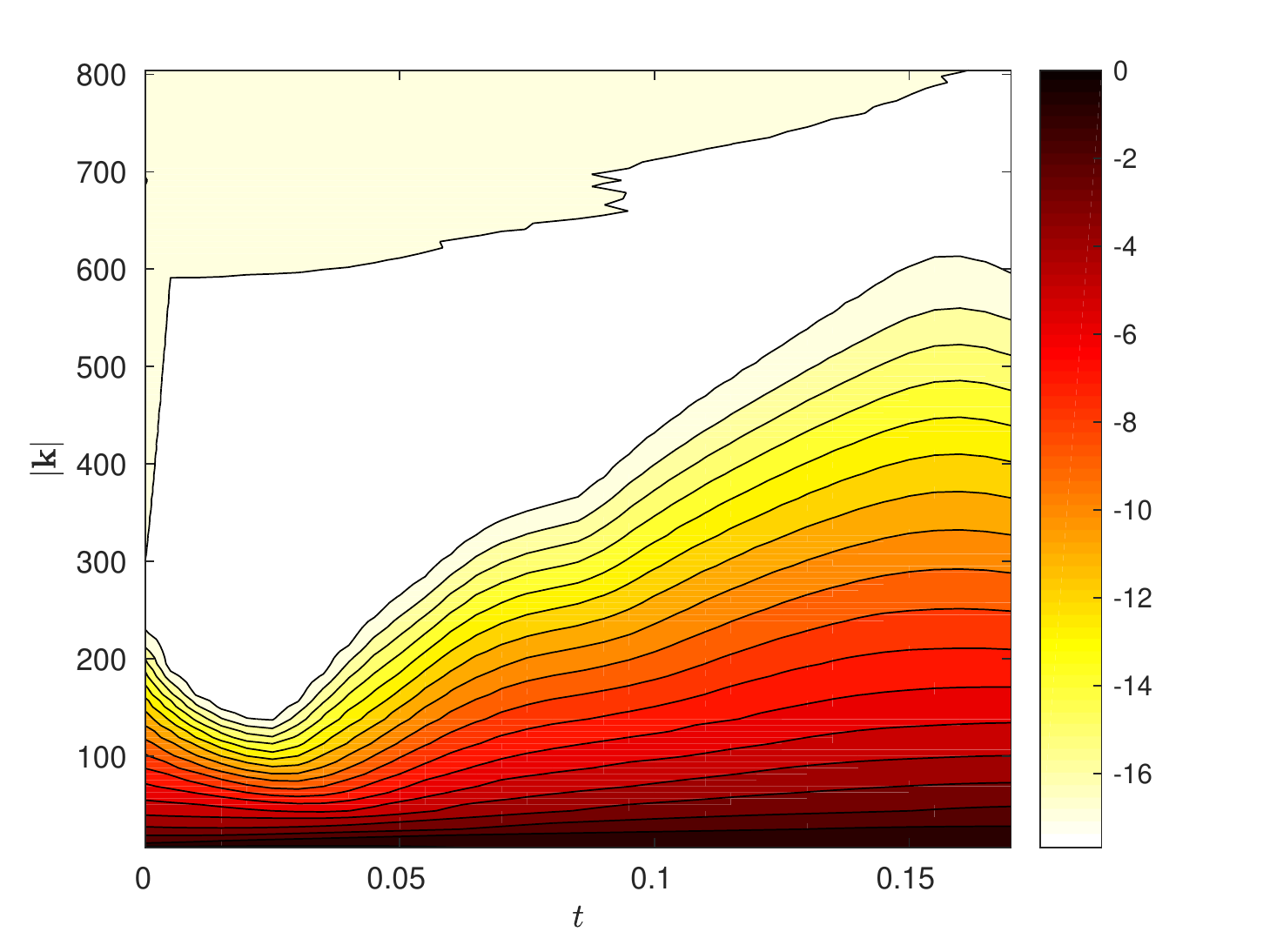}
\caption{Time evolution of the energy spectrum in the solution of the
  Navier-Stokes system \eqref{eq:NSE3D} with the optimal
  {asymmetric} initial condition $\tuET$ obtained by solving the
  the finite-time optimization problem \ref{pb:maxET} for the initial
  enstrophy $\E_0 = 500$ and $\tTE = 0.17$. The contour lines
  represent the level sets of the function $\log_{10} e(|\k|,t)$ with
  the distance between two nearby lines corresponding to one order of
  magnitude.}
\label{fig:spectra_cascade}
\end{center}
\end{figure}

Next, we analyze details of the flow evolution corresponding to the
optimal initial conditions $\tuET$. In figure
\ref{fig:spectra_cascade} we illustrate the evolution of the energy
spectrum $e(|\k|,t)$, cf.~\eqref{eq:e}, in time in the solution of the
Navier-Stokes system \eqref{eq:NSE3D} with the optimal {asymmetric}
initial condition $\tuET$ obtained by solving the the finite-time
optimization problem \ref{pb:maxET} with $\E_0 = 500$ and $\tTE =
0.17$, {cf.~figure \ref{fig:tubes}} (qualitatively similar behavior is
also observed for {other values of $\E_0 > 100$}). We notice that at
early stages of the flow evolution the energy spectrum recedes which
corresponds to the initially very slow growth of enstrophy already
noted in figure \ref{fig:Et}.  Then, the energy begins to flow towards
larger wavenumbers such that the energy spectrum becomes the most
developed {close to} the final time $t = \tTE$ when the enstrophy
maximum is achieved. {As regards this second stage, there are clearly
  two phases ending at instances of time corresponding approximately
  to the times when the ordering of $\E_1(\u(t))$, $\E_2(\u(t))$ and
  $\E_3(\u(t))$ changes, cf.~figure \ref{fig:E123}(c). This indicates
  that there are in fact two reconnection events occurring during the
  flow evolution: one approximately in the middle of the time interval
  $[0,\tTE]$ and another one close to its end.}

\begin{figure}
\begin{center}
\mbox{\subfigure[$t = 0.0$]{\includegraphics[width=0.45\textwidth]{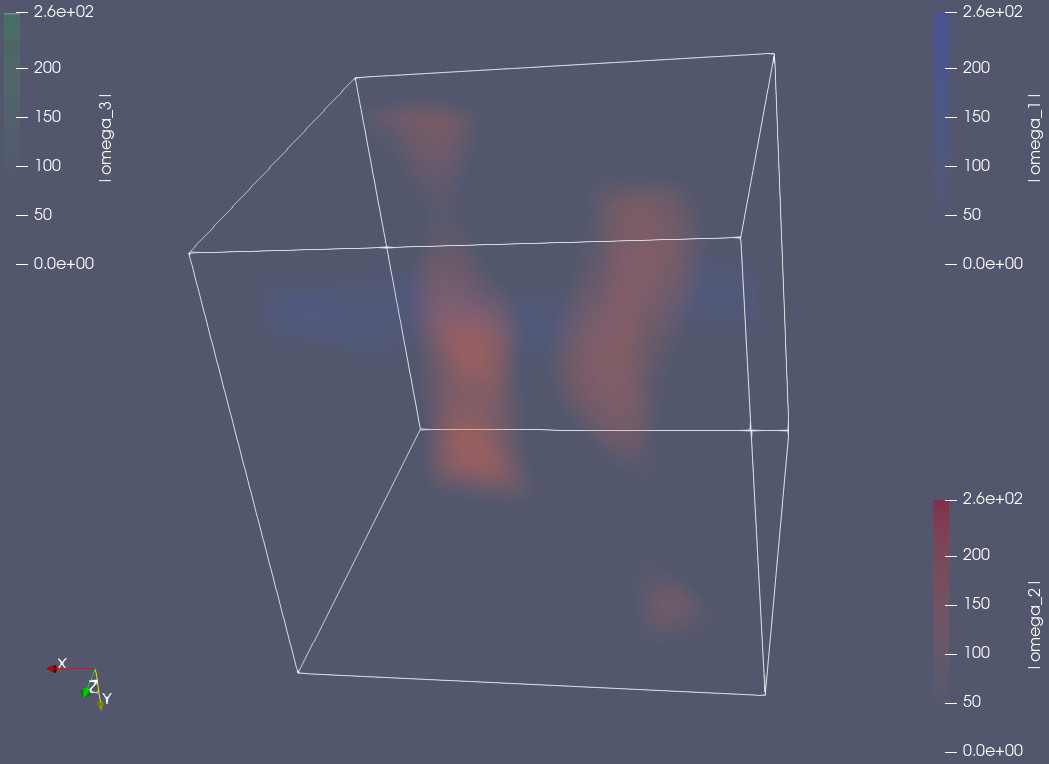}}\qquad
\subfigure[$t = 0.04$]{\includegraphics[width=0.45\textwidth]{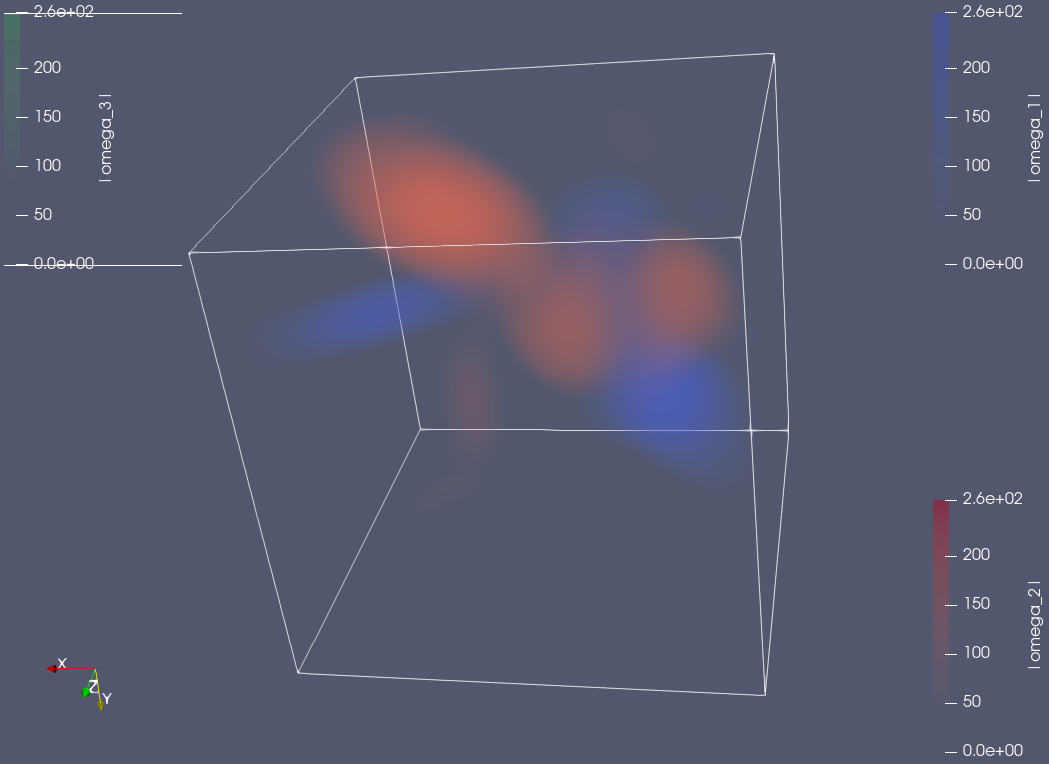}}}
\mbox{\subfigure[$t = 0.09$]{\includegraphics[width=0.45\textwidth]{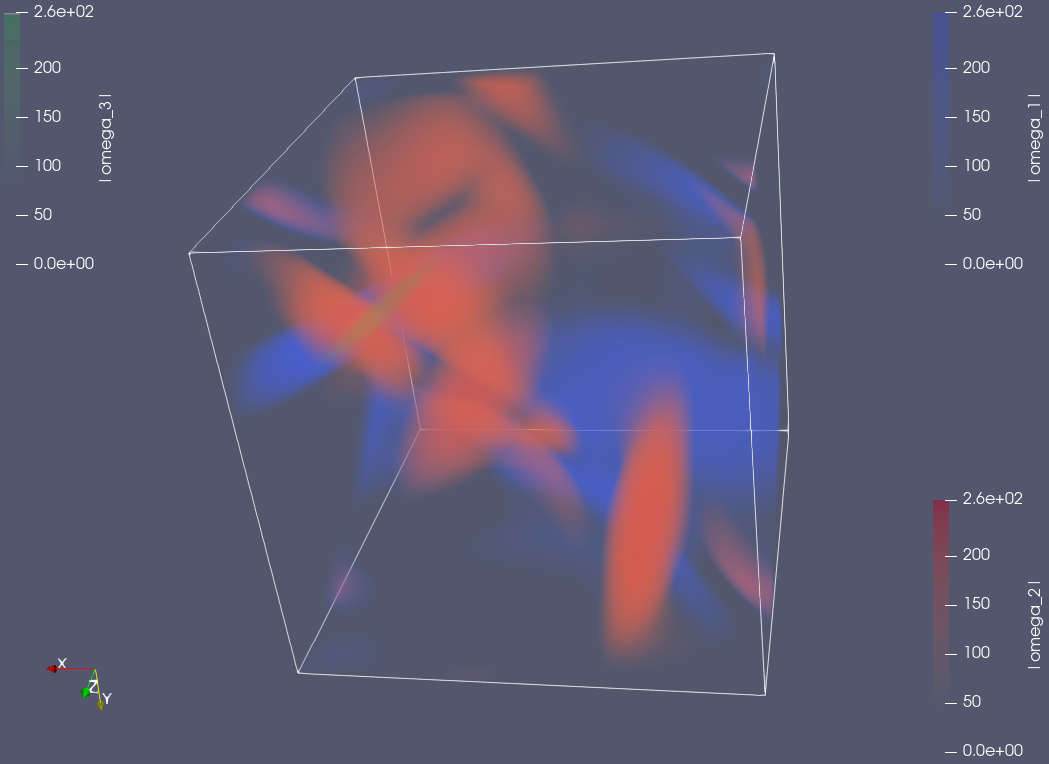}}\qquad
\subfigure[$t = 0.11$]{\includegraphics[width=0.45\textwidth]{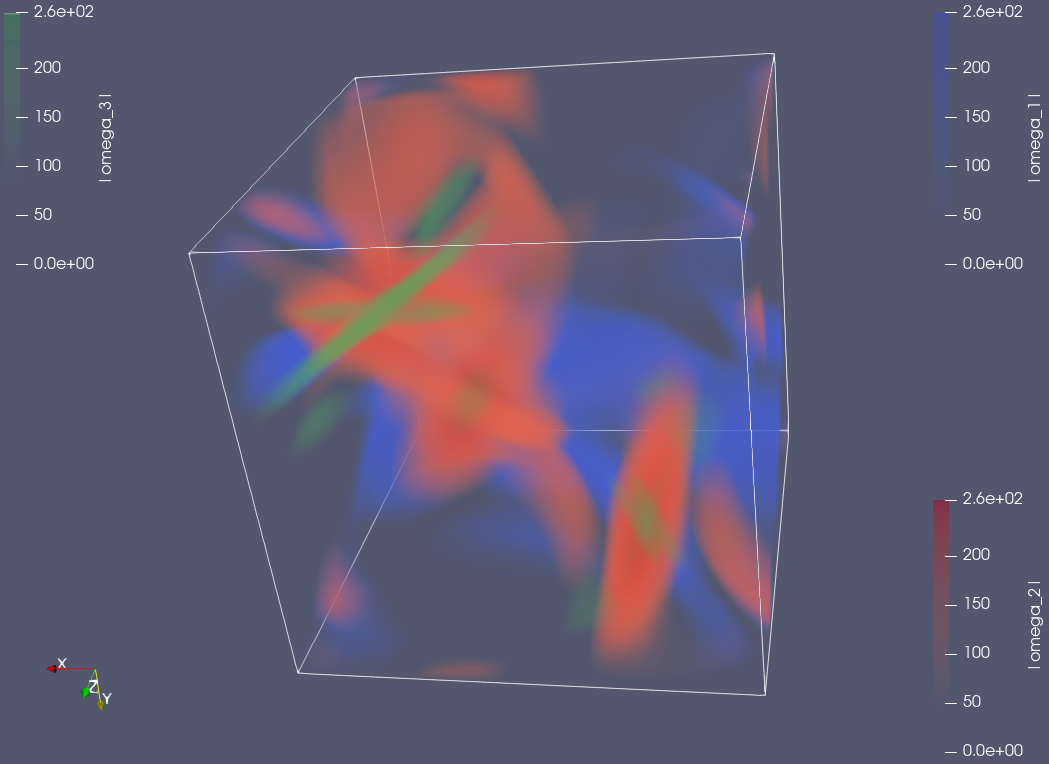}}}
\mbox{\subfigure[$t = 0.14$]{\includegraphics[width=0.45\textwidth]{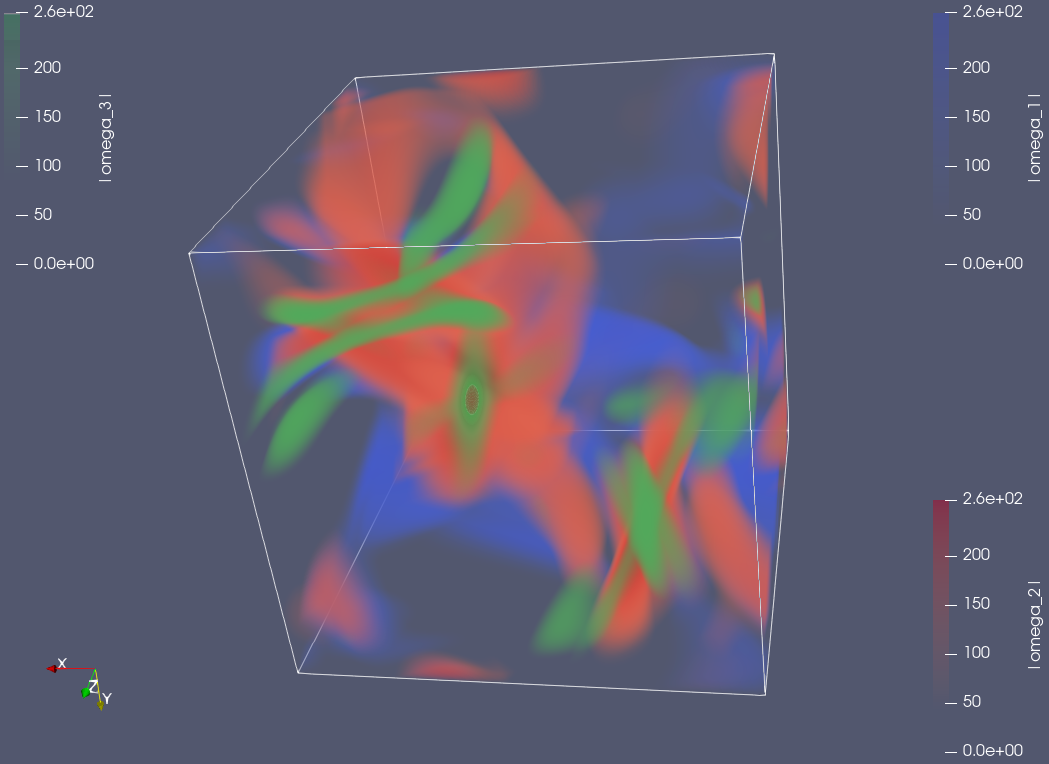}}\qquad
\subfigure[$t = 0.17 = \tTE$]{\includegraphics[width=0.45\textwidth]{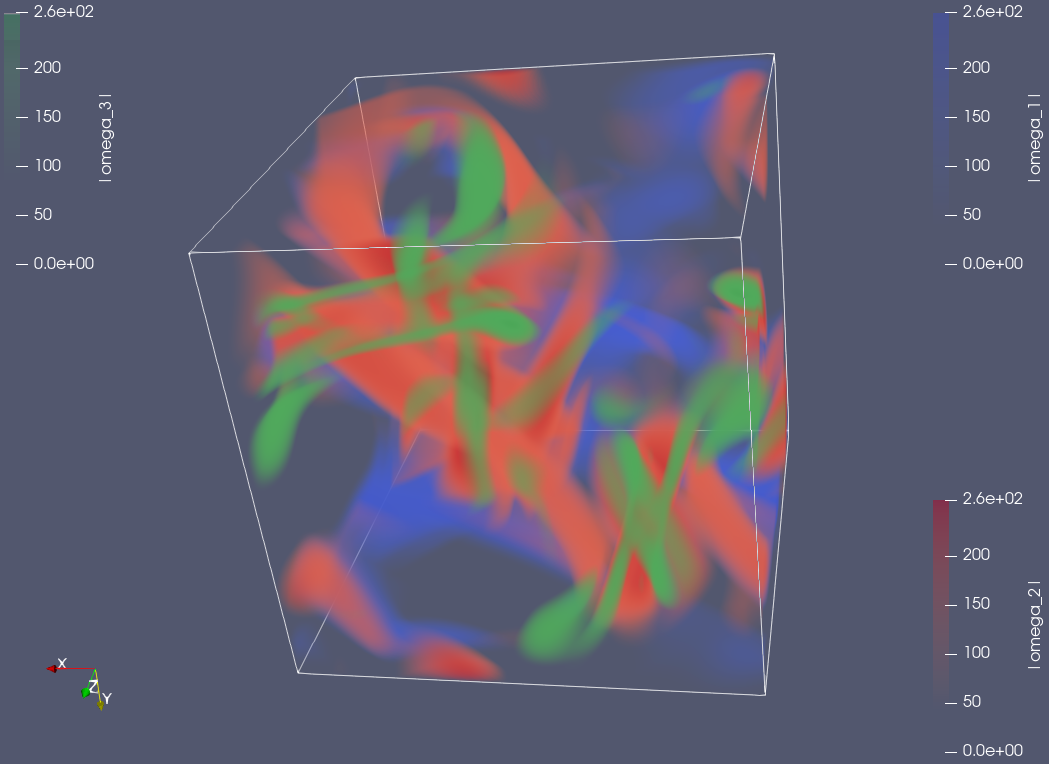}}} 
\caption{{Vorticity components (blue) $\tomega_1$, (red) $\tomega_2$
    and (green) $\tomega_3$ at the indicated time instances in the
    solution of the Navier-Stokes system \eqref{eq:NSE3D} with the
    optimal {asymmetric} initial condition $\tuET$ obtained by solving
    the finite-time optimization problem \ref{pb:maxET} for the
    initial enstrophy $\E_0 = 500$ and $\tTE = 0.17$. Supporting Movie
    1 is available on-line.}}
\label{fig:evolution}
\end{center}
\end{figure}
\begin{figure}
\begin{center}
\includegraphics[width=0.6\textwidth]{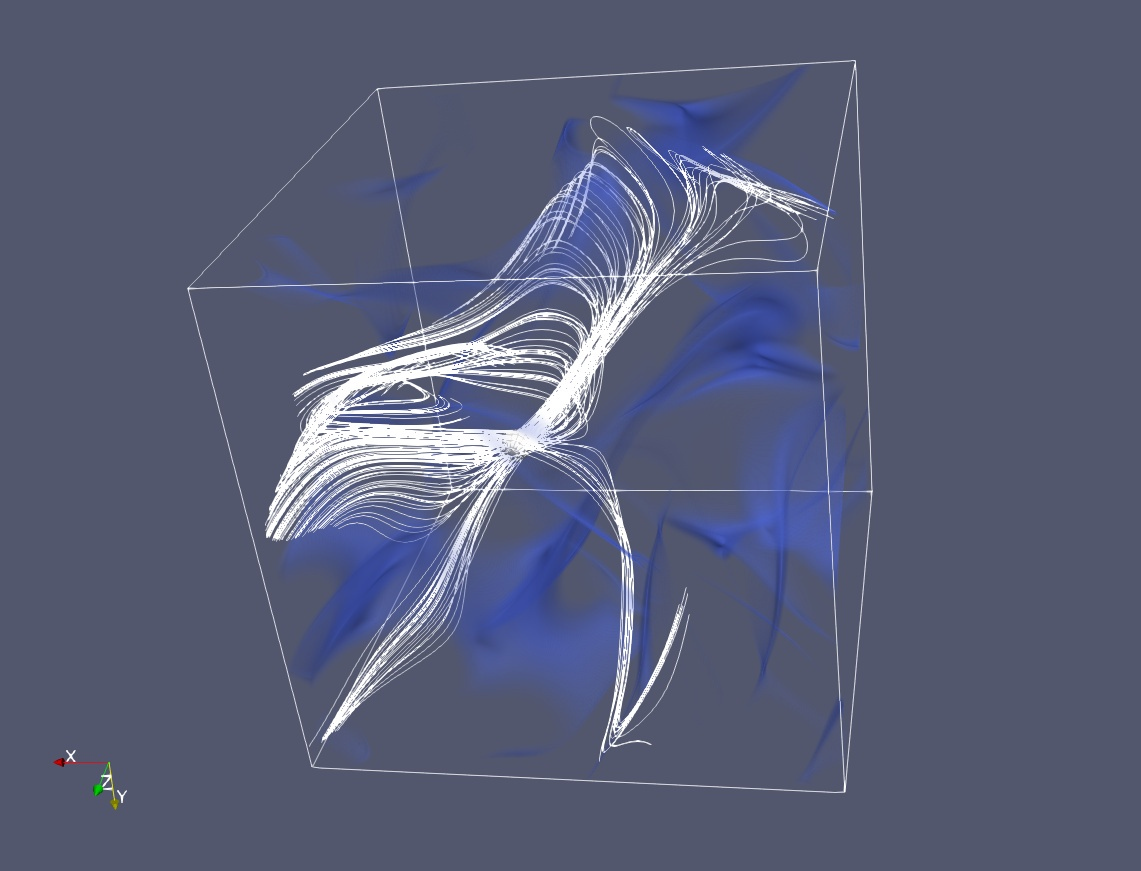}
\caption{Reconnection event occurring at the time $t = 0.09$ in the
  solution of the Navier-Stokes system \eqref{eq:NSE3D} with the
  optimal initial conditions $\tuET$ obtained by solving the
  finite-time optimization problem \ref{pb:maxET} for $\E_0 = 500$ and
  $\tTE = 0.17$, cf.~figure \ref{fig:evolution}(c). White lines
  represent the vortex lines in the neighborhood of the reconnection
  point whereas the blue regions correspond to locations where
  $|(\bnabla \times \u)(\x)| \approx 0$.}
\label{fig:reconnection}
\end{center}
\end{figure}

Finally, we discuss the flow evolution in the physical space and the
three vorticity components $\tomega_1$, $\tomega_2$ and $\tomega_3$
are shown in figure \ref{fig:evolution} at various times $t \in
[0,\tTE]$.  Since the same color scale is used in all panels in this
figure, some features of the optimal initial condition $\tuET$ evident
in figure \ref{fig:tubes} are not visible in figure
\ref{fig:evolution}(a).  In agreement with figure
\ref{fig:spectra_cascade}, we observe that at the early stages of the
flow evolution the vorticity field becomes less localized, cf.~figure
\ref{fig:evolution}(b), which is followed by a rapid development of
small scales, cf.~figures \ref{fig:evolution}(d,e).  The final state
$\bnabla \times \u(\tTE)$ is characterized by a fairly complicated
structure with turbulent-like spatial complexity, cf.~figure
\ref{fig:evolution}(f).  In particular, it features a combination of
tube-shaped and pancake-shaped regions of concentrated vorticity
(animations corresponding to the vorticity fields shown in figure
\ref{fig:evolution} and to optimal flow evolutions obtained for other
values of $\E_0$ and $\tTE$ are available on-line as Movies 1, 2 and
3). To provide more insight into what drives this flow evolution, in
figure \ref{fig:reconnection} we illustrate a typical reconnection
event occurring approximately at the time when the ordering of
$\E_1(\u(t))$, $\E_2(\u(t))$ and $\E_3(\u(t))$ changes, cf.~figures
\ref{fig:E123}(c) and \ref{fig:evolution}(c). We see that the bundle
of vortex lines aligned with the direction $x_2$ and visible in the
middle of the figure is split into two parts going in opposite
directions, one of which attaches to a perpendicular bundle of vortex
lines.  In figure \ref{fig:reconnection} we also visualize regions
where $|(\bnabla \times \u)(\x)| \approx 0$, which is a necessary
condition for vortex reconnection to occur
\citep{HussainDuraisamy2011,VelascoFuentes2017}. These regions are
therefore potential loci of other reconnection events.

\begin{figure}
\begin{center}
\mbox{
\Bmp{0.5\textwidth}
\subfigure[]{\includegraphics[width=1.0\textwidth]{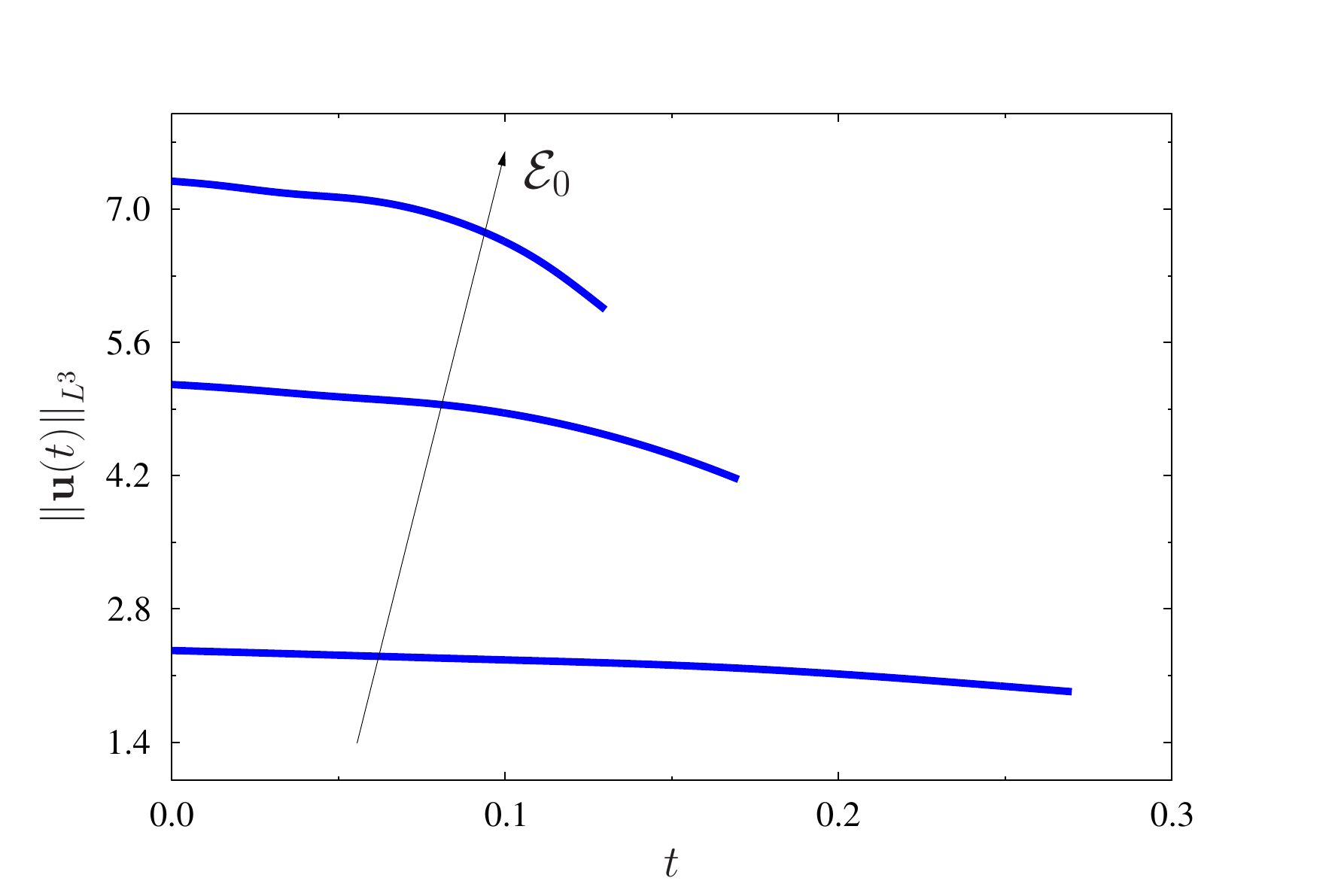}}
\Emp
\Bmp{0.5\textwidth}
\subfigure[]{\includegraphics[width=1.0\textwidth]{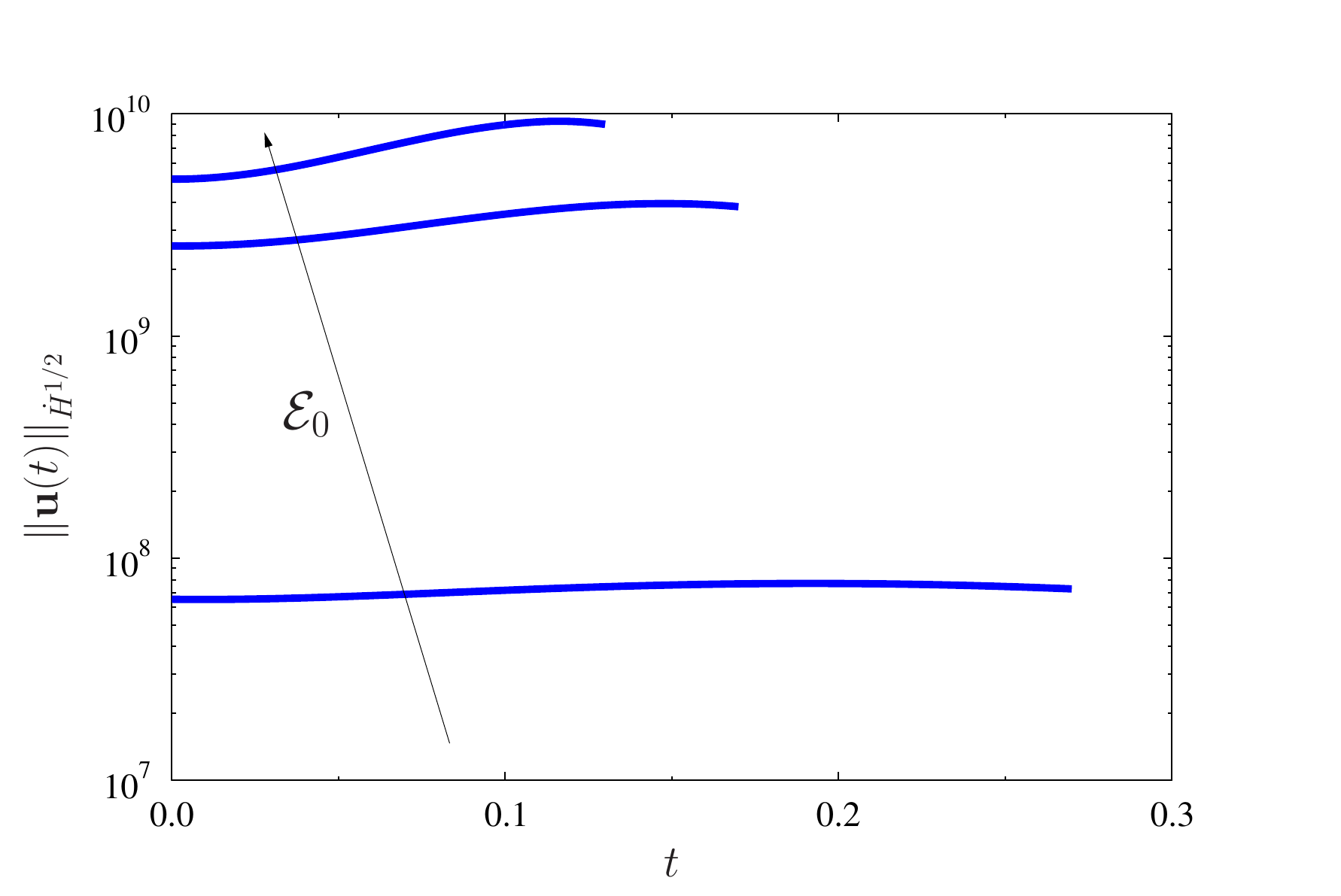}}
\Emp
}
\caption{Evolution of {(a)} the norm $\|\u(t)\|_{L^3}$ {and
    (b) the seminorm $\|\u(t)\|_{\dot{H}^{1/2}}$} of the solutions of
  the Navier-Stokes system \eqref{eq:NSE3D} with the optimal
  {asymmetric} initial conditions $\tuET$ obtained by solving the
  finite-time optimization problem \ref{pb:maxET} for $\E_0 = 100$ and
  $\tTE = 0.27$, $\E_0 = 500$ and $\tTE = 0.17$ and $\E_0 = 1000$ and
  $\tTE = 0.13$. Arrows indicates the trends with the increase of
  $\E_0$.}
\label{fig:cnorms}
\end{center}
\end{figure}
To close this section, in figure \ref{fig:cnorms} we analyze the time
evolution of the $L^3$ {norm and the $H^{1/2}$ seminorm} of the
solutions of the Navier-Stokes system \eqref{eq:NSE3D} with the
optimal {asymmetric} initial conditions $\tuET$ obtained by solving
the finite-time optimization problem \ref{pb:maxET} for $\E_0 = 100,
500, 1000$ and with the corresponding optimal optimization time
intervals $\tTE$. {These quantities, defined as}
\begin{subequations}
\label{eq:cnorms}
\begin{align}
\|\u(t)\|_{L^3} & := \left( \int_{\Omega} |\u(t, \x)|^3  \, d\x \right)^{1/3}, 
\label{eq:L3} \\
{\|\u(t)\|_{\dot{H}^{1/2}}} & := {\left[ \sum_{\k \in \ZZ^3} |\k| \left|\left[\widehat{{\mathbf{u}}}(t)\right]_{\k}\right|^2 \right]^{1/2}}
\label{eq:H05}
\end{align}
\end{subequations}
{are important, because they are critical norms} in the analysis
of the Navier-Stokes system \eqref{eq:NSE3D}
\citep{RobinsonRodrigoSadowski2016}. As we can see in figure
{\ref{fig:cnorms}(a)}, the $L^3$ norm decays monotonically in
each case. {On the other hand, the $H^{1/2}$ {seminorm}
  exhibits a significant transient growth, cf.~figure
  \ref{fig:cnorms}(b), and, interestingly, its maxima are in each case
  achieved at intermediate times $0 < t < \tTE$, i.e., before the
  enstrophy maximum is attained at $\tTE$.}

\bigskip

\noindent
\Bmp{\textwidth} \small
\href{https://www.youtube.com/watch?v=tHU6gRNrVdo}{{\sc{Movie 1.}} (available on-line)} Time
evolution of the vorticity components (blue) $\tomega_1$, (red)
$\tomega_2$ and (green) $\tomega_3$ in the solution of the
Navier-Stokes system \eqref{eq:NSE3D} with the optimal {asymmetric}
initial condition $\tuET$ obtained by solving the finite-time
optimization problem \ref{pb:maxET} for the initial enstrophy $\E_0 =
500$ and $\tTE = 0.17$, cf.~figure \ref{fig:evolution}.  The animation
covers the time interval $[0,\tTE]$.  \Emp

\bigskip

\noindent
\Bmp{\textwidth} \small
\href{https://www.youtube.com/watch?v=gsUhK7eifhU}{{\sc{Movie 2.}} (available on-line)} Time
evolution of the vorticity components (blue) $\tomega_1$, (red)
$\tomega_2$ and (green) $\tomega_3$ in the solution of the
Navier-Stokes system \eqref{eq:NSE3D} with the optimal {asymmetric}
initial condition $\tuET$ obtained by solving the finite-time
optimization problem \ref{pb:maxET} for the initial enstrophy $\E_0 =
300$ and $\tTE = 0.21$.  The animation covers the time interval
$[0,\tTE]$.  \Emp

\bigskip

\noindent
\Bmp{\textwidth} \small
\href{https://www.youtube.com/watch?v=o-Z1iEoJf5E}{{\sc{Movie 3.}} (available on-line)} Time
evolution of the vorticity components (blue) $\tomega_1$, (red)
$\tomega_2$ and (green) $\tomega_3$ in the solution of the
Navier-Stokes system \eqref{eq:NSE3D} with the optimal {asymmetric}
initial condition $\tuET$ obtained by solving the finite-time
optimization problem \ref{pb:maxET} for the initial enstrophy $\E_0 =
1000$ and $\tTE = 0.12$.  The animation covers the time interval
$[0,\tTE]$.  \Emp

\section{Discussion and Conclusions}
\label{sec:final}

In this study we have considered the question of the largest possible
growth of enstrophy in finite time in 3D Navier-Stokes flows starting
from initial data with enstrophy $\E_0$. This problem is motivated by
the open question concerning the global-in-time existence of classical
(smooth) solutions of the 3D Navier-Stokes system \citep{d09}. One of
the landmark results in this context is the conditional regularity
result due to \citet{ft89} asserting that solutions to the
Navier-Stokes system \eqref{eq:NSE3D} remain smooth and satisfy this
system in the classical sense as long as the enstrophy remains finite.
To probe this condition we have considered a family of optimization
problems for the Navier-Stokes system \eqref{eq:NSE3D}, cf.~Problem
\ref{pb:maxET}, in which initial data $\u_0$ with prescribed enstrophy
$\E_0$ is sought such that the enstrophy at the given time $T$ is
maximized. Thus, {in each case the problem is solved on a fixed
  time domain and there are two parameters, $\E_0$ and $T$. For each
  value of $\E_0$ the optimal time window $\tTE$ is then determined by
  solving problem \ref{pb:maxET} for several different values of $T$,
  cf.~figure \ref{fig:maxEt_vs_T}.  In principle, this last step could
  be eliminated by finding the optimal time window $\tTE$ as a part of
  the solution of a suitably modified optimization problem
  \ref{pb:maxET} in which the final enstrophy would be {\em
    simultaneously} maximized with respect to the initial condition
  $\u_0$ and the length $T$ of the time window.} However, such a
modified optimization problem would then be of the free-boundary type,
because its time domain $[0,T]$ would {not be a priori known,}
leading to significant complications in its numerical solution.
Therefore, it is preferable to solve Problem \ref{pb:maxET} for
different values of the two parameters $\E_0$ and $T$, which is
facilitated by the continuation approach. For given values of $\E_0$
and $T$, Problem \ref{pb:maxET} is solved computationally with a
state-of-the-art adjoint-based gradient ascent technique. Our approach
is formulated in the continuous (``optimize-then-discretize'') setting
with gradients defined in a suitable Sobolev space,
cf.~\eqref{eq:riesz} and \eqref{eq:gradH1}, which allows us to ensure
that the optimal initial {conditions possess} the required minimum
regularity.  The governing and adjoint systems, \eqref{eq:NSE3D} and
\eqref{eq:aNSE3D}, are solved efficiently with a massively parallel
pseudo-spectral approach.  Being based on the first-order optimality
conditions, this gradient optimization approach {makes it possible} to
find local maximizers only and it is not possible to guarantee that
these maximizers are global.  In order to partially address this
issue, in addition to the continuation approach described in
\$\ref{sec:optim}, cf.~Algorithm \ref{alg:optimAlg}, we have also
undertaken an extensive search for other maximizing branches using
suitably randomized initial guesses $\u^0$ in \eqref{eq:desc} which
however did not yield any maximizers distinct from the {two branches,
  a symmetric and an asymmetric one,} already reported in \S
\ref{sec:results}.  {Nevertheless, the existence of other branches of
  maximizers cannot be ruled out.}

For every considered value of the initial enstrophy $\E_0$ {on both
  the symmetric and asymmetric branch} there exists a well-defined
time $\tTE$ such that the optimized growth of the enstrophy at this
time is maximal. The structure of the optimal initial condition
producing this growth {is quite different on the symmetric and the
  asymmetric branch, and} changes as the initial enstrophy increases,
cf.~figure \ref{fig:tuE0T}.  {In the limit of large $\E_0$ the
  asymmetric branch dominates in terms of the maximum enstrophy growth
  and the corresponding initial conditions have} the form of three
perpendicular pairs of anti-parallel vortex tubes, cf.~figure \ref
{fig:tubes}. {Thus, our subsequent analysis has focused on this family
  of initial data.} Interestingly, as is evident from figure
\ref{fig:spec_u0_T_E0_100}, these optimal initial conditions $\tuET$
are smoother than the solutions $\tuE$ of the instantaneous
optimization optimization problem \ref{pb:maxdEdt_E} obtained for the
same value of $\E_0$ (to be precise, the velocity fields are
real-analytic in both cases, but in the latter case the {energy
  spectra} decay at a slower, though still exponential, rate).
Consequently, solution of the finite-time optimization problem
\ref{pb:maxET} requires a lower numerical resolution than the
corresponding instantaneous problem \ref{pb:maxdEdt_E} and, as a
result, we have been able to solve the finite-time problem for values
of the initial enstrophy $\E_0$ an order of magnitude higher than it
was possible for the instantaneous maximization problem in
\citet{ap16}.  Further on this note, we also observe that the extremal
flow trajectories originating from the optimal initial conditions
$\tuET$ involve velocity fields which become even smoother at early
stages of their evolution, cf.~figures \ref{fig:spectra_cascade} and
\ref{fig:evolution}, before developing small-scale components at later
stages when the maximum enstrophy is achieved. In other words, in
these optimal flow evolutions the energy flux is initially from small
to large scales, and then in the opposite direction near the end of
the optimization interval $[0,\tTE]$.  This behavior can be understood
by considering system \eqref{eq:dKdtdEdt} from which it is evident
that large enstrophy implies rapid energy dissipation. At the same
time it is known that a 3D Navier-Stokes flow may not generate
significant enstrophy unless its energy is sufficiently large
\citep{ap16}. Thus, in order to maximize its enstrophy at some large
time $T$, an optimal flow seeks to reduce its production at early
stages in order to conserve energy while rearranging itself
strategically such that a {sequence of bursts} of energy to small
scales will produce maximum enstrophy at the final time $T$.

The main finding of the present study is that the worst-case
enstrophy-maximizing Navier-Stokes flow evolutions constructed by
solving a suitable optimization problem always lead to a finite only
growth of enstrophy. This suggests that unbounded growth of enstrophy
allowed by estimate \eqref{eq:Et_estimate_E0} cannot, in fact, be
realized and, in the light of the conditional regularity result
\eqref{eq:RegCrit_FoiasTemam}, even {under} the extremal flow
evolution there is no evidence for singularity formation in finite
time. This also suggests the possibility of improving estimate
\eqref{eq:Et_estimate_E0} {with the power-law relation
  \eqref{eq:maxEt_vs_E0} serving as a potential target}.  As a matter
of course, the validity of these findings is restricted by the
possibility that we may not have found true global maximizers of
Problem \ref{pb:maxET}. To be more specific, the maximum enstrophy
produced in flows with initial data with enstrophy $\E_0$ was found to
scale in proportion to $\E_0^{3/2}$, {cf.~\eqref{eq:maxEt_vs_E0},}
which is the same power-law relation as observed by \citet{ap11a} in
solutions of an analogous finite-time optimization problem for the 1D
Burgers equation. This scaling can be in fact predicted based on
dimensional analysis. Noting that the {largest} flow structures have
characteristic dimension comparable to the size of the domain $L$,
{cf.~figures \ref{fig:tuE0T} and \ref{fig:evolution}}, there are three
physical parameters defining the problem, namely, $\E_0$, $L$ and the
kinematic viscosity $\nu$. The only nontrivial way to combine them
gives $\E_T \sim \left(L / \nu^2 \right)^{1/2} \E_0^{3/2}$. In
addition, we can deduce that $T \sim L^{3/2} \E_0^{-1/2}$ which
represents the scaling for $\tTE$ observed in figure
\ref{fig:Tmax_vs_E0}, {cf.~\eqref{eq:Tmax_vs_E0}}.  {In this context
  it would be interesting to know what form the optimal initial
  conditions would take if Problem \ref{pb:maxET} were solved on an
  unbounded domain $\Omega = \RR^3$. Since for a fixed initial
  enstrophy $\E_0$ the optimal initial conditions discussed in \S
  \ref{sec:results} would vanish in the limit $L \rightarrow \infty$,
  the solutions of this problem would likely have a completely
  different form.}  We add that in the instantaneous optimization
problem in which the characteristic dimension of the flow structures
{\em does not} scale with $L$, dimensional analysis also correctly
predicts that $d\E / dt \sim \E^3 / \nu^3$,
cf.~\eqref{eq:dEdt_estimate_E} \citep{ld08}.

Two distinct branches of maximizing initial conditions were found with
the one {consisting of} asymmetric states dominating in terms of
the maximum growth of enstrophy in the limit of large $\E_0$. The
corresponding initial conditions have the form of three perpendicular
pairs of anti-parallel vortex tubes and the flow evolution resulting
from this initial data is quite complex, cf.~figure
\ref{fig:evolution}, {involving} an initial flux of energy from
small to large scales followed by a sequence of reconnection events,
cf.~figure \ref{fig:reconnection}.  At its final stages the flow
evolution has the spatio-temporal complexity of turbulent flows,
although there is no evidence of a Kolmogorov-type -5/3 energy
spectrum in figure \ref{fig:spec_u0_T_E0_100}. This can be explained
by the fact that we consider an initial-value problem for system
\eqref{eq:NSE3D} and no statistically steady state is attained during
the evolution. More work is still needed to understand in detail the
physical mechanisms responsible for the extreme flow evolutions
illustrated in figure \ref{fig:evolution}, for example, in terms of
changes of the topology of vortex lines. {In particular, the fact
  that helicity remains zero during all flow evolutions starting from
  the optimal initial conditions $\tuEtT$ implies the presence of some
  symmetries in the velocity field.} We add that no evidence for
reconnection was found in the flow evolutions starting from the
optimal initial conditions on the symmetric branch.

{In involving pairs of anti-parallel vortex tubes the optimal
  initial conditions $\tuEtT$ we found resemble the configurations
  investigated in several earlier studies of extreme flow behavior
  \citep{MelanderHussain1989,k93,k13b,k13,Kerr2018}. In particular,
  \citet{opc12} used two perpendicular pairs of anti-parallel vortex
  tubes (referred to as ``dipoles'' in that study). Although in the
  case of Navier-Stokes flows none of these studies showed evidence of
  singularity formation in finite time, they all reported various
  forms of viscous vortex reconnection. More precisely, while the
  enstrophy would increase during the reconnection events, it would in
  all cases remain bounded. Based on three values of the Reynolds
  number defined as\footnote{{We note that this definition of the
      Reynolds number is different from the one used by
      \citet{opc12}.}} $Re := \left( \K(\u_0)\, \E_0 \right)^{1/4} /
  \nu$, the results reported by \citet{opc12} exhibit a linear
  dependence of the maximum attained enstrophy on the Reynolds number,
  i.e., $\max_t \E(t) \sim Re$. For comparison, noting that in our
  case {$\K(\tuEtT) \sim \E_0$ and therefore} $Re \sim
  \E_0^{1/2}$, the power-law relation \eqref{eq:maxEt_vs_E0} takes the
  form $\max_{T > 0} \E_T(\tuET) \sim Re^3$ when expressed in terms of
  the Reynolds number defined above.  Thus, even though the range of
  the Reynolds numbers explored in our study was approximately two
  orders of magnitude lower than in \citet{opc12}, in the flows
  studied here the maximum attained enstrophy grows much more rapidly
  when the Reynolds number is increased. We add that the ``trefoil''
  initial condition which recently received attention \citep{Kerr2018}
  is in fact fundamentally different from the optimal initial data
  studied here, because it is meant to be defined on an unbounded,
  rather than periodic, domain. Moreover, unlike the optimal initial
  conditions $\tuEtT$ and the corresponding flows, it is characterized
  by nonzero helicity.}

{Our results indicate that even in the worst-case scenario the
  nonlinear mechanisms of vortex stretching are significantly
  depleted, This observation is consistent with the findings of
  \cite{dggkpv13,gdgkpv14} who studied the behavior of suitably-scaled
  higher-order Lebesgue norms of vorticity in a number of different
  numerical simulations of turbulent Navier-Stokes flows, both forced
  and decaying. They provided evidence for a significant depletion of
  the rate of growth of enstrophy as compared to
  \eqref{eq:dEdt_estimate_KE}. More precisely, they demonstrated that
  in their simulations enstrophy amplification tends to occur at the
  rate $d\E/dt \sim \E^{\alpha}$ with $\alpha \in [1.575,1.75]$ (we
  recall from our discussion in \S \ref{sec:bounds} that $\alpha < 2$
  implies a regular behavior of the flow). These findings are also
  consistent with our observations, cf.~figure \ref{fig:dEvsE},
  indicting that while in the flow evolutions corresponding to the
  optimal initial data $\tuEtT$ the enstrophy may occasionally be
  amplified at the much higher rate, the sustained rate is
  proportional to $\E^{\alpha}$ with $0 < \alpha < 2$.}

The results summarized above represent a final stage of the research
program outlined in Table \ref{tab:estimates} which aimed to
characterize the largest possible growth of enstrophy and
enstrophy-like quantities in 1D Burgers and 2D \& 3D Navier-Stokes
flows. In particular, we sought to understand whether the sharpest
estimates {on the growth of these quantities} obtained using
{energy-type} methods can be realized in actual flows. Somewhat
paradoxically, the situation concerning 1D Burgers flows, where the
best finite-time estimate {obtained using energy methods} was found
{\em not} to be sharp \citep{ap11a}, is less satisfactory than in the
case of 2D Navier-Stokes flows where the estimates for both the
instantaneous and finite-time growth of palinstrophy were shown to be
sharp (and, interestingly, they are realized by the same field
\citep{ap13a,ayala_doering_simon_2018}). Moving forward, a natural
next step is to consider other conditional regularity results
complementary to condition \eqref{eq:RegCrit_FoiasTemam}. More
specifically, we will probe the family of the
Ladyzhenskaya-Prodi-Serrin conditions asserting that Navier-Stokes
flows $\u(t)$ are smooth and satisfy system \eqref{eq:NSE3D} in the
classical sense provided that \citep{KisLad57,Prodi1959,Serrin1962}
\begin{equation}
\u \in L^p([0,T];L^q(\Omega)), \quad 2/p+3/q \le 1, \quad q > 3.
\label{eq:LPS}
\end{equation}
These conditions were recently generalized by \citet{Gibbon2018} to
include norms of the derivatives of the velocity field.  {The
  main technical difficulty in testing conditions \eqref{eq:LPS} is
  that some of the corresponding variational optimization problems
  will be formulated on function spaces without the Hilbert structure
  and will therefore require more specialized approaches
  \citep{protas2008}.  Moreover, some of these optimization problems
  may also be nonsmooth.}  In addition, there are questions concerning
potential finite-time singularity formation in inviscid Euler flows
which can also be framed in terms of the extreme growth of
enstrophy-like quantities. We intend to investigate both these
questions in the near future.

\section*{Acknowledgments}

The authors wish to express sincere thanks to Diego Ayala for his help
with software implementation of the approach described in Section
\ref{sec:numer}, and to Miguel Bustamante, {Sergei Chernyshenko,
  Charles Doering, David Goluskin, Luo Guo and Keith Moffatt} for
enlightening discussions. {They also thank Paolo Orlandi for sharing
  data from his paper \citep{opc12}.}  DY was partially supported
through a Fields-Ontario Post-Doctoral Fellowship and BP acknowledges
the support through an NSERC (Canada) Discovery Grant. Computational
resources were provided by Compute Canada under its Resource
Allocation Competition.



\end{document}